\pdfoutput=1 

\documentclass[%
aip, 
numerical,
amsmath,
amssymb,
reprint,
shortbibliography
]{revtex4-1}

\usepackage[utf8]{inputenc}    
\usepackage[T1]{fontenc}       
\usepackage[abbreviations]{glossaries-extra}
\usepackage{mathptmx}          
\usepackage{amssymb}
\usepackage{amsmath}
\usepackage{amsthm}
\usepackage{pgfmath}\pgfmathsetseed{12345}
\usepackage{bm}                
\usepackage{siunitx}           
\usepackage{CJKutf8}           
\usepackage{booktabs}          
\usepackage{tabularx}          
\usepackage{longtable}         
\usepackage{array}             
\usepackage{makecell}          
\usepackage{multirow}          
\usepackage{threeparttable}    
\usepackage{dcolumn}           
\usepackage{graphicx}          
\usepackage[export]{adjustbox} 
\usepackage{float}             
\usepackage{subfigure}         
\usepackage{tikz}              
\usepackage{pgfplots}          
\usetikzlibrary{arrows,decorations.markings,shapes,positioning,arrows.meta}

\usepackage{hyperref}          
\usepackage{cleveref}          
\usepackage{color}             
\usepackage[dvipsnames, table]{xcolor} 
\usepackage{soul}              
\usepackage[most]{tcolorbox}   
\usepackage{lineno}
\usepackage{enumitem}          
\usepackage{everypage}         
\usepackage{etoolbox}          
\usepackage{comment}           
\usepackage{blindtext}         
\usepackage{todonotes}         
\usepackage{cancel}

\makeatletter \@ifclassloaded{elsarticle}
{ 
    \bibliographystyle{elsarticle-num-names}
    \biboptions{sort&compress}
}{ 
    \usepackage[sort&compress]{natbib}
} \makeatother

\apptocmd{\thebibliography}{\setlength{\itemsep}{0pt}}{}{}

\usepackage{acro}
\acsetup{
  list/display = used,
  list/sort = true,
  list/template = longtable, 
  list/heading = section*,
  list/name = Nomenclature,
  list/preamble = List of acronyms used in this paper:,
}

\DeclareAcronym{HPC}{
  short = HPC ,
  long  = high performance computing 
}
\DeclareAcronym{NS}{
  short = NS ,
  long  = Navier--Stokes 
}
\DeclareAcronym{NSF}{
  short = NSF ,
  long  = Navier--Stokes--Fourier 
}
\DeclareAcronym{NSFK}{
  short = NSFK ,
  long  = Navier--Stokes--Fourier--Korteweg 
}
\DeclareAcronym{NSFE}{
  short = NSFE,
  long =  Navier--Stokes--Fourier equation,
  short-plural-form = NSFEs,
  long-plural-form = Navier--Stokes--Fourier equations
}
\DeclareAcronym{PDE}{
  short = PDE,
  long = partial differential equation,
  short-plural-form = PDEs,
  long-plural-form = partial differential equations
}
\DeclareAcronym{EoS}{
  short = EoS,
  long = equation of state,
  short-plural-form = EoS,
  long-plural-form = equations of states
}
\DeclareAcronym{w.r.t.}{
  short = w.r.t. ,
  long  = with respect to 
}
\DeclareAcronym{BE}{
  short = BE ,
  long  = Boltzmann transport equation 
}
\DeclareAcronym{BGK}{
  short = BGK ,
  long  = Bhatnagar--Gross--Krook
}
\DeclareAcronym{BGK-BE}{
  short = BGK-BE ,
  long  = Bhatnagar-Gross-Krook--Boltzmann equation 
}
\DeclareAcronym{MB}{
  short = MB ,
  long  = Maxwell--Boltzmann 
}
\DeclareAcronym{GH}{
  short = GH ,
  long  = Grad--Hermite
}
\DeclareAcronym{DF}{
  short = DF ,
  long  = distribution function
}
\DeclareAcronym{PDF}{
  short = PDF ,
  long  = probability distribution function
}
\DeclareAcronym{EDF}{
  short = EDF ,
  long  = equilibrium distribution function
}
\DeclareAcronym{QE}{
  short = QE ,
  long  = quasi-equilibrium
}
\DeclareAcronym{SE}{
  short = SE ,
  long  = shifted-equilibrium
}
\DeclareAcronym{QEDF}{
  short = QEDF ,
  long  = quasi-equilibrium distribution function
}
\DeclareAcronym{SEDF}{
  short = SEDF ,
  long  = shifted-equilibrium distribution function
}
\DeclareAcronym{DDF}{
  short = DDF ,
  long  = double-distribution function
}
\DeclareAcronym{TRT}{
  short = TRT ,
  long  = two-relaxation-time
}
\DeclareAcronym{CE}{
  short = CE ,
  long  = Chapman--Enskog
}
\DeclareAcronym{CFD}{
  short = CFD ,
  long  = computational fluid dynamics
}
\DeclareAcronym{CFL}{
short = CFL ,
long = Courant--Friedrichs--Lewy
}
\DeclareAcronym{FV}{
  short = FV ,
  long  = finite volume
}
\DeclareAcronym{FD}{
  short = FD ,
  long  = finite difference
}
\DeclareAcronym{AMR}{
  short = AMR ,
  long  = adaptive mesh refinement
}
\DeclareAcronym{AAR}{
  short = AAR ,
  long  = adaptive algorithm refinement
}
\DeclareAcronym{AMAR}{
  short = AMAR ,
  long  = adaptive mesh and algorithm refinement
}
\DeclareAcronym{SAMR}{
  short = SAMR ,
  long  = block-structured adaptive mesh refinement
}
\DeclareAcronym{LBM}{
  short = LBM ,
  long  = lattice Boltzmann method
}
\DeclareAcronym{LBE}{
  short = LBE ,
  long  = lattice Boltzmann equation
}
\DeclareAcronym{LB}{
  short = LB ,
  long  = lattice Boltzmann
}
\DeclareAcronym{DVBM}{
  short = DVBM ,
  long  = discrete velocity Boltzmann method
}
\DeclareAcronym{UGKS}{
  short = UGKS ,
  long  = unified gas kinetic scheme
}
\DeclareAcronym{DUGKS}{
  short = DUGKS ,
  long  = discrete unified gas kinetic scheme
}
\DeclareAcronym{ELBM}{
  short = ELBM ,
  long  = entropic lattice Boltzmann method
}
\DeclareAcronym{LBGK}{
  short = LBGK ,
  long  = lattice Boltzmann Bhatnagar--Gross--Krook
}
\DeclareAcronym{IC}{
  short = IC ,
  long  = initial condition
}
\DeclareAcronym{BC}{
  short = BC ,
  long  = boundary condition
}
\DeclareAcronym{MUSCL}{
  short = MUSCL ,
  long  = monotonic upstream-centered scheme for conservation Laws
}
\DeclareAcronym{KL}{
  short = KL ,
  long  = Kullback-Leibler
}
\DeclareAcronym{2D}{
  short = 2D ,
  long  = two-dimensional
}
\DeclareAcronym{1D}{
  short = 1D ,
  long  = one-dimensional
}
\DeclareAcronym{MRT}{
  short = MRT ,
  long  = multiple relaxation time
}
\DeclareAcronym{MRT-LB}{
  short = MRT-LB ,
  long  = multiple relaxation time lattice Boltzmann
}
\DeclareAcronym{PonD}{
  short = PonD ,
  long  = particles on demand
}

\newcommand{\tbd}[1]{{\color{orange}{#1}}}

\newcommand{\mcirc}{\mathbin{\raisebox{-0.25ex}{\scalebox{1.8}{$\circ$}}}}
\newcommand{\mtriangledown}{\mathbin{\raisebox{-0.0ex}{\scalebox{1.2}{$\triangledown$}}}}

\tikzset{
  block_m/.style ={rectangle, draw=black, thick, fill=white,
    text width=14em, text centered, minimum height=3em},
  block_s/.style ={rectangle, draw=black, thick, fill=white,
    text width=8em, text centered, minimum height=3em},
  >={Stealth}
}
\tikzset{
  |-|/.style={
    to path={
      -- ++(0,-0.5)      
      -- ++(\x,0)        
      -- ++(0,0.5)       
      -- (\tikztotarget) 
    }
  }
}
\tikzset{
  -|-/.style args={#1 and #2}{
    to path={
      -- ++(#1,0) -- ++(0,\y) -- ++(#2,0) -- (\tikztotarget)
    }
  }
}

\begin{document}


\title[]{
    Consistent lattice Boltzmann model with body force and heat source \\for compressible flows of generic fluids
}
\date{\today}
\affiliation{Computational Kinetics Group, Department of Mechanical and Process Engineering, ETH Zürich, 8092 Zürich, Switzerland}
\altaffiliation{Group Website: \url{https://ckg.ethz.ch/}}
\author{S.~Chatterjee}
\email{schatterje@ethz.ch} 
\affiliation{Computational Kinetics Group, Department of Mechanical and Process Engineering, ETH Zürich, 8092 Zürich, Switzerland}
\author{R.M.~Strässle}
\email{rubenst@ethz.ch}
\thanks{Corresponding author.}
\affiliation{Computational Kinetics Group, Department of Mechanical and Process Engineering, ETH Zürich, 8092 Zürich, Switzerland}
\author{S.A.~Hosseini}
\email{shosseini@ethz.ch}
\affiliation{Computational Kinetics Group, Department of Mechanical and Process Engineering, ETH Zürich, 8092 Zürich, Switzerland}
\author{I.V.~Karlin}
\email{ikarlin@ethz.ch} 
\affiliation{Computational Kinetics Group, Department of Mechanical and Process Engineering, ETH Zürich, 8092 Zürich, Switzerland}


\begin{abstract}

    This work presents a consistent formulation of body-force and heat-source terms within a double-distribution-function lattice Boltzmann framework for simulating compressible flows of generic fluids.
    The proposed approach extends a recently developed kinetic framework for non-ideal compressible fluid dynamics by incorporating external forcing and volumetric heating through shifted quasi-equilibrium states. 
    The standard lattice Boltzmann discretization is applied with a second-order accurate integration along characteristics and product-form equilibria on nearest-neighbor lattices  are employed, supplemented by necessary correction terms. This formulation ensures the consistent recovery of the Navier–Stokes–Fourier equations, including the Korteweg stress tensor, across arbitrary equations of state, while maintaining independently controllable thermodynamic and transport coefficients.
    The methodology is rigorously validated and applied for two different flow regimes, with ideal-gas compressible flows and non-ideal/multiphase compressible flows, respectively, where a broad hierarchy of benchmarks is employed, including non-classical shock tubes, thermal Couette flows, and force-driven Poiseuille and Womersley flows. The framework's ability to capture complex thermodynamic processes using forcing and heating is further demonstrated through Rayleigh and Fanno flows, Joule–Thomson effects, and throttling processes. Furthermore, the model is validated for multiphase phenomena, including interface consistency and liquid–vapor co-existence. The results demonstrate excellent agreement with analytical solutions and reference data, while spatio-temporal grid-refinement studies confirm the expected second-order accuracy of the scheme.
    This establishes the model as a robust and efficient foundation for simulating highly compressible flows of generic fluids in the presence of numerical and physical source terms, which are omnipresent in numerical methods, nature and technology.

    \vspace{0.5cm}
    
\end{abstract}
\maketitle

\section{Introduction\label{sec:Introduction}}

    Body forces and heat sources play a significant role in both natural phenomena and engineering applications. Examples include gravity-driven convection, such as Rayleigh–Bénard flows, rotational flows influenced by centrifugal and Coriolis forces, and magneto-fluid dynamics affected by electromagnetic forces, among others. In addition, proper implementation of forcing terms in discrete solvers for partial differential equations become singularly important when complex geometries are modeled using approaches such as the immersed boundary methods. Despite the seemingly straightforward treatment of body forces in hydrodynamic and even Boltzmann equations, their integration into discrete velocity Boltzmann models remains quite challenging with numerous proposed schemes, such as \cite{huang2011forcing, guo2002discrete, kupershtokh2004new, kupershtokh2009equations}. 
    
    While challenges still remain, introduction of body forces into the isothermal \ac{LB} equations targeting the incompressible flow limit have made significant progress in recent years and been used for complicated configurations such as non-ideal fluid simulations where non-ideal contributions to the pressure along with surface tension are introduced through a body force term, see these reviews for more details \cite{mohamad2010critical, chen2014critical, bawazeer2021critical, hosseini2023lattice}. On the other side, extension to compressible flows has been quite limited with only a few publications, e.g., \cite{li2023body, Li2024self}, treating the topic in the literature. 
    
    Moving to the compressible flow regime, a number of major hurdles are brought in as energy balance equations must also be taken into account. The extension of the \ac{LBM} and other discrete velocity Boltzmann-based numerical methods to high-Mach number flows has been a topic of intensive research over the past decade. Various strategies have been developed to extend the original isothermal weakly compressible \ac{LBM} to handle compressible flows with energy balance. These advances have been, in part, motivated by insights from kinetic theory, allowing for, e.g., proper thermal diffusivity in non-unity Prandtl number flows via models such as Shakhov \cite{Shakhov}, Holway \cite{HolwayES} or the \ac{QE} approach \cite{gorban1994, gorban2005invariant, QE_ansumali-2007}. Inspired by Rykov's original work ~\cite{rykov_model_1976}, the introduction of \ac{DDF} models~\cite{He_LBMcharacteristics} had considerable impact on the development of efficient compressible \ac{LBM}-based solvers capturing variable specific heat capacities. This approach splits the distribution function into two reduced distribution functions, where the second one carries some form of energy, e.g., the total or internal non-translational energy~\cite{KarlinTwoPop, ProbingDoubleDist2024, strässle2025consistent-compressible}.
    This idea has gained popularity in discrete kinetic theory of gases, particularly in combination with the Bhatnagar–Gross–Krook (BGK) and similar collision operators, to extend isothermal solvers to compressible flows. 
    The class of \ac{DDF}-\ac{LBM} solvers relying on standard lattices~\cite{Prasianakis2007, Saadat2019, hosseini2020compressibility, FengCorrection, li2012coupling} and hybrid solvers, modeling the energy balance equation via a FD or FV solver~\cite{FENG_FV_DBM} are practical illustrations. Models relying on higher-order lattices properly recovering the energy balance equation~\cite{ChikatamarlaMultispeed, FrapolliMultispeed, strässle2025a-fully-conservative, strässle2026localkineticsensorsadaptive} without a need for correction terms were also developed in that context. While all these approaches relied on static quadratures and as a result were limited in terms of maximum achievable Mach number, the introduction of dynamic quadratures, starting with shifted lattices~\cite{Frapolli2016b, Ali_shiftedStencils, Coreixas2020adaptivevelocity} and culminating with the \ac{PonD} method~\cite{PonD18}, opened the door for hypersonic flow simulations~\cite{sawant2022detonation, Kallikounis23, Bhaduria23, Ji24}.

    The use and extension of the class of thermodynamically consistent models to compressible non-ideal fluid dynamics is a largely under-explored area that can considerably impact research on non-ideal compressible fluid dynamics. Early attempts at developing \ac{LBM} for compressible non-ideal flows were documented in \cite{HeDollen_2002_foundationsMultiphase}. Since then, thermal multi-phase models based on the pseudo-potential approach have witnessed steady growth. However, the vast majority of the models and studies in the literature have been tailored to boiling applications, see for instance \cite{Saito2021Lattice}. To the authors' knowledge, the only documented attempts at modeling compressible non-ideal flows beyond evaporation are \cite{Vienne2024}, where a hybrid \ac{LB} scheme with a \ac{FV} discretization of the energy equation was proposed, and \cite{Reyhanian20, Reyhanian21}, where the authors demonstrated a numerical model based on the \ac{PonD} realization of the semi-Lagrangian \ac{LBM} \cite{PonD18}. Recently, this gap in the literature was addressed with a novel kinetic model and its \ac{LB} realization for non-ideal fluids in the compressible regime \cite{Karlin2025Practical, hosseini2026latticeboltzmannmodelnonideal}. To retain the main advantages of the \ac{LBM}, the proposed model relied on classical nearest-neighbor lattices, taking advantage of a second distribution function for the energy balance equation \cite{KarlinTwoPop, saadat2021extended}. The kinetic model relies on a novel set of BGK-like collision operators with local- and shifted-equilibrium attractors ensuring recovery of target hydrodynamics. For instance, independent control over the bulk viscosity is guaranteed through the shifted equilibrium, which is a critical point, as the bulk viscosity dictated by the BGK structure can take on negative values in the hydrodynamic limit for equilibria with pressures other than the ideal gas pressure, see \cite{Hosseini2022Towards, hosseini2023lattice}. 

    In the present contribution we treat the next fundamental step in developing solvers for highly compressible flows of generic fluids, including non-ideal fluids, namely development of a consistent discrete velocity realization of body force and heat source terms. As discussed above, this is of utmost importance for body forces of physical motivation, e.g., gravity-driven flows, and of numerical motivation, e.g., forces and heat fluxes in the context of the immersed boundary method for simulations in complex geometries.    
    The topic of body force and heat source for compressible \ac{LB} simulations has long been neglected, until only recently in \cite{li2023body, Li2024self}. Since these proposed models are derived from the Boltzmann--BGK equation for monatomic gases, their applicability is restricted to fluids with a fixed specific heat ratio. Moreover, the works consider only the incorporation of body forces and do not address the modeling of heat sources.
    Here we will present a consistent approach to introduce body forces and heat sources in the context of a double distribution function lattice Boltzmann realization for compressible fluid dynamics with arbitrary specific heat, bulk viscosity and Prandtl number, which, being based on \cite{hosseini2026latticeboltzmannmodelnonideal}, is capable of capturing non-ideal and multiphase flows via an appropriate equation of state and body force for the Korteweg stresses.
    
    The paper is organized as follows:
    In Sec.~\ref{sec:Methodology}, the methodology of incorporating consistent body forces and heat sources is outlined, starting with a summary of the kinetic model, discretization in space-time and phase-space, and a discussion of the consistent hydrodynamic limit for compressible flows of generic fluids as well as the algorithm and implementations.
    In Sec.~\ref{sec:Applications}, the resulting lattice Boltzmann realization is then validated and applied to two different flow regimes, namely ideal-gas compressible flows and non-ideal and multiphase compressible flows, respectively.
    Lastly, in Sec.~\ref{sec:conslusions}, conclusions and outlooks are provided. 

\section{Methodology\label{sec:Methodology}}

    \subsection{Kinetic model\label{sec:Methodology:kinmodel}}
In the present work, we propose an extension of the kinetic model introduced in \cite{hosseini2026latticeboltzmannmodelnonideal}. It is a double distribution function kinetic model where
the $f$-distribution function defines the fluid density and momentum, while the $g$-distribution defines the bulk energy,
\begin{align}
    &\int \{m,m\bm{v}\} f d\bm{v}= \{\rho, \rho\bm{u}\},\label{eq:fcons}\\
    &\int mg d\bm{v}= \rho E,\label{eq:gcons}
\end{align}
where $m$ is the mass of the particle with a velocity $\bm{v}$.
The kinetic model is defined by the coupled kinetic equations,
\begin{align}
    &\partial_t f + \bm{v}\cdot\bm{\nabla} f = \Omega_{f},\label{eq:f_boltz}\\
    &\partial_t g + \bm{v}\cdot\bm{\nabla} g = \Omega_{g}\label{eq:g_boltz},
\end{align}
where the collision terms on the right-hand side are sought in the following form
\begin{align}
    & \Omega_{f} = \frac{1}{\tau}\left(f^{\rm eq} - f\right) + \frac{1}{\lambda}\left(f_\lambda^\star-f^{\rm eq}\right),\label{eq:f_coll}\\
    & \Omega_{g} = \frac{1}{\tau}\left(g^{\rm eq} - g\right) + \frac{1}{\lambda}\left(g_\lambda^\star-g^{\rm eq}\right).\label{eq:g_coll}
\end{align}
Here, the pair $\{f^{\rm eq}, g^{\rm eq}\}$ represents a local equilibrium attractor, while $\{f_\lambda^\star, g_\lambda^\star\}$ represents an intermediate quasi-equilibrium attractor. Furthermore, $\tau$ and $\lambda$ are the corresponding relaxation times.

Proceeding to describe the equilibrium distribution functions, we introduce the parameter $\theta$, which describes a reference temperature. This parameter fixes the width of the Maxwellian equilibria independent of the thermodynamic internal energy. The equilibrium associated with mass and momentum is then written as
\begin{equation}\label{eq:f_eq_cont}
    f^{\rm eq}(\rho,\bm{u},\theta)
    =
    \frac{\rho}{m{(2\pi \theta)}^{D/2}}
    \exp\left[-\frac{|\bm{v}-\bm{u}|^2}{2\theta}\right],
\end{equation}
where $\rho$ and the $D$ (number of dimensions) components of $\bm{u}$ are fixed by the local conservation constraints \eqref{eq:fcons}, while $\theta$ is specified independently. Thus, $f^{\rm eq}$ forms a ${D+2}$-parameter family.
The equilibrium distribution associated with the energy distribution, $g$, is defined as
\begin{equation}\label{eq:geq_cont}
    g^{\rm eq}(\rho,\bm{u},T,\theta)
    =
    \left(
    \frac{\bm{v}^2}{2}
    +e(\rho,T)
    -\frac{D\theta}{2}
    \right)
    f^{\rm eq}(\rho,\bm{u},\theta),
\end{equation}
so that the thermodynamic temperature $T$ enters through the specific internal energy, whereas $\theta$ enters through the Maxwellian kernel. 
Note that the specific internal-energy $e(v,T)=e(\rho,T)$, with the specific volume $v=1/\rho$, is defined by the familiar thermodynamic relation for its differential,
\begin{align}
de
&=c_v\,dT+\left[T\left(\dfrac{\partial P}{\partial T}\right)_v-P\right]{dv}\label{eq:de_v}
\\ 
&=c_vdT-\left[T\left(\dfrac{\partial P}{\partial T}\right)_\rho-P\right]\frac{d\rho}{\rho^2}\label{eq:de_rho},
\end{align}
where $P(\rho,T)=\left.P(v,T)\right|_{v={1}/{\rho}}$ is the thermodynamic pressure specified by the generic equation of state and $c_v$ is the specific heat at constant volume
\begin{equation}
    c_v=\left(\dfrac{\partial e}{\partial T}\right)_v.
\end{equation}
The equilibrium distributions \eqref{eq:f_eq_cont} and \eqref{eq:geq_cont} satisfy the conservation laws of mass, momentum, and bulk energy,
\begin{align}
    &\int \{m,m\bm{v}\} f^{\rm eq}\, d\bm{v}= \{\rho, \rho\bm{u}\},\label{eq:fcons_app}\\
    &\int mg^{\rm eq}\, d\bm{v}= \rho E.\label{eq:gcons_app}
\end{align}

The quasi-equilibrium state $\{f_\lambda^\star, g_\lambda^\star\}$ is constructed using shifted values of the flow velocity, reference temperature, and thermodynamic temperature,
\begin{align}
    \bm{u}^\star_\lambda
    &= \bm{u}+\lambda\frac{\bm{F}}{\rho}
    ,\label{eq:sh_vel}\\
    \theta^\star_\lambda
    &= \theta + \lambda\alpha \theta
    \left(\bm{\nabla}\cdot\bm{u}\right) 
        { 
        + \lambda (\gamma-1) \frac{Q}{\rho}
        }
    ,\label{eq:sh_temp} \\
    T_\lambda^\star
    &= T 
        { 
        + \lambda \frac{Q}{\rho c_v}} - \lambda^2\frac{\bm{F}\cdot\bm{F}}{2\rho^{2} c_v
        } 
    ,\label{eq:sh_T}
\end{align}
where $\bm{F}$ and $Q$ denote the body force and heat source, and the non-dimensional parameters $\alpha$  and $\gamma$ shall be specified below. 
The quasi-equilibrium distribution $f_\lambda^\star$ is a shifted Maxwellian,
\begin{equation}\label{eq:fstar_cont}
    f_\lambda^\star(\rho,\bm{u}_\lambda^\star,\theta_\lambda^\star)
    =
    f^{\rm eq}(\rho,\bm{u}_\lambda^\star,\theta_\lambda^\star).
\end{equation}
The corresponding quasi-equilibrium for the energy population is defined as
\begin{multline}\label{eq:gstar_cont}
    g^\star_\lambda(\rho,\bm{u},\theta,\bm{u}_\lambda^\star, T_\lambda^\star,\theta_\lambda^\star)
    =\\ \Bigg( \frac{\bm{v}^2}{2} + e(\rho, T_\lambda^\star) - \frac{D\theta_\lambda^\star}{2}
    + \frac{\bm{q}^{\rm c}_\lambda\cdot(\bm{v}-
        \bm{u} 
    )}{\rho \theta}
    \Bigg)
    f^{\rm eq}(\rho,\bm{u}_\lambda^\star,\theta_\lambda^\star).
\end{multline}
The vector $\bm{q}^{\rm c}_\lambda$ that appears in the last term is defined by
\begin{equation}\label{eq:qc}
    \bm{q}^{\rm c}_\lambda = \lambda P\left(\bm{\nabla}h - \frac{k}{\mu}\bm{\nabla}T\right),
\end{equation}
where $h$ is the specific enthalpy
\begin{equation}\label{eq:enthalpy}
h = e + \frac{P}{\rho},
\end{equation}
$\mu$ is the dynamic shear viscosity and $k$ the thermal conductivity.

Upon performing a Chapman–Enskog analysis to obtain the hydrodynamic limit of the kinetic equations, see \cite{hosseini_2025_compressiblenonideal}, the reference temperature $\theta$ can be found as
\begin{equation}
    \label{eq:theta}
    \theta=\frac{P}{\rho}.
\end{equation}
Recovery of the Navier--Stokes viscous stress tensor requires the relaxation time $\tau$ to be
\begin{equation}
    \tau=\frac{\mu}{P},
\end{equation}
and the parameters $\alpha$ and $\gamma$ to have the form
\begin{equation}
    \label{eq:corr_bulk}
    \alpha=\left(\frac{D+2}{D}-\frac{\rho c_s^2}{P}-\frac{\eta}{\mu}\right),
\end{equation}
{
\begin{equation}
    \label{eq:corr_heat_source}
    \gamma = 1 + \frac{1}{\rho c_v} \left(\frac{\partial P}{\partial T}\right)_\rho,
\end{equation}
}{
where $c_s$ is the speed of sound, given by the thermodynamic relation
\begin{equation}
    c_s^2 = \left(\frac{\partial P}{\partial \rho}\right)_T + \frac{T}{\rho^2c_v}\left(\frac{\partial P}{\partial T}\right)_\rho^2,
\end{equation}
and $\eta$ denotes the dynamic bulk viscosity.
For an ideal gas, the parameter $\gamma$ simplifies to the ratio of specific heats at constant pressure, $c_P$, and constant volume, $c_v$.}
Note that the relaxation time $\lambda$ does not affect the hydrodynamic limit, thus remaining a free parameter that can be conveniently specified in the subsequent discretization.

\subsection{Second-order-in-time discretization\label{sec:int_chars}}
We follow a procedure first introduced by \cite{He_LBMcharacteristics} to discretize the kinetic models introduced here,
\begin{align}
    \partial_t f &+\bm{v}\cdot\bm{\nabla}f=-\frac{1}{\tau}(f-f^{\rm eq})+\frac{1}{\lambda}(f_\lambda^\star-f^{\rm eq})
    ,\\
    \partial_t g &+\bm{v}\cdot\bm{\nabla}g=-\frac{1}{\tau}(g-g^{\rm eq})+\frac{1}{\lambda}(g_\lambda^\star-g^{\rm eq})
    .
\end{align}
The main ingredient in space/time discretization is the integration along characteristics, here the velocities $\bm{v}$, over a time $\delta t$ that leads to
\begin{align}
    f(\bm{x}+\bm{v}\delta t, t+\delta t) &- f(\bm{x}, t) 
        \\ \nonumber
        &= \int_t^{t+\delta t}\left[\frac{1}{\tau}\left(f^{\rm eq} - f\right) + {\frac{1}{\lambda}}(f_\lambda^\star - f^{\rm eq})\right]dt' 
    ,\\
    g(\bm{x}+\bm{v}\delta t, t+\delta t) &- g(\bm{x}, t) 
        \\ \nonumber
        &= \int_t^{t+\delta t}\left[\frac{1}{\tau}\left(g^{\rm eq} - g\right) + {\frac{1}{\lambda}}(g_\lambda^\star - g^{\rm eq})\right]dt'
    .
\end{align}
The integrals on the right hand sides are approximated using a trapezoidal rule,
\begin{multline}
    \int_t^{t+\delta t}\left[\frac{1}{\tau}\left(f^{\rm eq} - f\right) +{\frac{1}{\lambda}} (f_\lambda^\star - f^{\rm eq})\right]dt' 
    \\
    = \frac{\delta t}{2\tau}\left(f^{\rm eq}(\bm{x},t) - f(\bm{x},t)\right) + \frac{\delta t}{2{\lambda}}\left(f_\lambda^\star(\bm{x},t) - f^{\rm eq}(\bm{x},t)\right) \\
    + \frac{\delta t}{2\tau}\left(f^{\rm eq}(\bm{x}+\bm{v}\delta t, t+\delta t) - f(\bm{x}+\bm{v}\delta t, t+\delta t)\right) \\ 
    + \frac{\delta t}{2{\lambda}}\left(f_\lambda^\star(\bm{x}+\bm{v}\delta t, t+\delta t) - f^{\rm eq}(\bm{x}+\bm{v}_i\delta t, t+\delta t)\right) + \mathcal{O}(\delta t^3)
    .
 \end{multline}
A similar equation can be written for $g$ that we will omit for readability. The resulting system is implicit in time. To remove the implicitness, the following transformations of variables are introduced \citep{He_LBMcharacteristics},
\begin{align}
    \bar{f}(\bm{x},t) = f(\bm{x},t) &- \frac{\delta t}{2\tau}\left(f^{\rm eq}(\bm{x},t) - f(\bm{x},t)\right) 
        \nonumber \\ 
        &- \frac{\delta t}{2{\lambda}}\left(f_\lambda^\star(\bm{x},t) - f^{\rm eq}(\bm{x},t)\right)
    ,\\
    \bar{g}(\bm{x},t) = g(\bm{x},t) &- \frac{\delta t}{2\tau}\left(g^{\rm eq}(\bm{x},t) - g(\bm{x},t)\right) 
        \nonumber  \\ 
        &- \frac{\delta t}{2{\lambda}}\left(g_\lambda^\star(\bm{x},t) - g^{\rm eq}(\bm{x},t)\right)
    .
\end{align}
Introducing this transformation back into the integrated time-evolution equations, we get
\begin{align}
 \bar{f}(\bm{x}+\bm{v}\delta t, t+\delta t) = \bar{f}(\bm{x}, t) &+ 2\beta\left(f^{\rm eq}(\bm{x},t) - \bar{f}(\bm{x},t)\right) 
    \\ \nonumber
    &+ {\frac{\delta t}{\lambda}}\left(1-\beta\right) \left(f_\lambda^\star(\bm{x},t) - f^{\rm eq}(\bm{x},t)\right)
 ,\\
 \bar{g}(\bm{x}+\bm{v}\delta t, t+\delta t) = \bar{g}(\bm{x}, t) &+ 2\beta\left(g^{\rm eq}(\bm{x},t) - \bar{g}(\bm{x},t)\right) 
    \\ \nonumber
    &+ {\frac{\delta t}{\lambda}}\left(1-\beta\right) \left(g_\lambda^\star(\bm{x},t) - g^{\rm eq}(\bm{x},t)\right)
 ,
\end{align}
where $\beta\in[0,1]$ is the relaxation parameter given by
\begin{equation}
    \beta=\frac{\delta t}{2\tau+\delta t}
    .\label{beta}
\end{equation}
The final step is to evaluate the moments of the distribution function that are needed to define $\{f^{\rm eq}, g^{\rm eq}\}$ and $\{f_\lambda^\star, g_\lambda^\star\}$ using the transformed distribution functions $\{\bar{f}, \bar{g}\}$. Integrating over $\{\bar{f}, \bar{g}\}$ and using the definitions for the transformed variables, it is readily shown that
\begin{align}
\int m \bar{f} d\bm{v} &= \int m f d\bm{v} = \rho,\\
\int m \bm{v} \bar{f} d\bm{v} &= \int m \bm{v} \left(f - \frac{\delta t}{2\lambda}\left(f_\lambda^\star-f^{\rm eq}\right)\right)d\bm{v} \nonumber \\&= \rho\bm{u}-\frac{\delta t}{2}\bm{F},\\
\int m \bar{g} d\bm{v} &= \int m\left(g - \frac{\delta t}{2\lambda}\left(g_\lambda^\star-g^{\rm eq}\right)\right)d\bm{v} \nonumber \\&= \rho E - \frac{\delta t}{2} \left(\bm{u}\cdot\bm{F} + Q\right).
\end{align}
Renaming the variables $\{\bar{f},\bar{g}\}\to \{f,g\}$ and dropping the dependence on the untransformed distribution function, we obtain the second-order-in-time accurate kinetic equations \begin{align}
    \label{eq:pre_LBM_f}
    f(\bm{x}+\bm{v}\delta t,t+\delta t)={f}(\bm{x},t)&+ 2\beta\left(f^{\rm eq}(\bm{x},t) - {f}(\bm{x},t)\right)    \\ \nonumber
    &+ {\frac{\delta t}{\lambda}}\left(1-\beta\right) \left(f_\lambda^\star(\bm{x},t)- f^{\rm eq}(\bm{x},t)\right),\\
    \label{eq:pre_LBM_g}
    g(\bm{x}+\bm{v}\delta t,t+\delta t)={g}(\bm{x},t)&+ 2\beta\left(g^{\rm eq}(\bm{x},t) - {g}(\bm{x},t)\right)    \\ \nonumber
    &+ {\frac{\delta t}{\lambda}}\left(1-\beta\right) \left(g_\lambda^\star(\bm{x},t) - g^{\rm eq}(\bm{x},t)\right),
\end{align}
along with the corresponding transformed expressions for the conserved hydrodynamic fields,
\begin{align}
\int m {f} d\bm{v} & = \rho,\label{eq:transform_rho}\\
\int m \bm{v} {f} d\bm{v} &= \rho\bm{u}-\frac{\delta t}{2}\bm{F},\label{eq:transform_u}\\
\int m {g} d\bm{v} &= \rho E - \frac{\delta t}{2} \left(\bm{u}\cdot\bm{F}+Q\right).\label{eq:transform_E}
\end{align}

\subsection{Phase-space discretization\label{sec:phase-space-disc}}

The phase-space is discretized with a discrete set of $Q=27$ particle velocities $\bm{v}_i=c\bm{c}_i$ in $D=3$ dimensions, where
\begin{equation}
    \label{eq:D3Q27c}
    \bm{c}_i=(c_{ix},c_{iy},c_{iz}),\ c_{i\alpha}\in\{-1,0,1\}.
\end{equation}
This standard nearest-neighbor $D3Q27$ lattice is characterized by the lattice speed of sound
\begin{equation}
    \label{eq:cs}
    \varsigma=\frac{1}{\sqrt{3}}c,
\end{equation}
where we use lattice units by setting $c=1$.  
The discrete velocity equations for the populations $f_i(\bm{x},t)$ and $g_i(\bm{x},t)$, $i=1,\dots, Q$, follow from \eqref{eq:pre_LBM_f} and \eqref{eq:pre_LBM_g} as
\begin{align}
    \label{eq:LBM_f}
    f_i(\bm{x}+\bm{c}_i\delta t,t+\delta t)=f_i(\bm{x},t)
    &+ 2\beta\left(f_i^{\rm eq}(\bm{x},t) - {f_i}(\bm{x},t)\right)    \\ \nonumber
    &+ \left(1-\beta\right) \left(f_i^\star(\bm{x},t)- f_i^{\rm eq}(\bm{x},t)\right),\\
    \label{eq:LBM_g}
    g_i(\bm{x}+\bm{c}_i\delta t,t+\delta t)={g_i}(\bm{x},t)&+ 2\beta\left(g_i^{\rm eq}(\bm{x},t) - {g_i}(\bm{x},t)\right)    \\ \nonumber
    &+ \left(1-\beta\right) \left(g_i^\star(\bm{x},t) - g_i^{\rm eq}(\bm{x},t)\right),
\end{align}
where we set $\lambda=\delta t$ for convenience (see last remark of Sec.\ref{sec:Methodology:kinmodel}).

The equilibrium populations $\{f_i^{\rm eq}, g_i^{\rm eq}\}$ and the shifted equilibrium populations $\{f_i^\star, g_i^\star\}$ remain to be defined hereafter.
Following the product-form formalism \citep{karlin2010factorization}, we introduce functions in two variables, $\xi_{\alpha}$ and $\zeta_{\alpha\alpha}$,
\begin{multline}
        \label{eq:phi}
 \Psi_{{i\alpha}}(\xi_{\alpha},\zeta_{\alpha\alpha})=1-c_{i\alpha}^2+ \frac{1}{2}\left[(3c_{i\alpha}^2-2)\zeta_{\alpha\alpha}+c_{i\alpha}\xi_{\alpha}\right], \\ \ i=1,\dots,Q,\ \alpha=x,y,z.
\end{multline}
The equilibrium populations $f_i^{\rm eq}$ are defined by setting the parameters in the functions \eqref{eq:phi} as
\begin{align}
    &\xi_{\alpha}^{\rm eq}=u_{\alpha},\label{eq:def_xi_eq}\\
    &\zeta_{\alpha\alpha}^{\rm eq}={\theta}
    +u_{\alpha}^2.\label{eq:def_zeta_eq}
\end{align}
With the definitions \eqref{eq:def_xi_eq} and \eqref{eq:def_zeta_eq} in the functions \eqref{eq:phi}, 
the local equilibrium populations are written in product-form,
    \begin{equation}\label{eq:LBMeq}
         f_i^{\rm eq}= \rho\prod_{\alpha}\Psi_{{i\alpha}}\left(u_\alpha,{\theta}+u_{\alpha}^2\right).
    \end{equation}
With $\lambda=\delta t$, the shifted flow velocity \eqref{eq:sh_vel} and shifted reference temperature  \eqref{eq:sh_temp} are
\begin{align}
    &\bm{u}^\star= \bm{u}+\delta t\left(\frac{\bm{F}}{\rho}\right),\label{eq:sh_vel_1}\\
    & \theta^\star = \theta + \delta t \alpha \theta \left(\bm{\nabla}\cdot\bm{u}\right) {+ \delta t (\gamma - 1) \frac{Q}{\rho}}.\label{eq:sh_temp_1}
\end{align}
For the shifted-equilibrium populations $f_i^\star$, the parameters $\xi_\alpha$ and $\zeta_{\alpha\alpha}$ in the functions  \eqref{eq:phi} are set as 
\begin{align}
	&\xi_{\alpha}^{\star} = {u_{\alpha}^\star},\label{eq:xistar}	\\
	  &\zeta_{\alpha\alpha}^{\star} = 
      {\theta^\star}
      +{{u_{\alpha}^\star}^2} + {\delta t}\Phi_{\alpha\alpha}.\label{eq:zetastar}
\end{align}
Importantly note that, compared to the continuous-velocity kinetic model described in the previous sections, the discrete-velocity formulation requires the additional correction term
\begin{equation}\label{eq:correction}
    \Phi_{\alpha\alpha} = -\frac{1}{\rho}\partial_{\alpha}\left(\rho u_{\alpha}^3 + {3\rho u_\alpha(\theta-\varsigma^2)} 
    \right),
\end{equation}
which is the standard correction necessary to restore Galilean invariance in the hydrodynamic limit on nearest-neighbor discrete-velocity lattices, see \citep{Prasianakis2007, li2012coupling, hosseini2020compressibility}.
Combining \eqref{eq:xistar} and \eqref{eq:zetastar} with the same product-form construction utilizing \eqref{eq:phi}, the shifted-equilibrium populations may be written as
\begin{equation}\label{eq:LBMstar}
     f_i^\star=\rho\prod_{\alpha}\Psi_{{i\alpha}}\left({u_{\alpha}^\star},
     {\theta^\star}
     {
     + \left({u_{\alpha}^\star}\right)^2}+{\delta t}\Phi_{\alpha\alpha}\right).
\end{equation}

For the $g$-populations, we follow the generating-function representation as in \citep{saadat2021extended, sawant2022detonation, strässle2025consistent-compressible}. 
The generating function is the bulk energy per unit mass,
\begin{equation}
    \label{eq:Egen}
    E(\rho,\bm{u},T) = e(\rho,T)+\frac{{u}^2}{2}.
\end{equation}
The corresponding equilibrium populations are constructed by repeated application of the operators
\begin{equation}\label{eq:Oa}
   { \mathcal{O}_\alpha (\theta)E = 
    \theta
        \dfrac{\partial E}{\partial u_\alpha}+ {u}_{\alpha} E,}
\end{equation}
whose dependence on the reference temperature is indicated explicitly.
The discrete equilibrium $g_i^{\rm eq}$ is defined by setting  $\xi_\alpha=\mathcal{O}_{\alpha}$ and $\zeta_{\alpha\alpha}=\mathcal{O}^2_{\alpha}$ in the functions \eqref{eq:phi} and interpreting the product-form as an operator acting on the generating function \eqref{eq:Egen},
    \begin{equation}\label{eq:LBMeqG}
         g_i^{\rm eq}\left(\rho,\bm{u},T,{\theta}\right)= \rho\prod_{\alpha}\Psi_{{i\alpha}}\left(\mathcal{O}_\alpha(\theta),{\mathcal{O}_\alpha(\theta)}^2\right)
         E(\rho,\bm{u},T).
    \end{equation}
Shifted-equilibrium populations $g_i^\star$ are defined using the equilibrium product-form \eqref{eq:LBMeqG} evaluated at shifted values \eqref{eq:sh_vel_1} and \eqref{eq:sh_temp_1}, and adding a correction,
\begin{equation}\label{eq:LBMgstar}
    g_i^\star = g_i^{\rm eq}\left(\rho,\bm{u}^\star, T^\star,
    {\theta^\star}
\right) 
    +
    \begin{cases}
    \dfrac{1}{2}\bm{c}_i\cdot\bm{q}^{\rm c}, & c_i^2=1,\\
    0, & \text{otherwise}.
\end{cases}
\end{equation}
where the non-equilibrium energy flux $\bm{q}^c$ and shifted temperature $T^\star$ are given by \eqref{eq:qc} and
\eqref{eq:sh_T}, respectively, with $\lambda=\delta t$, as
\begin{align}\label{eq:sh_T_1}
    &T^\star = T {+ \delta t \frac{Q}{\rho c_v}} - \delta t^2\left(\frac{\bm{F}\cdot\bm{F}}{2\rho^{{2}} c_v}\right),\\
    &{\bm{q}^{\rm c}=\delta t P\left(\bm{\nabla}h - \frac{k}{\mu}\bm{\nabla}T\right)}.\label{eq:qc_1}
\end{align}

The locally conserved fields entering the equilibrium and shifted-equilibrium populations, namely the density $\rho$, the momentum $\rho\bm{u}$ and the bulk energy $\rho E$, are defined from the zeroth- and first-order moments of the populations, as in Eqs. \eqref{eq:transform_rho}, \eqref{eq:transform_u} and \eqref{eq:transform_E}, with sums over discrete velocities replacing velocity-space integrals,
\begin{align}
    &\rho=\sum_{i=1}^Q f_i,\label{eq:def_h}\\
    &\rho\bm{u}=\sum_{i=1}^Q \bm{c}_i f_i + \frac{\delta t}{2} \bm{F},
    \label{eq:def_u}\\
    &\rho E =\sum_{i=1}^Q g_i + \frac{\delta t}{2}(\bm{u}\cdot\bm{F}+Q).
\label{eq:def_E}
\end{align}


\subsection{Hydrodynamic limit\label{sec:hydro-limit}}

The full Chapman-Enskog multiscale expansion of the specified lattice Boltzmann model can be found in appendix~\ref{app:CE}. Here, we summarize the key results.

In the hydrodynamic limit, the Navier–Stokes–Fourier equations, including body-force and heat-source contributions, are recovered as
\begin{gather}
    \partial_t\rho + \bm{\nabla}\cdot \left( \rho \bm{u} \right) = 0,
    \\
    \partial_t \left( \rho \bm{u} \right) + \bm{\nabla}\cdot \left(\rho \bm{u}\otimes\bm{u} + P\bm{I} + \bm{T}_{\rm NS} \right) - \bm{F}  = 0,
    \\
    \partial_t \left( \rho E\right)  + \bm{\nabla}\cdot\left( \rho \bm{u}\left(E + P/\rho\right) + \bm{u}\cdot\bm{T}_{\rm NS} +  \bm{q}_{\rm F} \right) - \bm{u}\cdot\bm{F} - Q= 0,
\end{gather}
where the Navier-Stokes stress tensor is
\begin{equation}
\bm{T}_{\rm NS} =  -\mu\left(\bm{\nabla}\bm{u} + \bm{\nabla}\bm{u}^\dagger - \frac{2}{D}\bm{\nabla}\cdot\bm{u} \bm{I}\right) - \eta \bm{\nabla}\cdot\bm{u}\bm{I},
\end{equation}
and the Fourier heat flux is 
\begin{equation}\label{eq:qF}
    {\bm{q}_{\rm F}} = -k\bm{\nabla}T.
\end{equation}
The dynamic shear viscosity is related to the relaxation parameter as
\begin{equation}
    \mu = \left(\frac{1}{2\beta}-\frac{1}{2}\right) P \delta t.
\end{equation}
Note that no equation of state had to be specified to obtain this result. 

This concludes the formulation of the proposed lattice Boltzmann model for compressible flow simulations of generic fluids in the presence of body-forces and heat sources. 
Some details of the present implementation are discussed next.

\subsection{Algorithm and implementations\label{sec:source-terms}}

\begin{figure}[b]
  \centering
  \begin{tikzpicture}[node distance=0.5cm and 1cm]
    \footnotesize
    \node[block_m, minimum width=6.5cm, text width=6.5cm] (ini) {
     Select $\mu$, $k$, $\eta$, $R$, $c_v$ \\
     select equation of state and parameters \\
     specify forms of body force and heat source \\
     specify initial conditions for macroscopic fields\\
     initialize $\{f_i,g_i\}\leftarrow\{f_i^{\rm eq},g_i^{\rm eq}\}$};
    \node[block_m, below=of ini] (forcing)
    {Compute $\rho$ \eqref{eq:def_h}, $\bm{F}$ \eqref{eqn:force_decomposition} and $Q$};
    \node[block_m, below=of forcing, minimum width=5cm, text width=5cm] (moments)
    {Compute $\bm{u}$ \eqref{eq:def_u}, $E$ \eqref{eq:def_E}, $T$ \eqref{eq:TdependsongenericEOS}, and $P$ \eqref{eq:genericEOS}};
    \node[block_m, below=of moments, minimum width=5cm, text width=5cm] (correction)
    {Compute $\bm{u}^*$ \eqref{eq:sh_vel_1}, $\theta^*$ \eqref{eq:sh_temp_1}, $\Phi_{\alpha\alpha}$ \eqref{eq:correction}, $T^*$ \eqref{eq:sh_T_1} and $\bm{q}^{\rm c}$ \eqref{eq:qc_1}};
    \node[block_s, below=of correction] (fext)
    {Compute $f_i^\star$ \eqref{eq:LBMstar} and $g_i^\star$ \eqref{eq:LBMgstar}};
    \node[block_s, right=0.5cm of fext] (feq)
    {Compute $f_i^{\rm eq}$ \eqref{eq:LBMeq} and $g_i^{\rm eq}$ \eqref{eq:LBMeqG}};
    \node[block_m, below=0.5cm of fext] (CollStream)
    {Compute $\beta$ \eqref{beta}, collide and stream (\eqref{eq:pre_LBM_f} and \eqref{eq:pre_LBM_g})};
    \node[block_s, left=0.5cm of fext] (next)
    {$t \leftarrow t + \delta t$};
    \draw[->] (ini) -- (forcing);
    \draw[->] (forcing) -- (moments);
    \draw[->] (moments) -- (correction);
    \draw[->] (correction) -- (fext);
    \draw[->] (fext.south) -- (CollStream.north);
    \draw[->] (moments.east) -| (feq.north);
    \draw[->] (feq.south) |- (CollStream.east);
    \draw[->] (CollStream.west) -| (next.south);
    \draw[->] (next.north) |- (forcing.west);
\end{tikzpicture}
  \caption{Structure of the algorithm.}
  \label{Fig:FlowChart}
\end{figure}
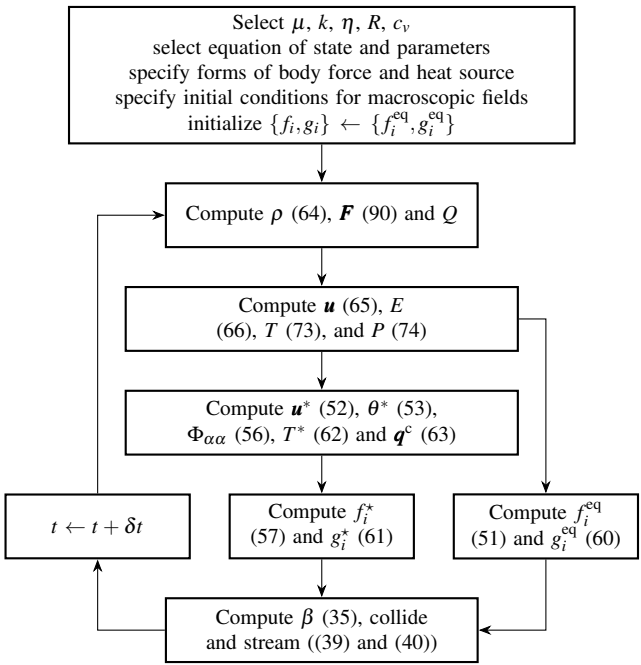

The overall structure of the proposed algorithm is shown in Fig.~\ref{Fig:FlowChart}.

Note that $\mu$, $k$, $R$, and $c_v$ are freely tunable parameters.
As a result of shifting the reference temperature with the parameter $\alpha$, the bulk viscosity $\eta$ is also an independently tunable parameter, with which, e.g., the Navier-Stokes bulk viscosity $\eta = \mu (2/D - R/c_v)$
can be achieved, or it can be selected to dampen spurious acoustic modes.
Lastly, as the relaxation time $\lambda$ does not affect the hydrodynamic limit, thus also remaining a free parameter, it was conveniently specified herein as $\lambda=\delta t$ for the lattice Boltzmann realization.

The bulk energy $E$ is obtained from Eq.~\eqref{eq:def_E} and the thermodynamic temperature is evaluated using the caloric relation derived from \eqref{eq:de_rho}. 
An explicit relation  for $T$ can be derived in the form of 
\begin{equation}
        T(\rho, \bm{u}, E) = \mathcal{F}(\rho, \bm{u}, E),\label{eq:TdependsongenericEOS}
\end{equation}
if an equation of state in the form of
\begin{equation}
    P(\rho,T) = \mathcal{F} (\rho,T),\label{eq:genericEOS}
\end{equation}
is a linear function, $\mathcal{F}$, in temperature.
Otherwise, e.g. in the Peng–Robinson model, the temperature must instead be determined by solving the corresponding nonlinear caloric relation.
Computation of thermodynamic temperature $T$  through \eqref{eq:TdependsongenericEOS} is then followed by evaluation of $P$ using \eqref{eq:genericEOS}.
The specific equation of state will be specified in later applications.

The evaluation of $\Phi_{\alpha\alpha}$, and $\bm{q}^{\rm c}$ requires spatial derivatives of $\rho$, $T$, $h$, and $\bm{u}$. In this work, all derivatives are computed using standard second-order central finite differences, except for the first-order derivative in Eq.~\eqref{eq:correction}, which is discretized using an upwind-biased scheme to enhance numerical stability at higher Mach numbers \citep{saadat2021extended, hosseini2020compressibility, renard2021improved}.

\section{Applications\label{sec:Applications}}

    In this section, the model with body force and heat source is thoroughly validated for various applications to compressible flows (Sec.~\ref{sec:idealgascompressible}), first for the ideal gas regime, followed by the non-ideal and multiphase regime (Sec.~\ref{sec:nonidealgascompressiblemultiphase}). 
    A brief validation of the model for dispersion- and dissipation-dominated benchmarks without application of body force and heat source can be found in appendix~\ref{app:Modeconsistency} for the sake of completeness.  

    \subsection{Ideal-gas compressible flows}\label{sec:idealgascompressible}

        We start the validation of body force and heat source in this section for the ideal gas regime, using the ideal gas equation of state, cf. \eqref{eq:genericEOS}, 
        \begin{equation}
            P = \rho R T\label{eq:idealEOS}.
        \end{equation}
        For an ideal gas, the thermodynamic temperature can be explicitly computed, cf. \eqref{eq:TdependsongenericEOS}, as
        \begin{equation}\label{eq:idealgasT}
            T = \frac{E - \frac{1}{2}\bm{u}^2}{c_v}.
        \end{equation}
        
        \subsubsection{Poiseuille flow}
            
           \begin{figure*}[t]
                \centering
                \includegraphics[width=0.49\linewidth]{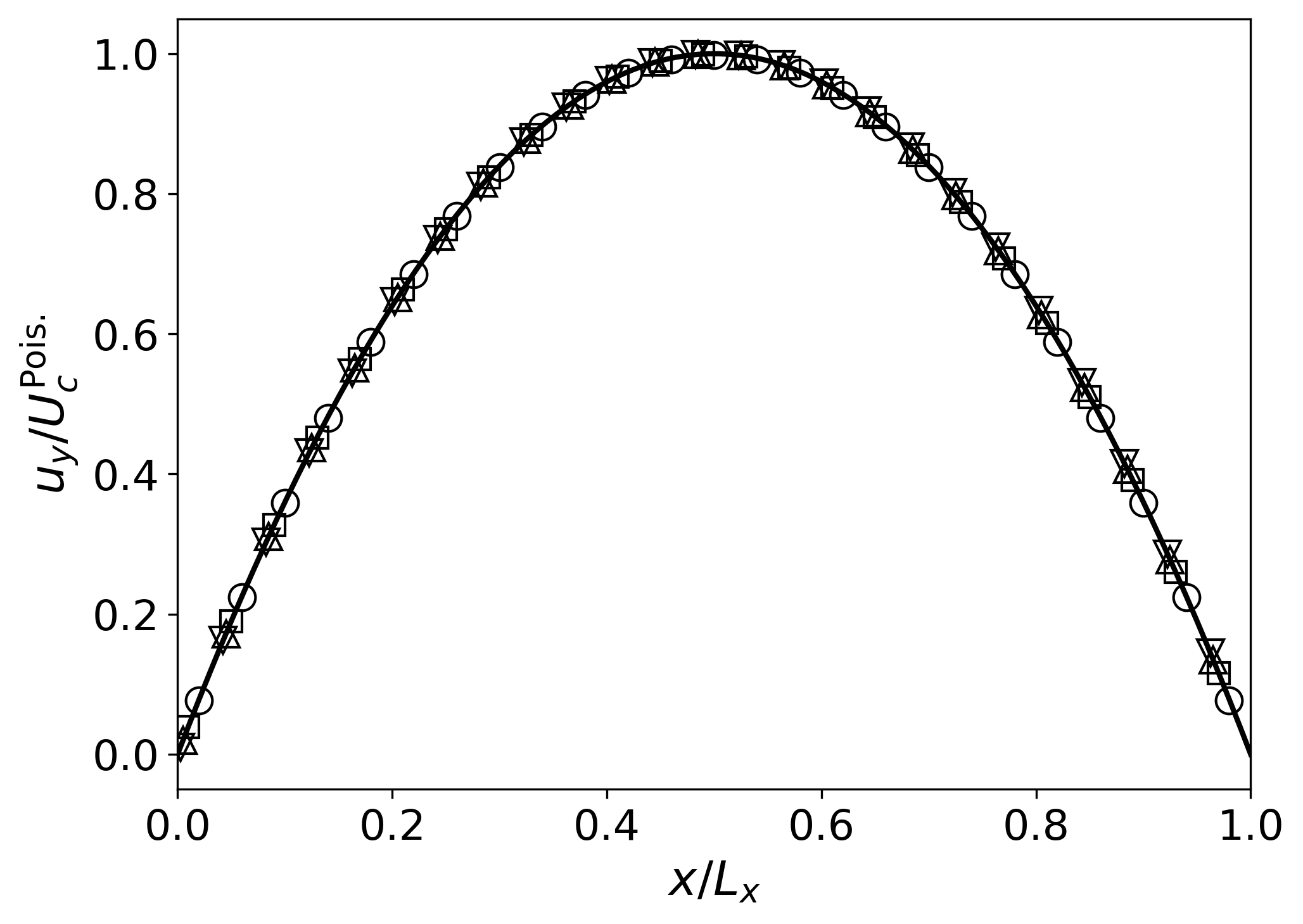}
                \includegraphics[width=0.49\linewidth]{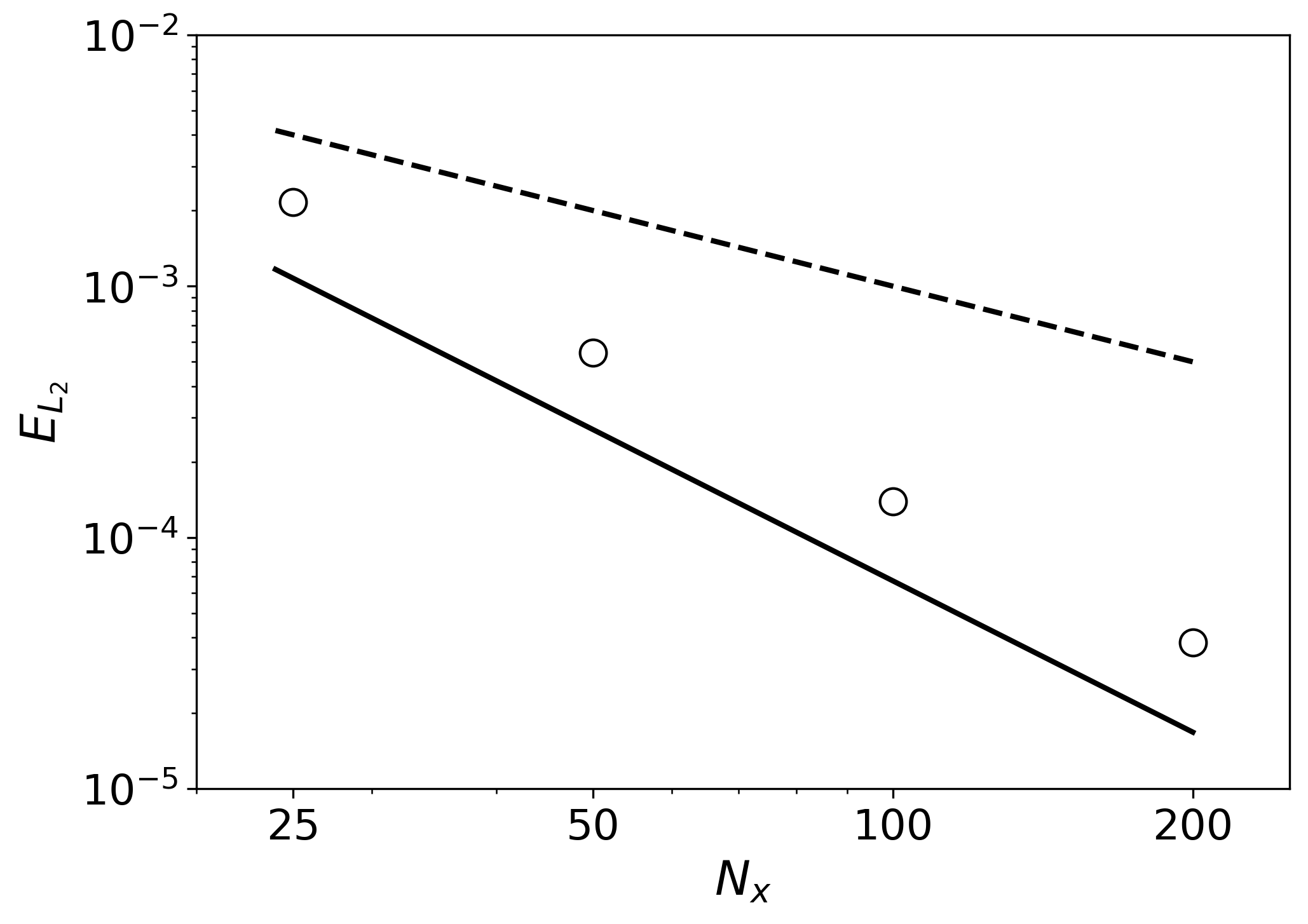}    
                \caption{Results for isothermal Poiseuille flow at Re = 100. (left) Numerical solutions (markers) compared to the analytical solution (solid line). Markers $\mcirc$, $\square$, $\triangle$ and $\mtriangledown$ represent the results for $N_x$ $\in \{25, 50, 100, 200\}$ respectively. (right) $L_2$ error shows second order convergence with grid size. Dashed and solid lines have slope $-1$ and $-2$, respectively.}
                \label{fig:Forcing_Ideal_IsothermalPoiseuille}%
            \end{figure*}
            
            We first examine flow driven by a steady body force by considering isothermal Poiseuille flow. To enforce isothermal conditions, the evolution equation for the $g$-populations (representing energy) is not solved. A one-dimensional domain of length $L_x = 1$ mm is discretized with $N_x$ lattice nodes and subjected to a constant uniform streamwise acceleration $g_y$, which results in a body force $F_y = \rho g_y$. Periodic boundary conditions are applied in the streamwise direction, while no-slip conditions at the walls are enforced using the half-way bounce-back scheme \cite{LBMBookKrueger}.

            The system is initialized at $\rho_0 = \rho_{cr}$ and $P_0 = P_{cr}$, and the dynamic viscosity is set to $\mu = 10^{-2}$ Pa$\cdot$s. The Reynolds number is defined as $\mathrm{Re} = \rho_0 U_c^{\text{Pois.}} L_x/\mu = 100$, where $U_c^{\text{Pois.}} = \rho_0 g_y L_x^2 / 8\mu$ is the analytical centerline velocity. Simulations are performed for $N_x \in \{25, 50, 100, 200\}$. Starting from a quiescent state with uniform density and zero velocity, the system is advanced until a steady velocity field is obtained.
            
            The resulting velocity profiles are compared with the analytical solution
            \begin{equation}
                u_y^{\rm an.} = 4 \ U_c^{\text{Pois.}} \frac{x}{L_x}\left(1 - \frac{x}{L_x}\right),
            \end{equation}
            as shown in Fig.~\ref{fig:Forcing_Ideal_IsothermalPoiseuille}. Excellent agreement is observed between the numerical and analytical solutions.

            To assess the spatial accuracy of the scheme, the relative $L_2-$norm error is defined as
            \begin{equation}
                E_{L_2} = \frac{\sqrt{\sum (u_y^{\rm sim.}- u_y^{\rm an.})^2}}{\sqrt{\sum (u_y^{\rm an.})^2}}.
            \end{equation}
            The variation of $E_{L_2}$ with grid refinement is also shown in Fig. ~\ref{fig:Forcing_Ideal_IsothermalPoiseuille}. The error decreases quadratically with the grid spacing, confirming the expected second-order spatial convergence of the numerical scheme.

        \subsubsection{Womersley flow}
        
            \begin{figure*}[htbp!]
                \centering
                \includegraphics[width=0.44\linewidth]{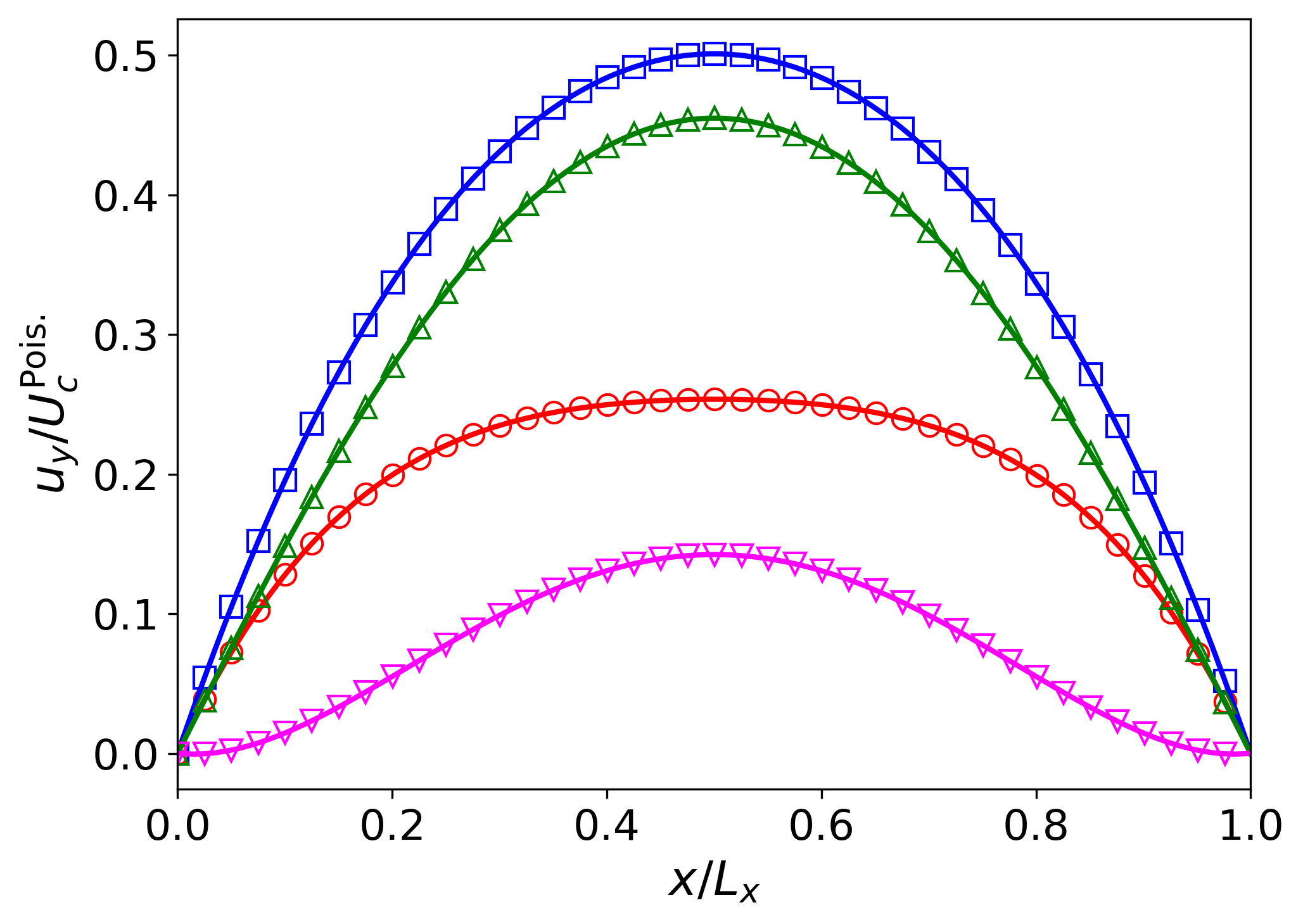}
                \includegraphics[width=0.44\linewidth]{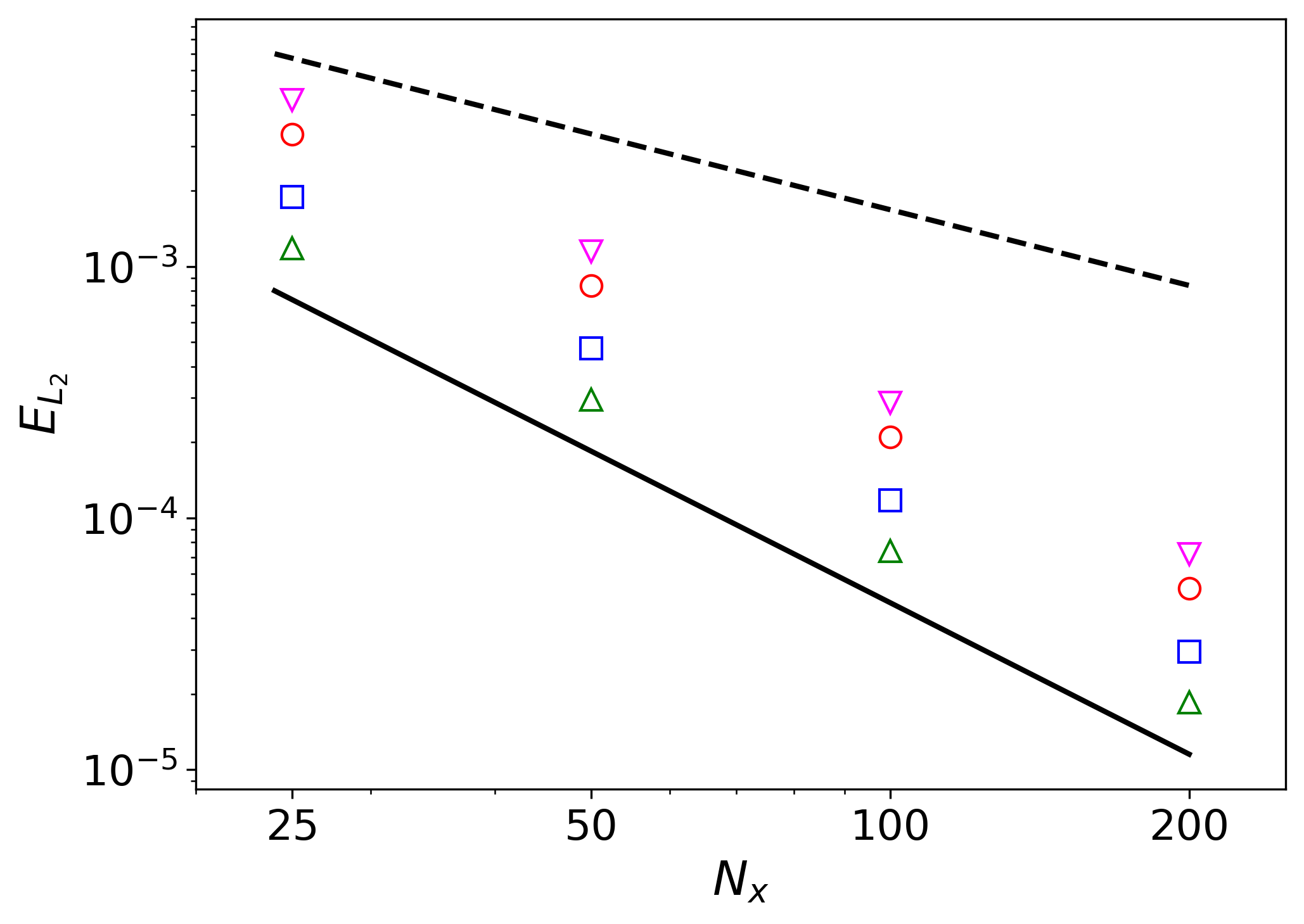}\\
                \includegraphics[width=0.44\linewidth]{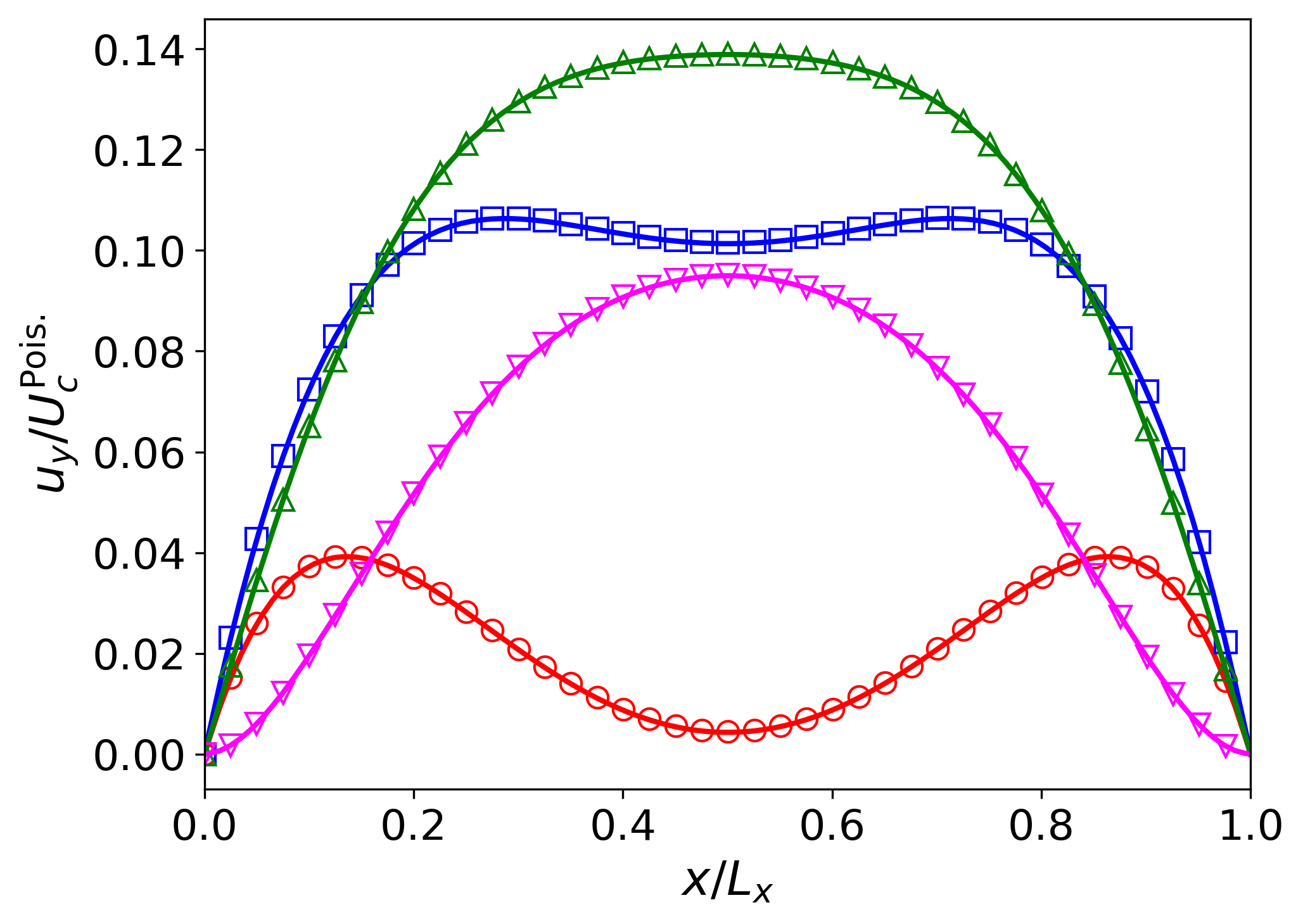}
                \includegraphics[width=0.44\linewidth]{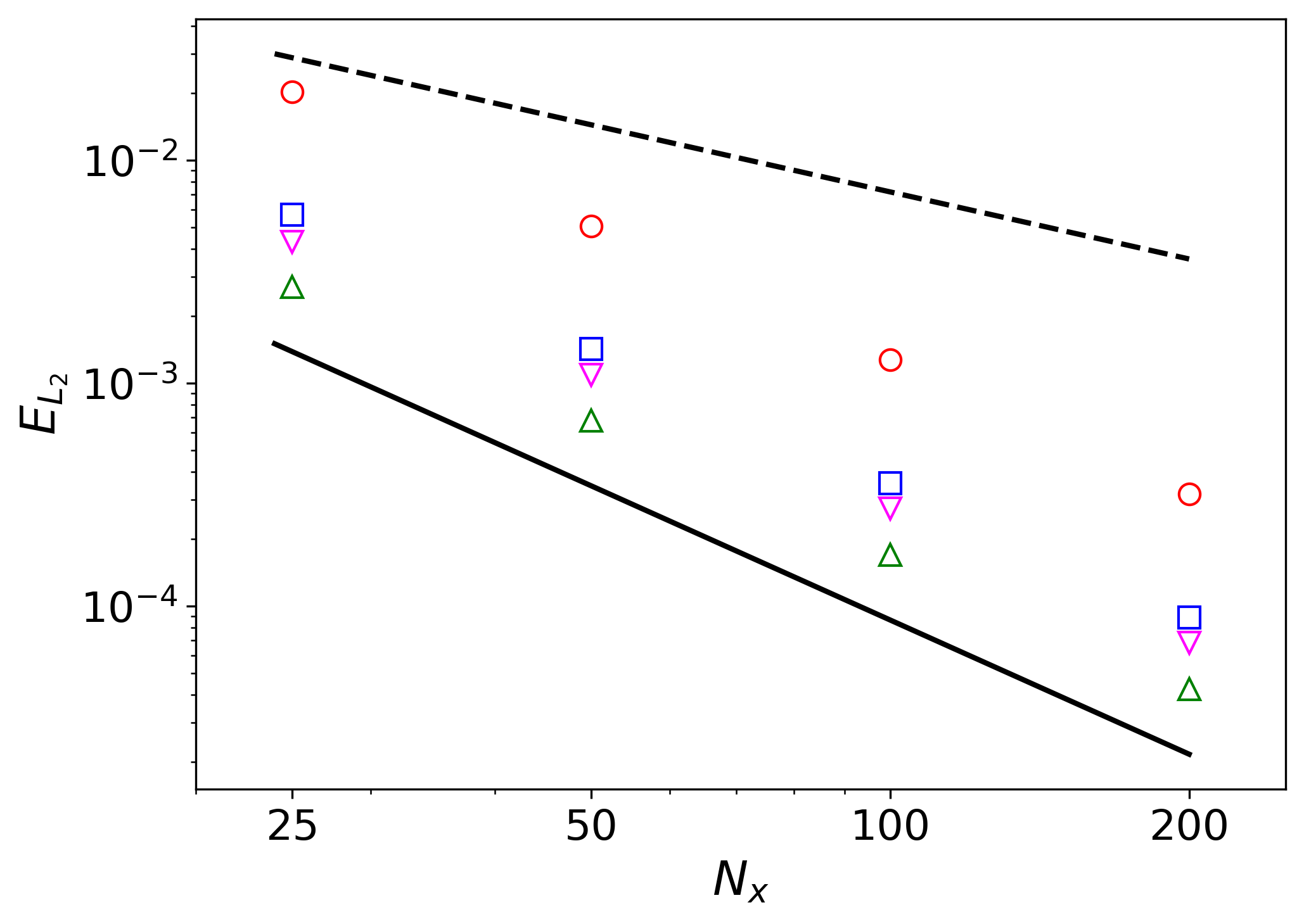}\\
                \includegraphics[width=0.44\linewidth]{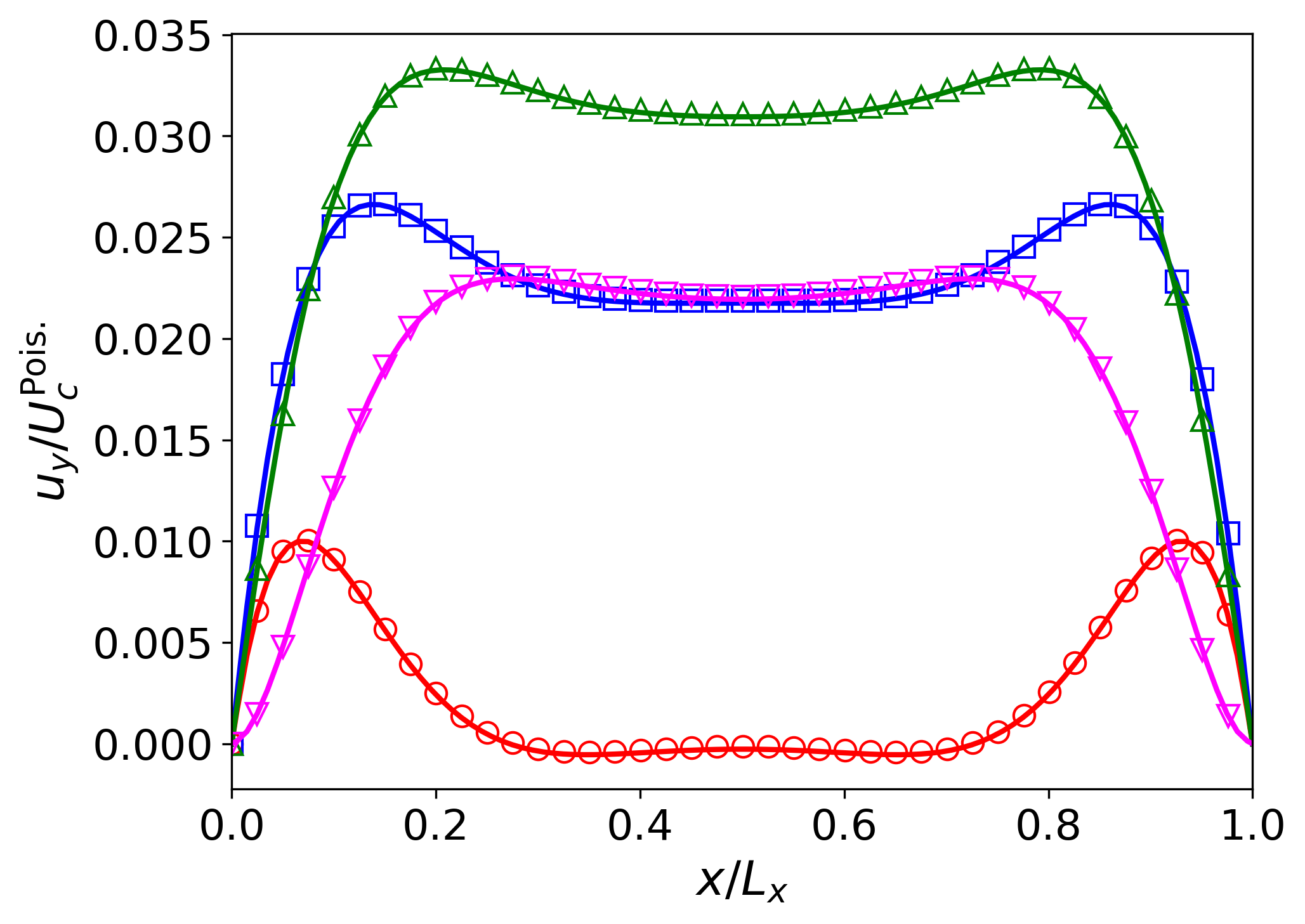}
                \includegraphics[width=0.44\linewidth]{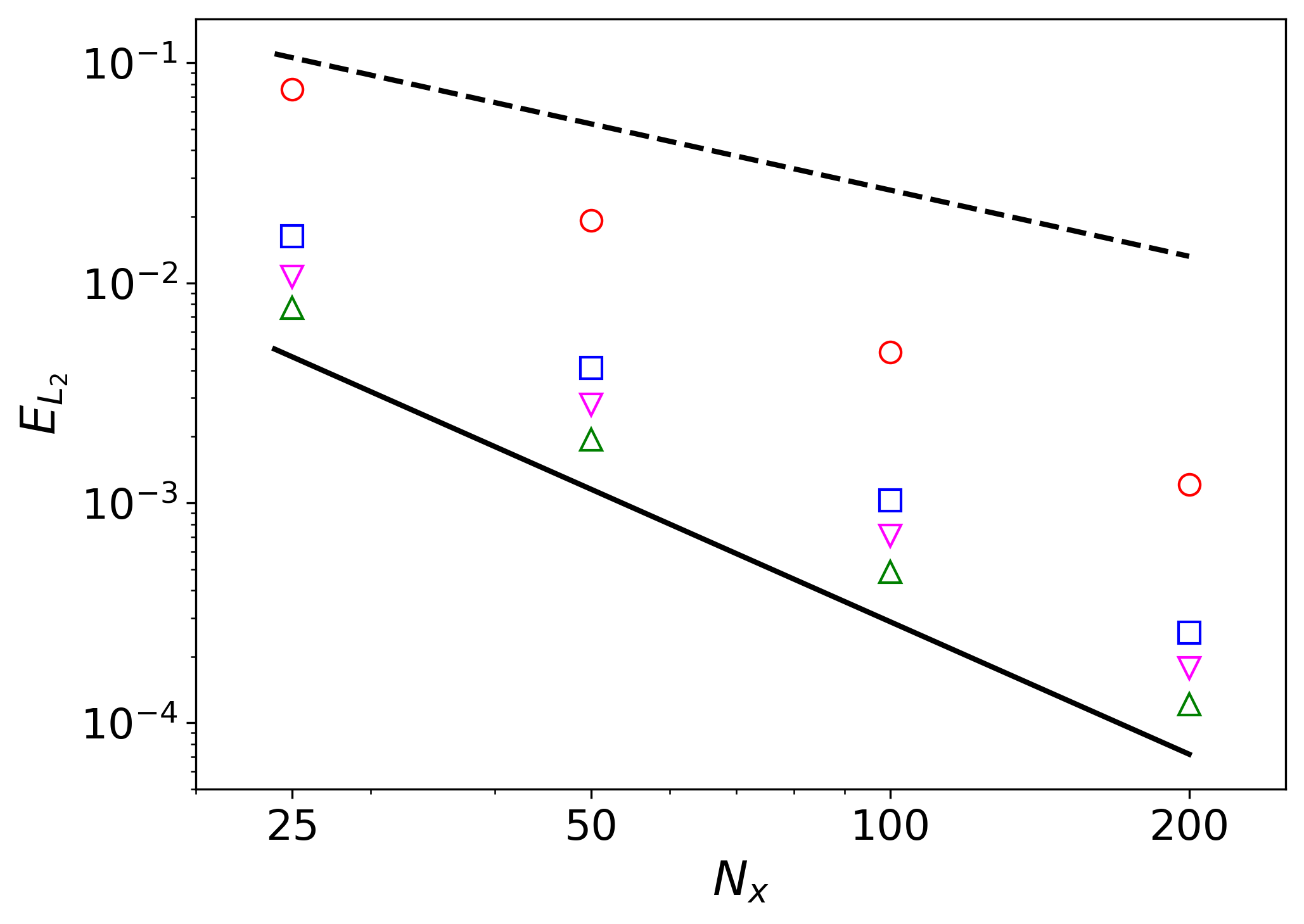}\\
                \includegraphics[width=0.44\linewidth]{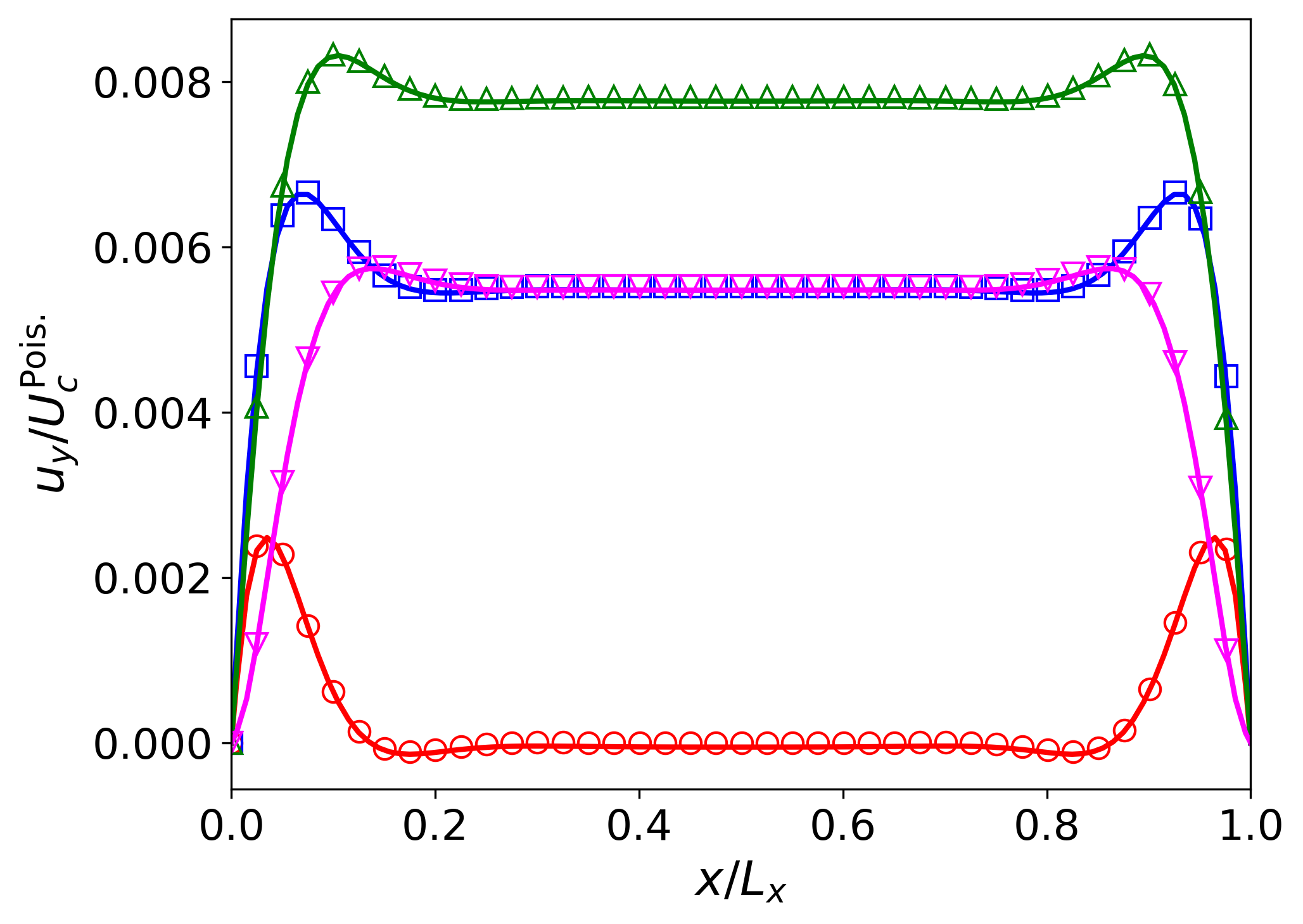}
                \includegraphics[width=0.44\linewidth]{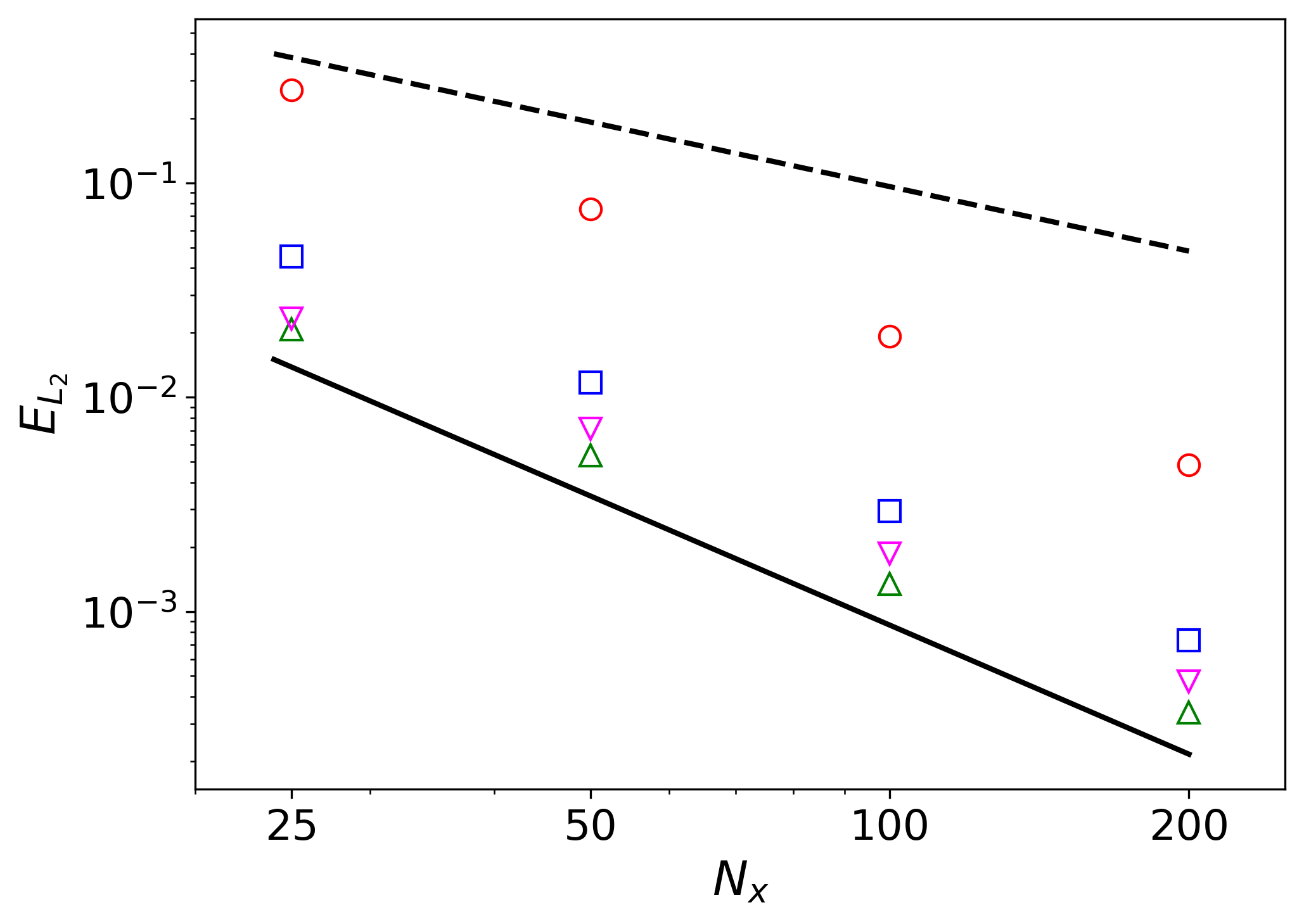}\\
                \caption{Solutions for pulsatile flow at Re = 100 and different Womersley numbers Wo $ \in \{4, 8, 16, 32\}$. (left) Markers $\mcirc$, $\square$, $\triangle$, and $\mtriangledown$ represent the analytical velocity profiles at times  $t/T_p \in \{0,\,1/8,\,1/4,\,3/8\}$  respectively and solid lines denote the numerical results. (right) $L_2$ error of the velocity fields. Dashed and solid lines have slope $-1$ and $-2$ respectively. }
                \label{fig:Forcing_Ideal_Womersley}%
            \end{figure*}
            
            Next, we consider isothermal pulsatile flow driven by a spatially uniform but temporally oscillatory body force. The numerical setup is identical to that of the previous case, except that the streamwise forcing is prescribed as $F_y= \rho g_y \cos(\omega t)$, where $\omega = 2\pi/T_p$ is the angular frequency and $T_p$ is the oscillation period. The Womersley number is defined as $\mathrm{Wo} = L_x \sqrt{\rho_0 \omega/\mu}$. The analytical solution for this flow, originally derived by \cite{Womersley}, is given by
            \begin{equation}
                u_y^{\rm an.}(x,t) =
                \mathcal{R}e\left\{
                \frac{g_y}{i\omega}
                \left[
                1 -
                \frac{
                \cosh\!\left(\sqrt{i}\,\mathrm{Wo}\left(\frac{x}{L_x}-\frac{1}{2}\right)\right)
                }{
                \cosh\!\left(\sqrt{i}\,\mathrm{Wo}/2\right)
                }
                \right]
                e^{i\omega t}
                \right\},
            \end{equation}
            where $\mathcal{R}e$ denotes the real part of a complex quantity. Simulations are performed for $\mathrm{Wo} \in [4, 8, 16, 32]$. The flow is initialized with a uniform density field, while the initial velocity field is prescribed using the analytical solution evaluated at $t = 0$. The system is then evolved for one complete oscillation period, and results are collected at different time intervals. This procedure is repeated for all combinations of Womersley numbers and grid sizes.

            Fig.~\ref{fig:Forcing_Ideal_Womersley} compares the numerical and analytical velocity profiles at different time instants. For clarity, only results corresponding to $N_x = 100$ are shown in the velocity profiles. The left panels show excellent agreement between numerical and analytical solutions across all Womersley numbers considered, demonstrating the ability of the method to accurately capture unsteady forcing.
            
            The right panels present the $L_2$ norm relative error in velocity as defined previously, shown for various grid resolutions, at each time interval and for each Womersley number case. For every combination of Womersley number and time interval, the error decreases quadratically with grid refinement, again confirming second-order convergence of the method.
            
        \subsubsection{Poiseuille flow with cross-flow acceleration}
        
            \begin{figure*}[htbp!]
                \centering
                \includegraphics[width=0.47\linewidth]{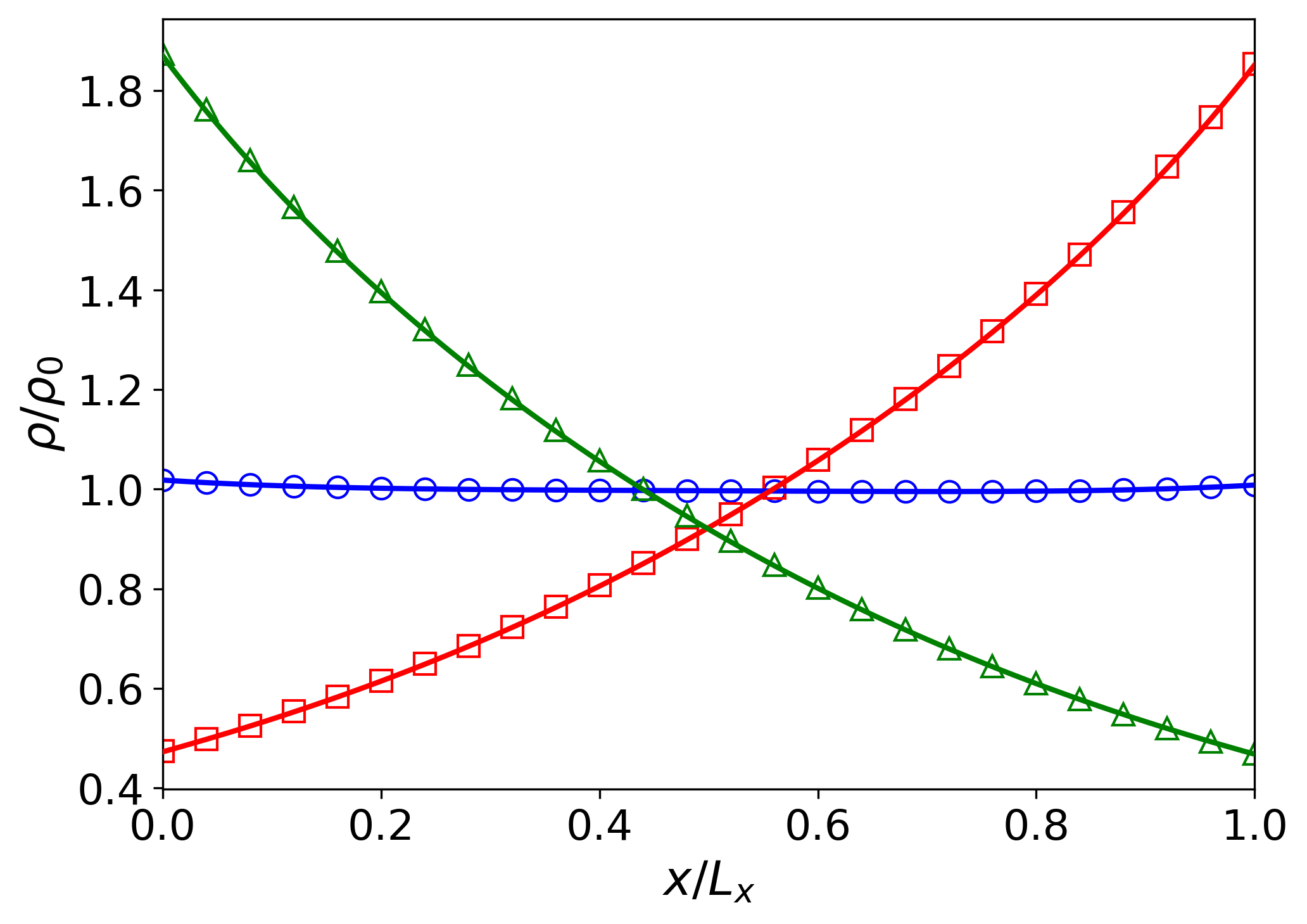}
                \includegraphics[width=0.47\linewidth]{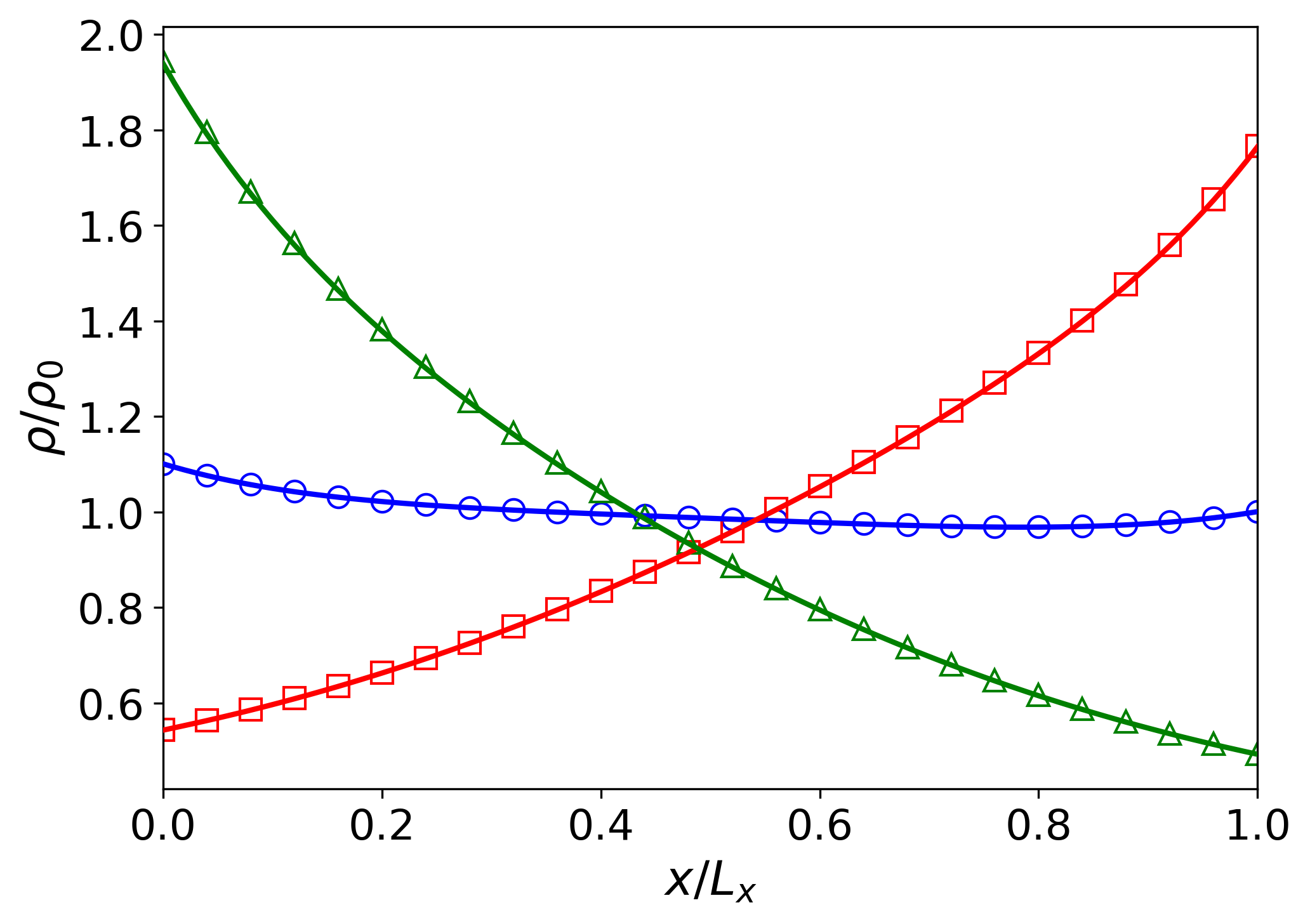}\\
                \includegraphics[width=0.47\linewidth]{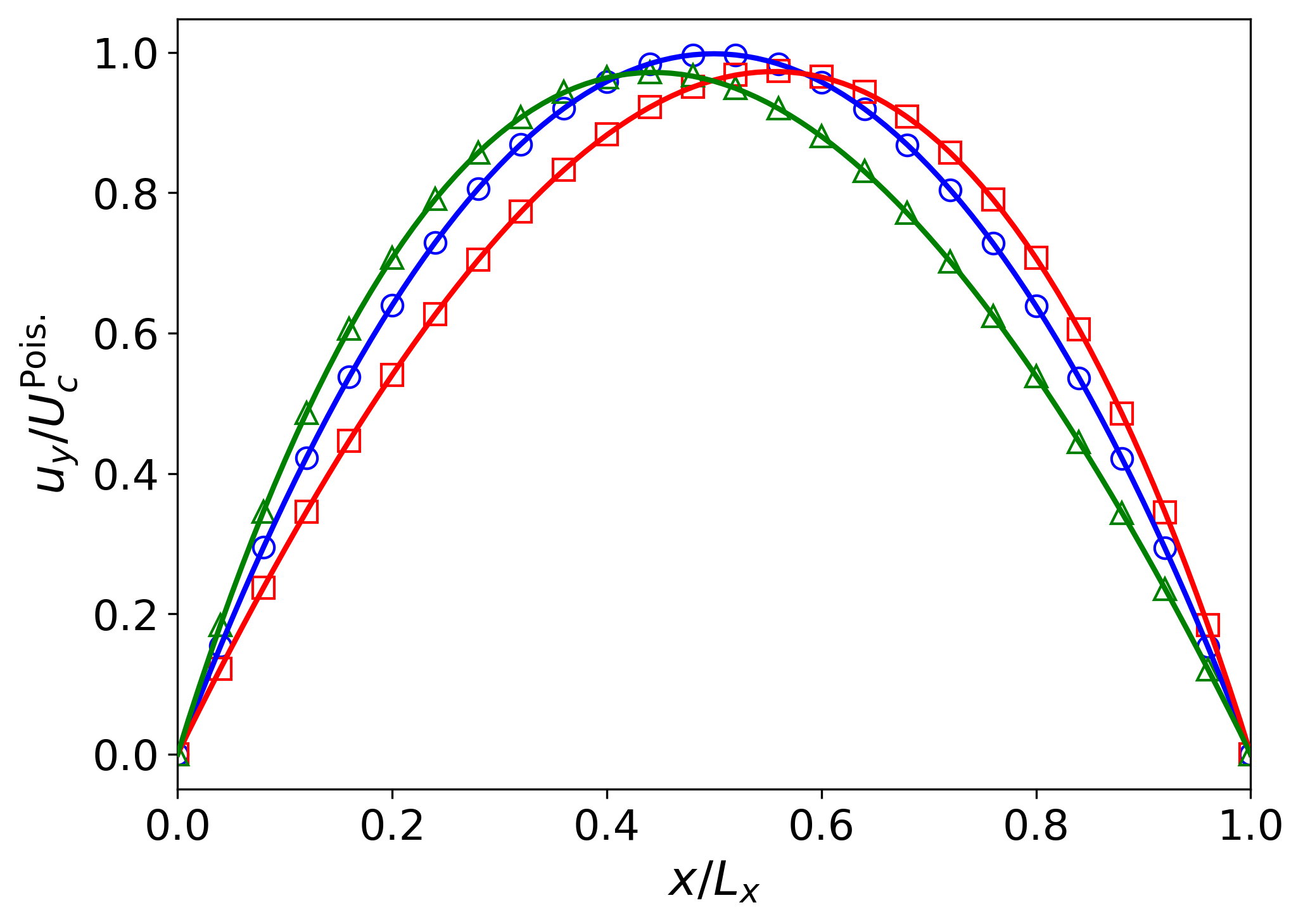}
                \includegraphics[width=0.47\linewidth]{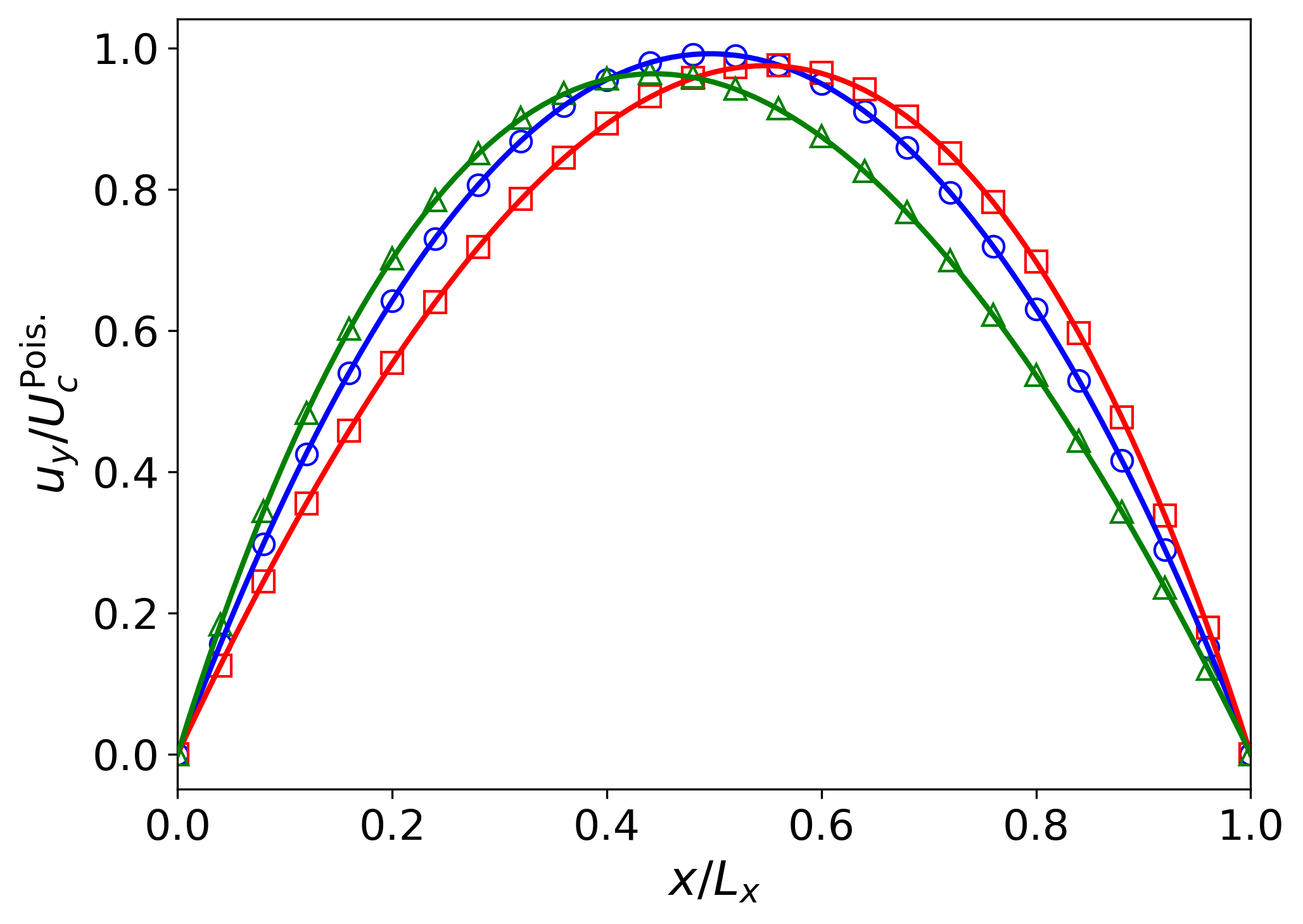}\\
                \includegraphics[width=0.47\linewidth]{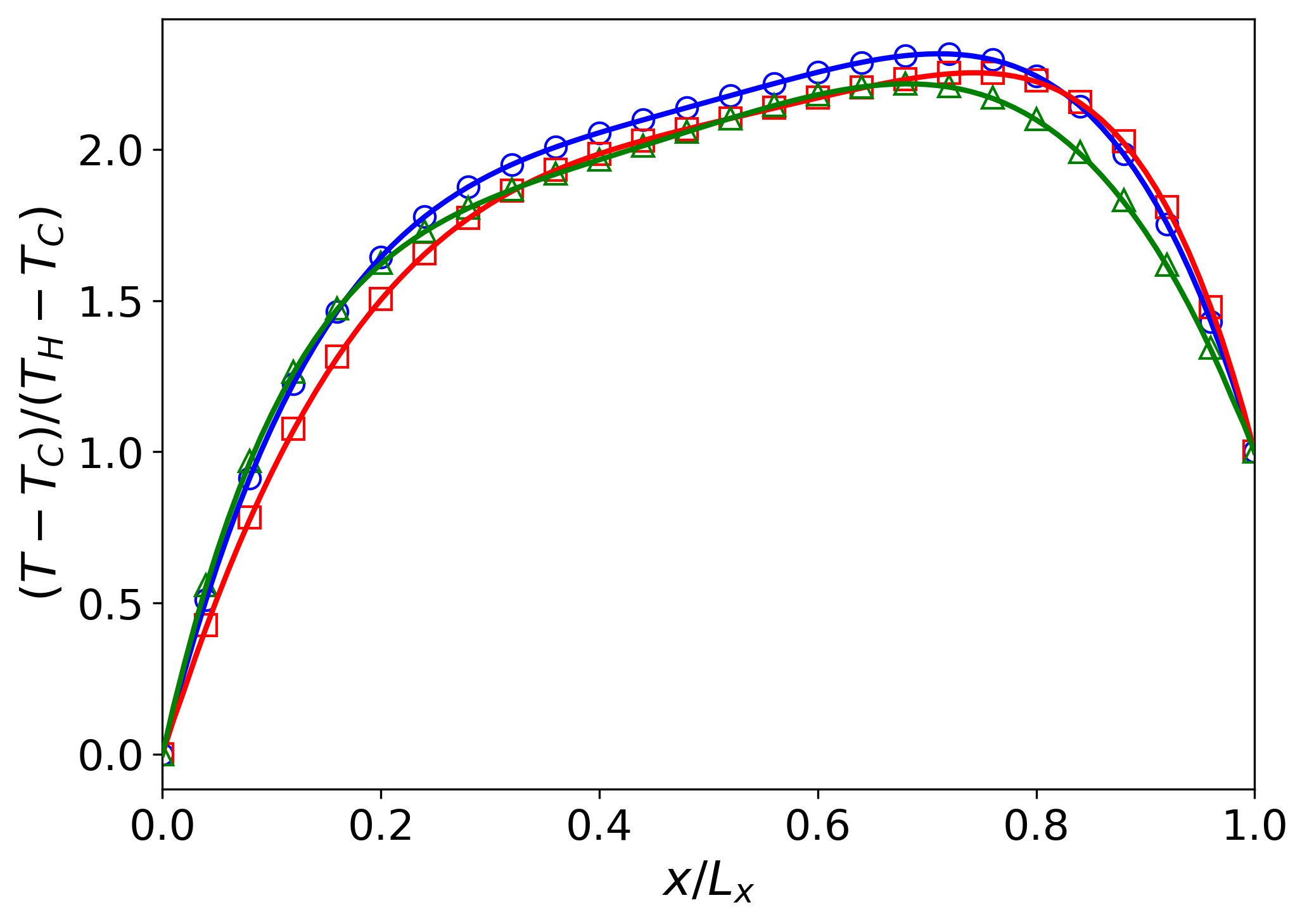}
                \includegraphics[width=0.47\linewidth]{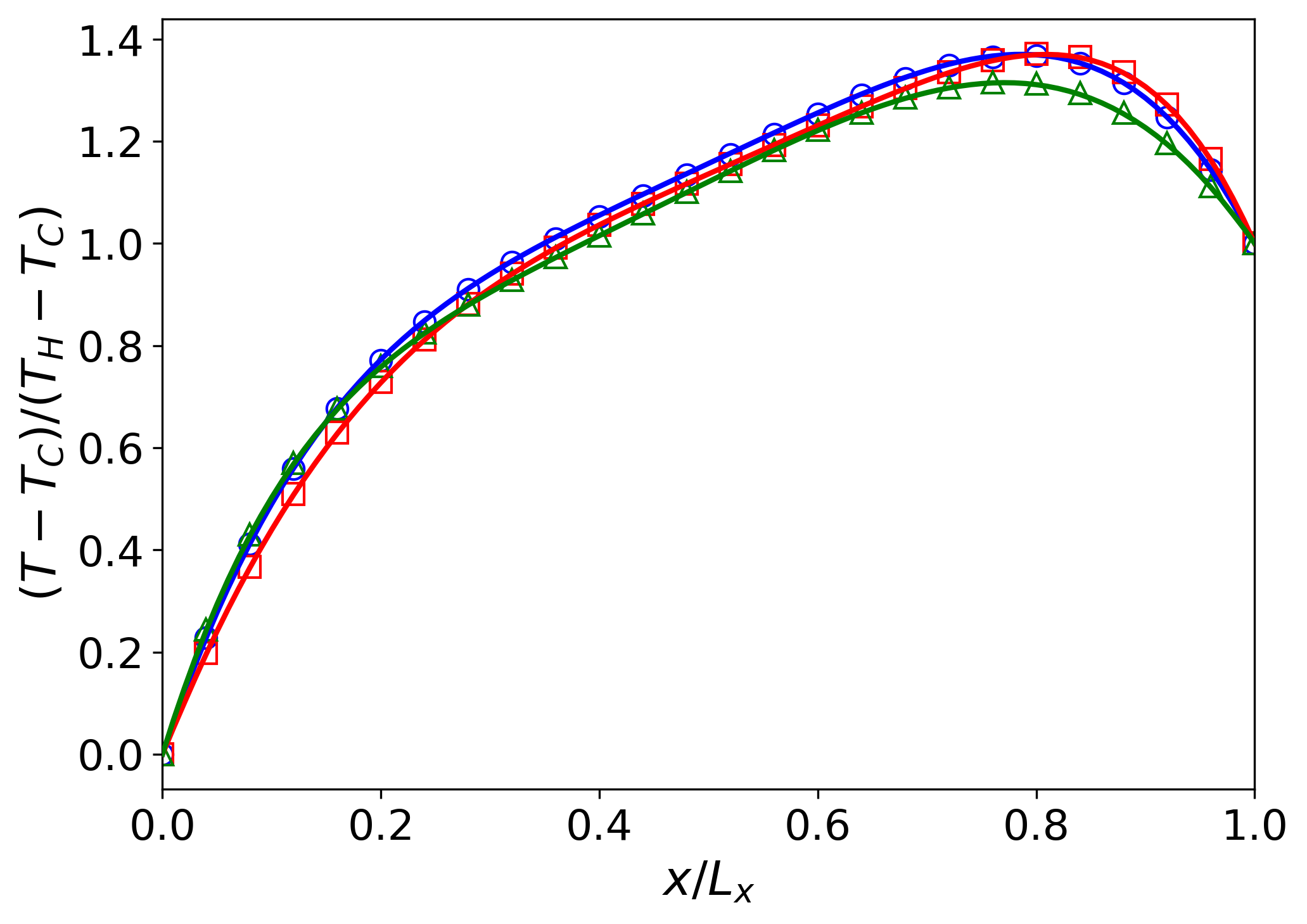}\\
                \caption{Solutions of density, velocity and temperature fields (solid lines) for compressible Poiseuille flow with cross-flow acceleration at Re = 100 and (left panels) Pr = 0.5, Ec = 10 and (right panels) Pr = 2, Ec = 1. Markers $\mcirc$, $\square$ and $\triangle$ represent the reference solution for $g_x/g_y \in \{0, 50, -50\}$ respectively, obtained using a high resolution finite difference method.}
                \label{fig:Forcing_Ideal_CompressiblePoiseuille}%
            \end{figure*}
            
            Next, we assess the performance of the forcing scheme in compressible thermal flows by considering compressible Poiseuille flow in the presence of a transverse body force. In this configuration, flow between two stationary walls is driven by a uniform streamwise acceleration $g_y$, resulting in $F_{y} = \rho g_y$. The left wall is maintained at a temperature $T_{\rm C}$, while the right wall is maintained at a higher temperature $T_{\rm H}$. In addition, a uniform cross-stream acceleration $g_x$ acts on the flow, which leads to $F_{x} = \rho g_x$.
            
            The computational domain is identical to that used in the previous test cases. The reference density is set to $\rho_0 = 5$ kg/m$^3$, and the dynamic viscosity is fixed at $\mu = 10^{-3}$ Pa$\cdot$s. The Reynolds and Mach numbers are prescribed as $\mathrm{Re} = \rho_0 U_c^{\mathrm{Pois.}} L_x/\mu = 100$, $\mathrm{Ma} = U_c^{\mathrm{Pois.}}/c_s(T_{\rm C}) = 0.5$, which determines the cold-wall temperature $T_{\rm C}$. The Eckert number is defined as $\mathrm{Ec} = (U_c^{\mathrm{Pois.}})^2/C_P [T_{\rm H} - T_{\rm C}]$.
            Simulations are performed for $(\mathrm{Pr}, \mathrm{Ec}) \in \{(0.5, 10), (2, 1)\}$ and $g_x/g_y \in \{0, 50, -50\}$ for each combination of Prandtl and Eckert numbers. The simulations are initialized with a uniform density field, zero velocity, and a linearly varying temperature profile, and are advanced in time until the solution converges.
            
            The reference solution is obtained using the finite-difference scheme described in \cite{BodyForceLBM_LiShan2023} on a sufficiently fine grid. Fig.~\ref{fig:Forcing_Ideal_CompressiblePoiseuille} compares the simulation results with this reference solution. It can be seen that the transverse acceleration modifies the velocity profile, reducing the peak velocity below $U_c^{\mathrm{Pois.}}$ and shifting its location away from the channel centerline. Furthermore, the pressure gradient induced by the transverse body force gives rise to significant density variations across the channel. Excellent agreement with the reference solutions is observed for all cases considered.

        \subsubsection{Rayleigh flow}
            
            \begin{figure*}[htbp!]
                \centering
                \includegraphics[width=0.49\linewidth]{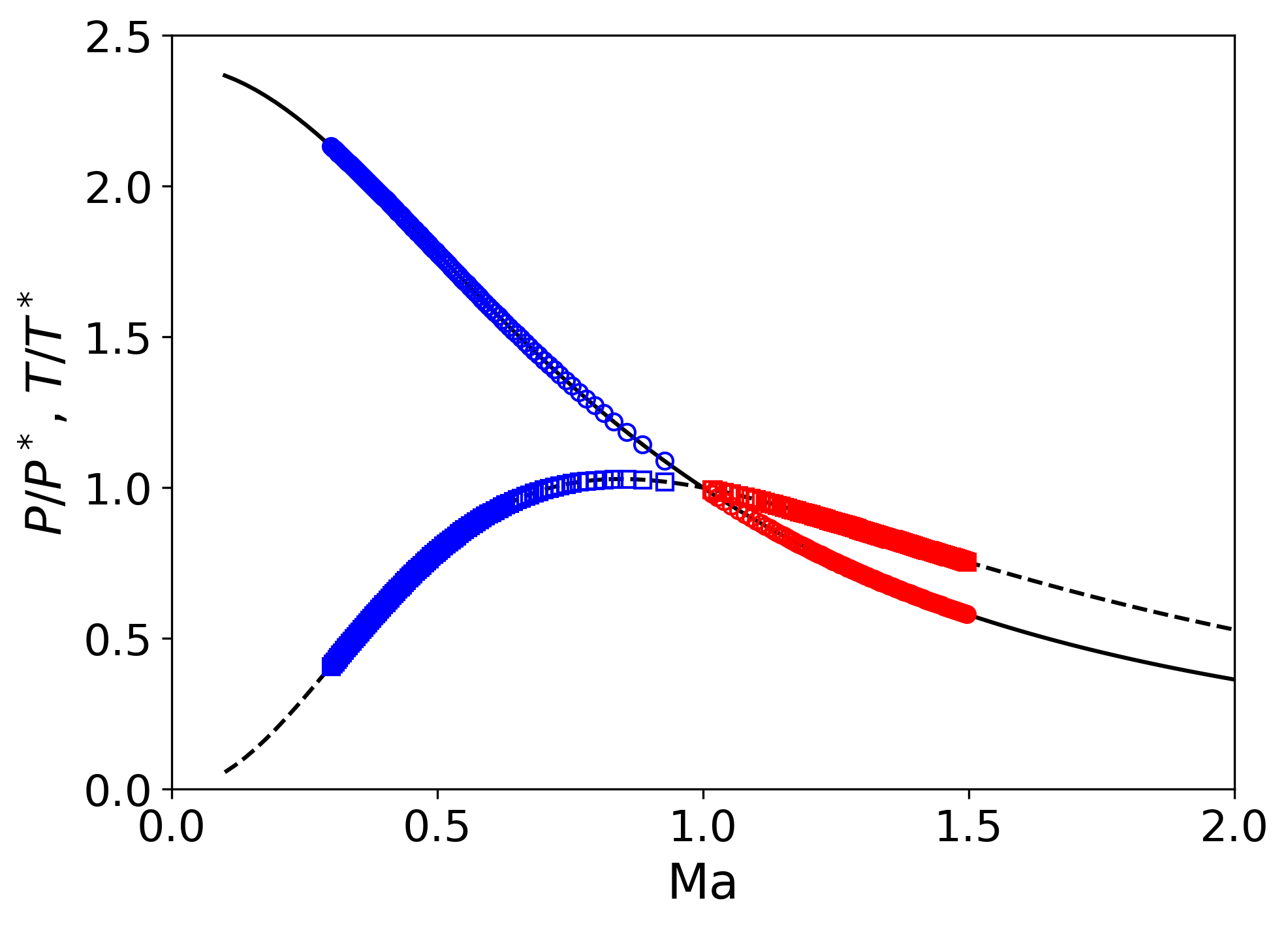}\phantom{a}%
                \includegraphics[width=0.49\linewidth]{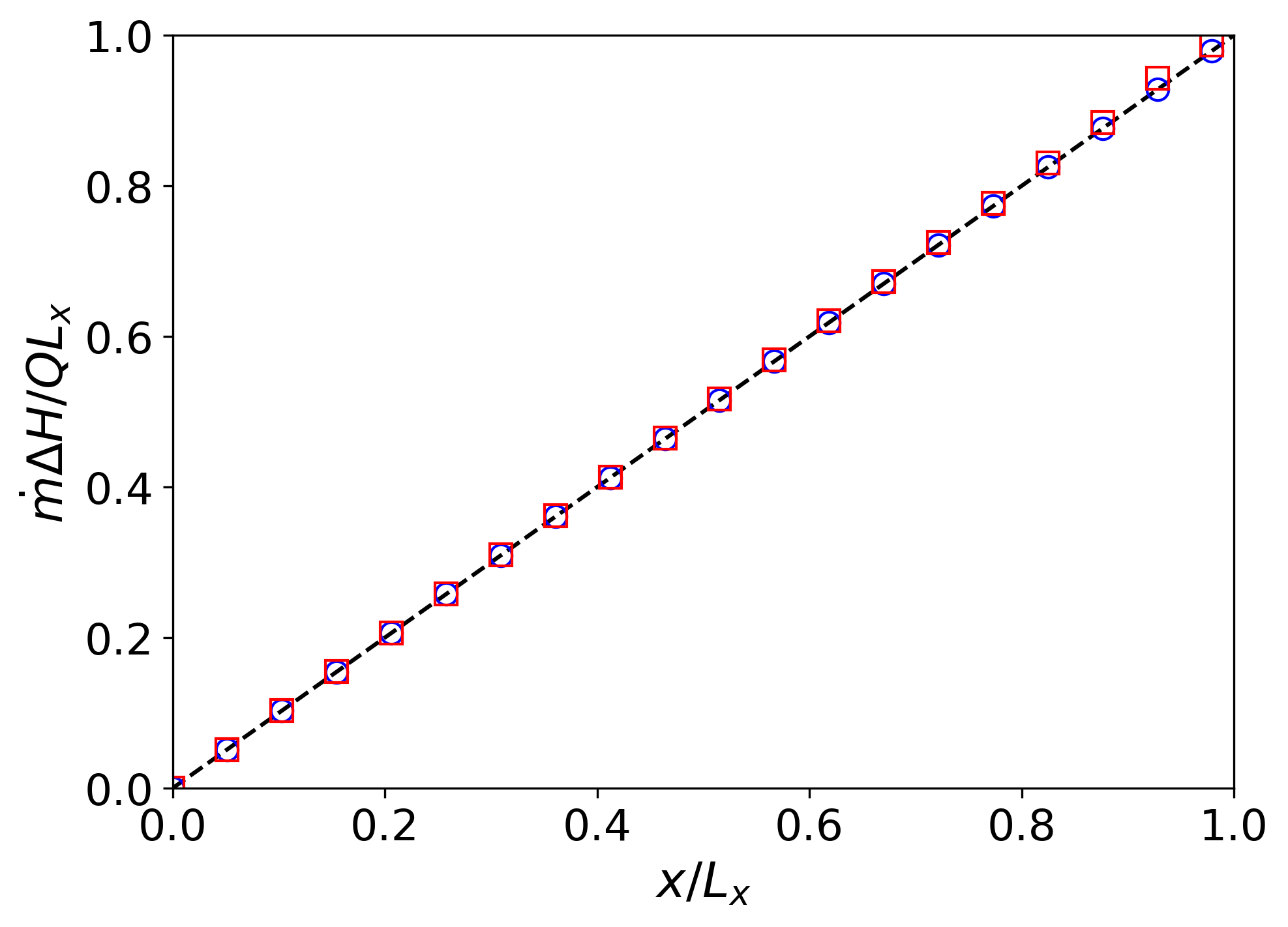}\\
                \includegraphics[width=0.49\linewidth]{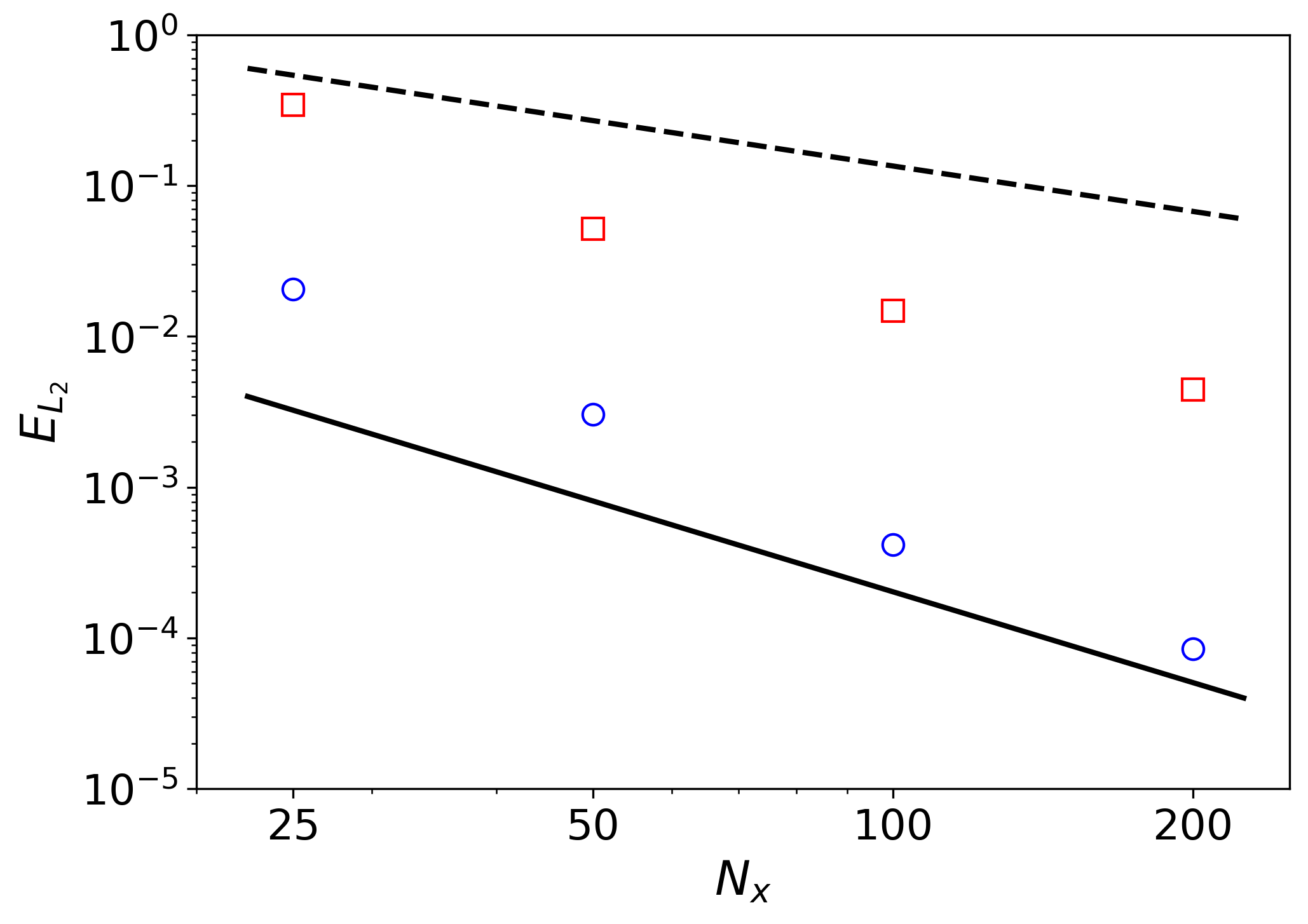}
                \caption{Results from Rayleigh flow simulations. Blue markers correspond to simulations with a subsonic inlet, while red markers correspond to simulations with a supersonic inlet. (Top left) Normalized pressure and temperature, $P/P^*$ and $T/T^*$, plotted as functions of Mach number Ma, where the superscript asterisk denotes the choked (sonic) state. Solid and dashed lines represent the analytical Rayleigh-flow solutions for pressure and temperature ratios, respectively. (Top right) Variation of the total enthalpy flow rate along the duct, normalized by the total heat input rate in the domain. The dashed line represents the analytical result \eqref{eq:ray_H_an}, where the change in total enthalpy flow rate equals the cumulative heat addition. (Bottom) $L_2$ error of the total enthalpy in the duct. Dashed and solid lines have slope $-1$ and $-2$ respectively.}
                \label{fig:Forcing_Ideal_RayleighFlow}%
            \end{figure*}

            Next, in order to probe the consistency of the heat source implementation in compressible thermal flow simulations, we consider the classical Rayleigh flow problem. The Rayleigh flow describes a one-dimensional, steady, inviscid flow of an ideal gas through a constant-area duct with volumetric heat addition. The flow is driven purely by thermodynamic effects due to heat input, without the influence of viscosity or external body forces, making it an ideal benchmark for energy-momentum coupling in compressible flow solvers.

            The governing equations for the Rayleigh flow are obtained from the steady one-dimensional Euler equations with a volumetric heat source $Q$. 
            Conservation of mass yields a constant flux of mass, $\rho u_x = \dot{m}$.
            The momentum equation in the absence of friction and area variation reduces to
            \begin{equation}
                \frac{d}{dx}(\rho u_x^2 + P) = 0.
            \end{equation}
            The energy balance is expressed in terms of the total enthalpy flux, which evolves due to heat addition:
            \begin{equation}
                \frac{d}{dx}(\rho u_x H) = Q,
            \end{equation}
            where $H = h + \frac{u^2}{2}$ denotes the total enthalpy. Thus, the spatial variation of the total enthalpy is directly governed by the prescribed volumetric heat source and is given by
            \begin{equation}
                H(x) = H_{\rm in} + \frac{Q x}{\dot{m}},
                \label{eq:ray_H_an}
            \end{equation}
            where $H_{\rm in}$ is the specific total enthalpy at the inlet.
            For an ideal gas with constant specific heat capacities, these equations admit a closed-form analytical solution in terms of the Mach number. Introducing the critical (sonic) reference state, denoted by a superscript asterisk, the non-dimensional temperature and pressure ratios are given by the Rayleigh relations as
            \begin{gather}
                \frac{T}{T^*} = \frac{(1+\gamma)^2 {\rm Ma}^2}{(1+\gamma {\rm Ma}^2)^2},
                \\
                \frac{P}{P^*} = \frac{1+\gamma}{1+\gamma {\rm Ma}^2}.
            \end{gather}
            When heat is added, the flow evolves towards a sonic state $\mathrm{Ma}=1$, which acts as a limiting (choked) condition for ducts with subsonic and supersonic inlets. 

            Simulations are performed in a one-dimensional domain of length $L_x = 1\,\mathrm{mm}$, discretized using $N_x$ lattice nodes for $N_x \in \{25, 50, 100, 200\}$. Two cases with subsonic and supersonic inlet conditions are considered, respectively.
            For the case with subsonic inlet conditions, two characteristics enter the domain and one leaves it. Accordingly, at the inlet, the total pressure and temperature are prescribed as $P_{0,{\rm in}} = P_{cr}$ and $T_{0,\rm in} = T_{cr}$. At the outlet, the static pressure is specified as $P_{\rm out} = 0.45\,P_{cr}$. The inlet velocity adjusts implicitly to satisfy the imposed stagnation conditions and the interior solution.
            For the case with supersonic inlet conditions, all characteristics enter the domain. Hence, the inlet static pressure, temperature, and Mach number are prescribed as $P_{\rm in} = P_{cr}$, $T_{\rm in} = T_{cr}$, and $\mathrm{Ma}_{\rm in} = 1.5$, respectively. At the outlet, no characteristics enter the domain; therefore, no physical boundary conditions are imposed, and all variables are extrapolated from the interior solution. 
            Wherever boundary data are not explicitly specified, variables are obtained through simple zero-order extrapolation from interior cells, corresponding to von Neumann boundary conditions.
            The volumetric heat source is adjusted so that the flow approaches sonic conditions near the outlet to capture the behavior close to choking. Simulations are initialized with uniform fields equal to the inlet conditions, and evolved until convergence, indicated by a constant mass flow rate throughout the domain. 
            
            Figure \ref{fig:Forcing_Ideal_RayleighFlow} depicts the results. For clarity, only the results corresponding to $N_x = 100$ are shown in the top panels.
            The top left panel demonstrates the variation of normalized pressure and temperature, $P/P^*$ and $T/T^*$, as functions of the Mach number. Both subsonic and supersonic inlet cases collapse onto the corresponding analytical Rayleigh-flow solutions. The top right panel shows the variation of the total enthalpy flow rate along the duct, normalized by the total heat input rate. The numerical results closely follow the analytical reference \eqref{eq:ray_H_an}. This agreement confirms the consistent formulation and implementation of the volumetric heat source. 
            
            The bottom panel presents the $L_2$-norm relative error in specific total enthalpy for various grid resolutions for the subsonic and supersonic simulations. 
            It can be seen that the expected quadratic decrease of the error with grid refinement is achieved, again confirming the second-order convergence of the method. 
            Note that in the context of this convergence study, acoustic scaling has been applied to the time-step size. Furthermore, to maintain stability and recover the proper convergence rate, the viscosity applied in simulation scale with $\delta x^2$. This choice ensure non-zero viscosity, maintaining stability, while ensuring that the error brought about by the viscosity scales with the Burnett level, i.e. maintains the order of convergence of the solver. 
            
        \subsubsection{Fanno flow}

            \begin{figure*}[htbp!]
                \centering
                \includegraphics[width=0.49\linewidth]{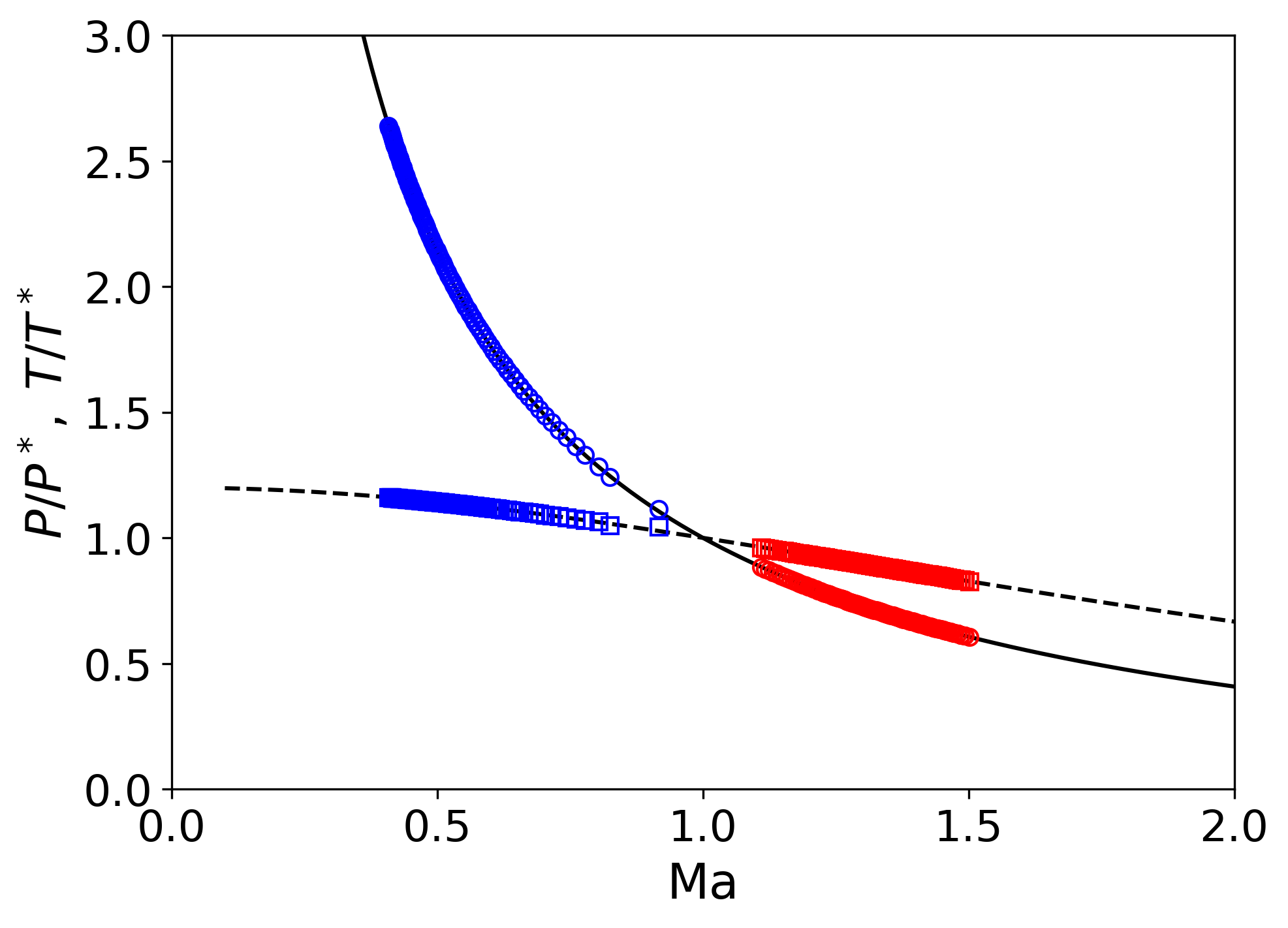}\phantom{a}%
                \includegraphics[width=0.49\linewidth]{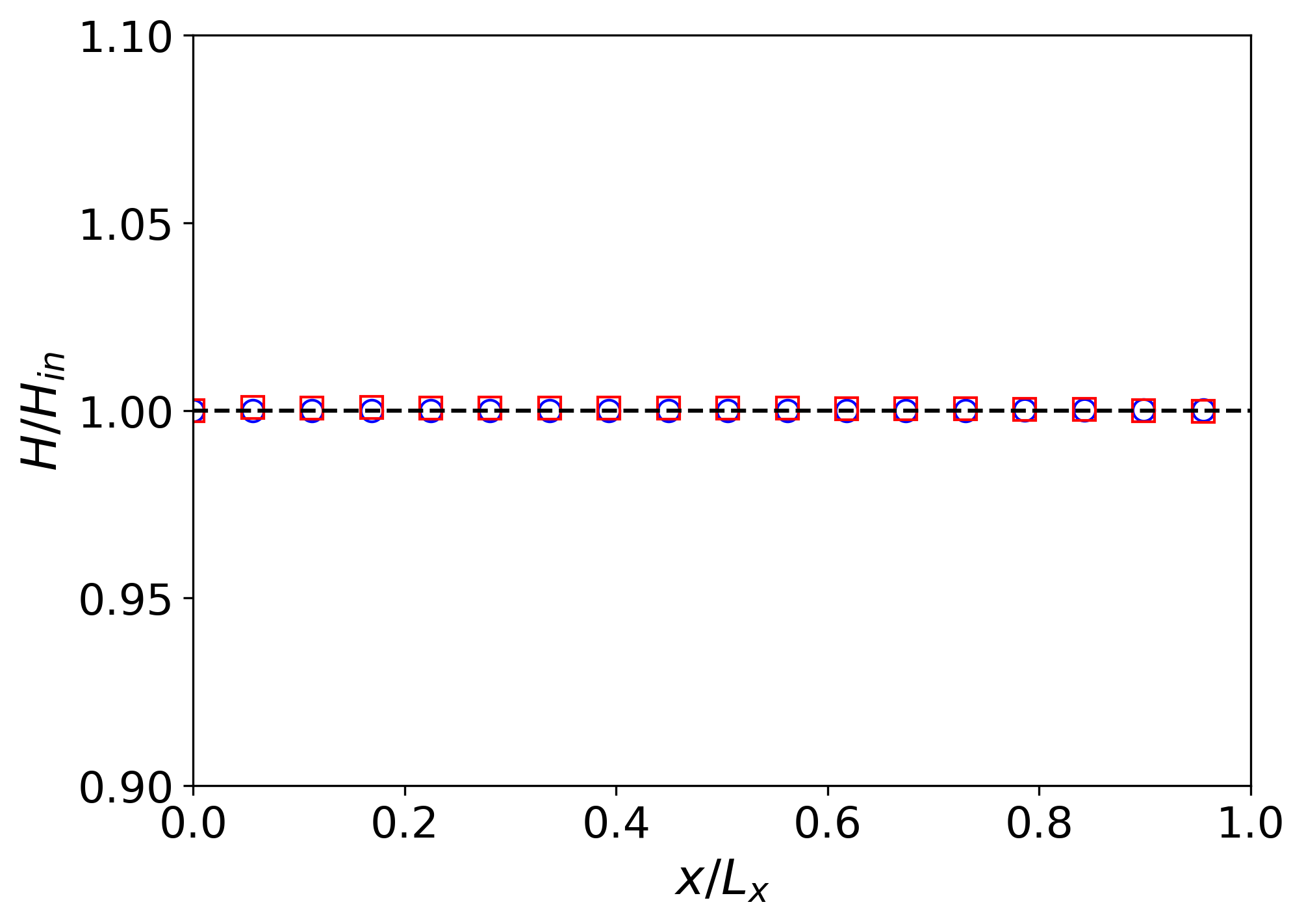}%
                \caption{Results from Fanno flow simulations. Blue markers correspond to simulations with a subsonic inlet, while red markers correspond to simulations with a supersonic inlet. (Left) Normalized pressure and temperature, $P/P^*$ and $T/T^*$, plotted as functions of Mach number Ma, where the superscript asterisk denotes the choked (sonic) state. Solid and dashed lines represent the analytical Fanno-flow solutions for pressure and temperature ratios, respectively. (Right) Variation of the total enthalpy along the duct, normalized by the inlet value. The dashed line represents the analytical result, where the total enthalpy remains constant along the duct.}
                \label{fig:Forcing_Ideal_FannoFlow}%
            \end{figure*}
        
            As a final benchmark for ideal compressible flows, we consider the Fanno flow problem, which describes steady, one-dimensional, adiabatic flow through a constant-area duct in the presence of wall friction. The problem is named after the Italian engineer Gino Girolamo Fanno, who developed the underlying analysis in his 1904 master's thesis at ETH Zürich \cite{FannoGinnoLIBETHLib, bar2004fundamentals}. Unlike Rayleigh flow, where energy exchange occurs through heat addition, Fanno flow is driven purely by viscous effects, which redistribute energy between kinetic and internal forms while maintaining adiabatic conditions.

            The governing equations consist of the one-dimensional Euler equations augmented by a friction forcing term. The momentum equation is written as
            \begin{equation}
                \frac{d}{dx}(\rho u_x^2 + P) = -\frac{f_D}{D_h} \frac{\rho u_x^2}{2},
            \end{equation}
            where $F_{x} = -f_D/D_h \ \rho u_x^2/2$ represents the streamwise friction force, with $f_D$ and $D_h$ being the Darcy friction factor and hydraulic diameter of the duct respectively. The flow remains adiabatic, and therefore the total enthalpy $H$ is conserved along the duct.
            Similar to the Rayleigh flow relations, these equations also admit a closed-form analytical solution in terms of the Mach number. Introducing the critical (sonic) reference state, denoted by a superscript asterisk, the non-dimensional temperature and pressure ratios can be expressed as
            \begin{gather}
                \frac{T}{T^*} = \frac{1+\gamma}{2+(\gamma-1) {\rm Ma}^2},
                \\
                \frac{P}{P^*} = \frac{1}{\rm Ma}\sqrt{\frac{\gamma + 1}{2 + (\gamma - 1) {\rm Ma}^2}}.
            \end{gather}
            Here as well, friction drives the flow towards the sonic state for both subsonic and supersonic inlets.
            
            Simulations are performed on the same domain as in the Rayleigh flow case, for the two inlet regimes. For the subsonic inlet case, the total pressure and temperature at the inlet are prescribed as $P_{0,\rm in} = P_{cr}$ and $T_{0,\rm in} = T_{cr}$ and the static pressure at the outlet is specified as $P_{\rm out} = 0.3\,P_{cr}$. For the supersonic inlet case, the inlet static pressure, temperature, and Mach number are prescribed as $P_{\rm in} = P_{cr}$, $T_{\rm in} = T_{cr}$, and ${\rm Ma}_{\rm in} = 1.5$, respectively. The hydraulic diameter is set to $D_h = 0.1 L_x$, and the friction factor is adjusted such that the flow approaches close to the sonic conditions at the outlet.
            In the 1D numerical formulation, friction is modeled through the body force term $F_x$ in the momentum equation. However, to ensure thermodynamic consistency with Fanno flow physics, the mechanical work extracted by this force must be converted into internal energy, which in reality would happen due to viscous dissipation. This is enforced by introducing a compensating energy source term, $Q = -\bm{F}\cdot \bm{u}$ in the total energy equation.
            
            Figure \ref{fig:Forcing_Ideal_FannoFlow} shows the results. The left panel illustrates the variation of the normalized pressure and temperature, $P/P^*$ and $T/T^*$, as functions of the Mach number. Both subsonic and supersonic inlet conditions collapse onto the corresponding analytical Fanno-flow solutions. The right panel presents the variation of the total enthalpy flux along the duct, normalized by its inlet value. It can be seen that the total enthalpy remains constant along the flow, validating the consistency of the coupling between the momentum and energy equations and the corresponding source terms.

    \subsection{Non-ideal and multiphase compressible flows}\label{sec:nonidealgascompressiblemultiphase}

            We continue the validation of body force and heat source in this section for the non-ideal and multiphase regime.
            
            Firstly, the van der Waals equation of state, cf. \eqref{eq:genericEOS}, is used,
            \begin{equation}\label{eq:vdw_eos}
                P = \frac{\rho R T}{1 - b\rho} - a\rho^2.
            \end{equation}
            The long-range molecular attraction parameter $a$ and excluded volume parameter $b$ are defined in terms of the critical-state thermodynamic variables, critical temperature $T_{cr}$ and critical pressure $P_{cr}$ as: $a = 27R^2 T_{cr}^2 / 64 P_{cr}$, $b = R T_{cr} / 8 P_{cr}$. 
            For all non-ideal and multiphase test cases in the present section, the critical properties of Nitrogen (N$_2$) listed in Table \ref{tab:nitrogen_properties} are used.
            For the van der Waals fluid, since the equation of state \eqref{eq:vdw_eos} is linear in temperature, the specific heat at constant volume $c_v$ is independent of density. Hence, integration of \eqref{eq:de_rho} leads to the explicit relation,  cf. \eqref{eq:TdependsongenericEOS}, 
            \begin{equation}\label{eq:lbm_temp}
                T = \frac{E - \frac{1}{2}\bm{u}^2 + a\rho}{c_v}.
            \end{equation}
            This closed-form inversion results from the simple caloric structure of the van der Waals model. 

            \begin{table}[h]
                \caption{Critical properties for Nitrogen. The critical density is calculated by fitting the critical temperature and pressure from \cite{NitrogenProperties_Jacobsen} to the van der Waal equation of state.}
                \centering
                \setlength{\tabcolsep}{10pt}
                \begin{tabular}{ccccc}
                \hline
                Fluid    & $R/c_v$ & $P_{cr}$ (Pa)     & $T_{cr}$ (K) & $\rho_{cr}$ (kg/m\textsuperscript{3}) \\ \hline
                Nitrogen & 0.4     & $3.4 \times 10^6$ & 126.2        & 241.96                                                 \\ \hline
                \end{tabular}
                \label{tab:nitrogen_properties}
            \end{table}

           \begin{figure*}[htbp!]
                \centering
                \includegraphics[width=0.49\linewidth]{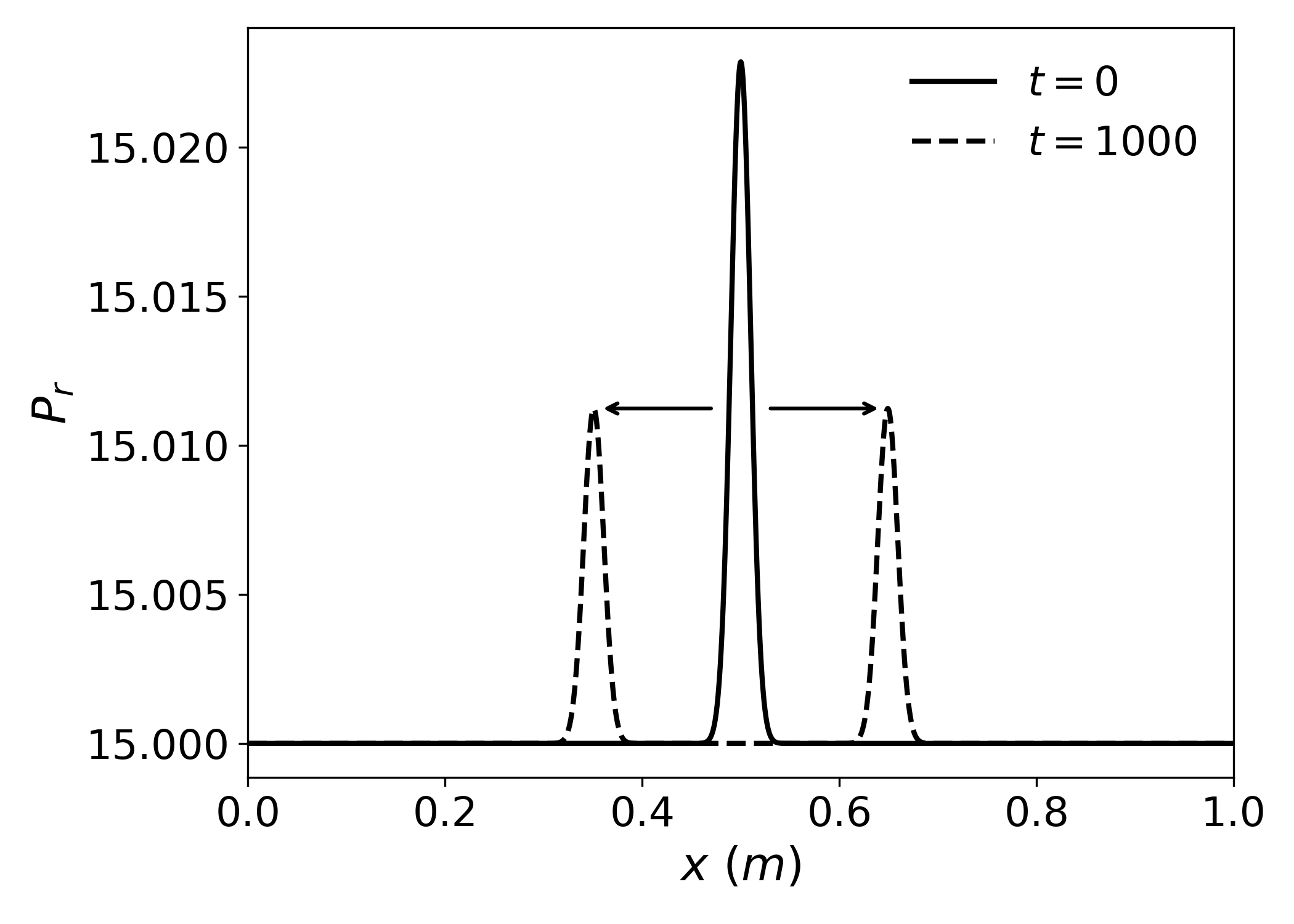}
                \includegraphics[width=0.49\linewidth]{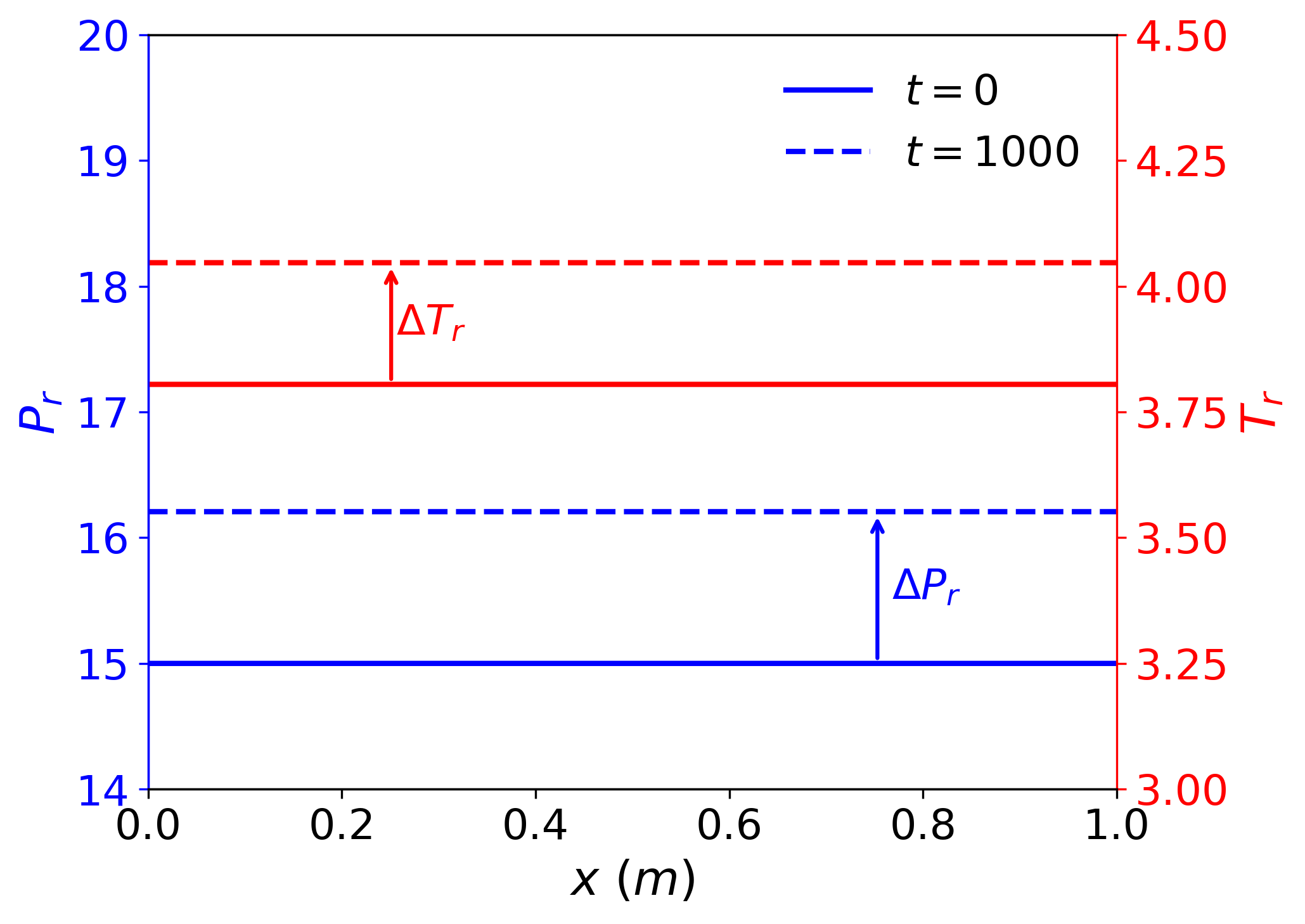}
                \caption{Setup for measuring the Joule-Thomson coefficient based on (left) the isothermal speed of sound and (right) isochoric heating for $P_{r} = 15$, $\hat{h} = 15$.}
                \label{fig:Forcing_JT_measurederivatives}
            \end{figure*}
            
            Secondly, for a non-ideal fluid with capillary effects in the multiphase regime, the total force is decomposed as
            \begin{equation}
                \bm{F}=\bm{F}_{\rm K}+\bm{F}_{\rm B},
                \label{eqn:force_decomposition}
            \end{equation}
            where \(\bm{F}_{\rm K}\) is the Korteweg capillary force and \(\bm{F}_{\rm B}\) is an external body force. The Korteweg force is defined as the negative divergence of the Korteweg stress tensor,
            \begin{equation}
                \bm{F}_{\rm K}=-\bm{\nabla}\cdot\bm{T}_{\rm K},
            \end{equation}
            where
            \begin{equation}
                \bm{T}_{\rm K} = \kappa\,\bm{\nabla}\rho\otimes\bm{\nabla}\rho -
                \kappa\left( \rho\nabla^{2}\rho +\frac{1}{2}|\bm{\nabla}\rho|^{2}\right)\bm{I},
                \label{eqn:korteweg_tensor}
            \end{equation}
            and \(\kappa\) denotes the capillarity coefficient, which is also a tunable parameter similar to similar to $\mu$, $\eta$, $k$, $R$, and $c_v$. 
            Taking the divergence of \(\bm{T}_{\rm K}\) yields 
            \begin{equation}
                \bm{F}_{\rm K} = \kappa \rho\,\bm{\nabla}\nabla^{2}\rho,
                \label{eqn:korteweg_force}
            \end{equation}
            which is the expression applied in all following benchmarks containing multiple fluid phases.

        \subsubsection{Inversion line and Joule--Thomson coefficient\label{sec:JTthrottling1}}
           
            To assess the thermodynamic consistency of the proposed model for non-ideal fluids in the presence of source terms, we start by evaluating its ability to reproduce the Joule--Thomson (JT) effect, similar to \cite{Reyhanian21}. During a throttling process, a fluid passes through a restriction without exchanging heat or performing external work. Under these conditions, the specific enthalpy remains constant, while the pressure decreases. The resulting temperature change is characterized by the Joule--Thomson coefficient, 
            \begin{equation}        
                \mu_{JT}=\left(\frac{\partial T}{\partial P}\right)_h = - \frac{1}{C_P} \left[\frac{1}{\rho} - \frac{T}{\rho^2} \frac{(\partial P/\partial T)_{\rho}}{ (\partial P/\partial \rho)_{T}}\right],
                \label{eqn:JT_coeff}
            \end{equation}
            The sign of $\mu_{JT}$ determines whether the fluid cools or heats during throttling. For an ideal gas, $\mu_{JT}=0$, implying no temperature change. Real fluids, however, exhibit positive and negative JT coefficients, separated by the inversion line in the $T-P$ plane.

            To verify that the present model correctly reproduces this behavior, the Joule--Thomson coefficient is evaluated numerically at several thermodynamic states in two steps. In the first step, the derivative $(\partial P/\partial \rho)_T$ is determined from the isothermal speed of sound. A small Gaussian pressure perturbation is introduced into an otherwise quiescent domain, and the propagation speed of the resulting pressure pulses is measured. Since the process is isothermal, the $g$-populations are not evolved during this simulation.
            In the second simulation, the derivative $(\partial P/\partial T)\rho$ is obtained through isochoric heating. A spatially uniform heat source $Q$ is applied to a periodic domain initially at rest. Owing to the periodic boundary conditions, the total volume remains constant throughout the simulation, thereby enforcing isochoric conditions. The measured values of the derivatives are the substituted in Eq. \eqref{eqn:JT_coeff}. Representative simulation results for both steps are shown in Fig. \ref{fig:Forcing_JT_measurederivatives}. The initially imposed Gaussian pressure pulse splits into two waves that propagate in opposite directions in the first step, while the second step shows the increase in temperature and pressure resulting from uniform isochoric heating.
            
            Simulations are performed for three different reduced enthalpies $\hat{h} = h / R T_{cr} \in \{5, 11.25, 15\}$ at reduced pressures $P_r \in \{3, 6, 9, 12, 15\}$, and the measured values of the Joule--Thomson coefficient are compared against the theoretical predictions from the van der Waal's EOS in Fig. \ref{fig:Forcing_NI_InversionLine}. Excellent agreement is observed between the theoretical and measured values.

        \subsubsection{Throttling process\label{sec:JTthrottling2}}

            \begin{figure*}[htbp!]
                \centering
                \includegraphics[width=0.49\linewidth]{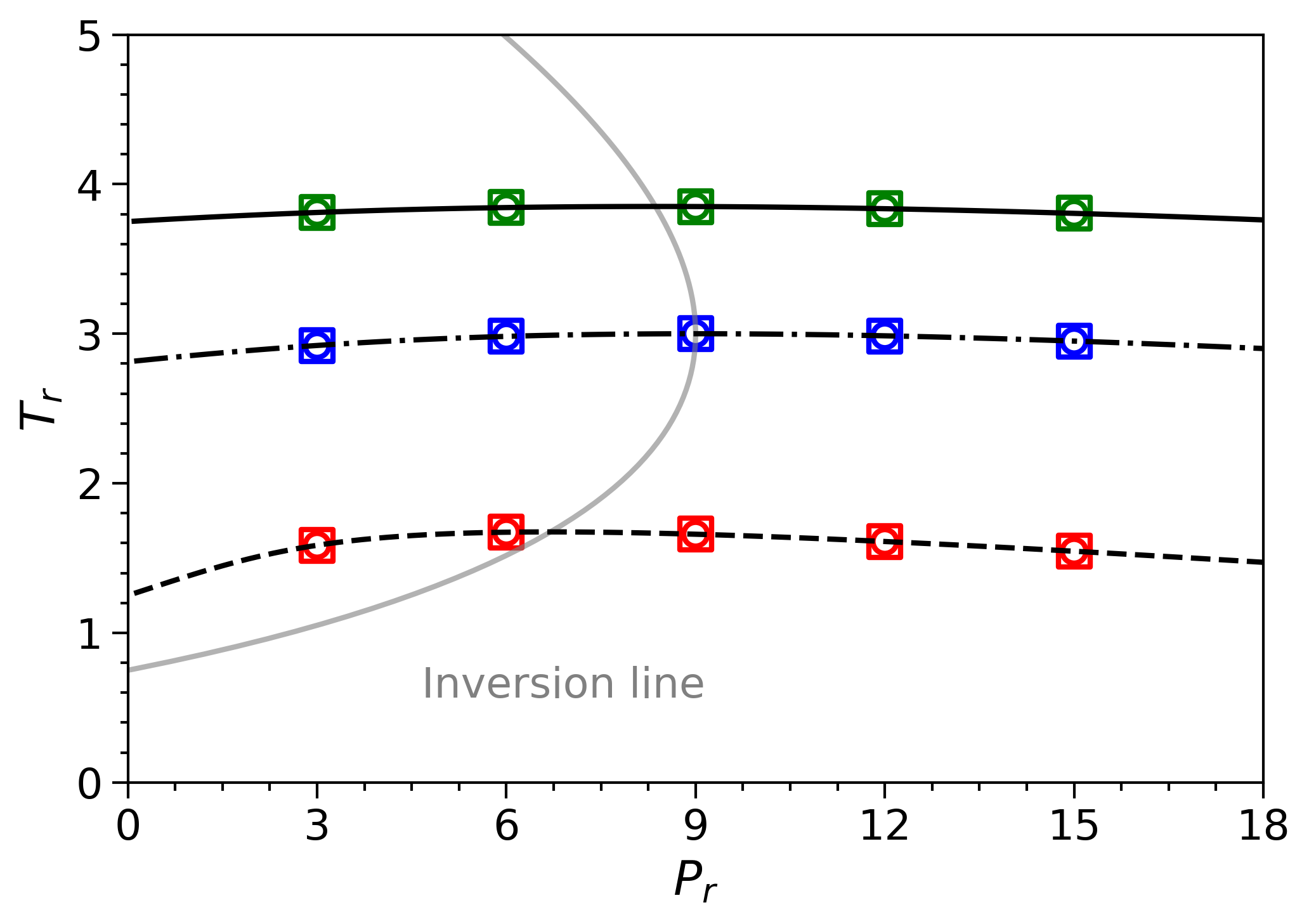}\phantom{a}%
                \includegraphics[width=0.49\linewidth]{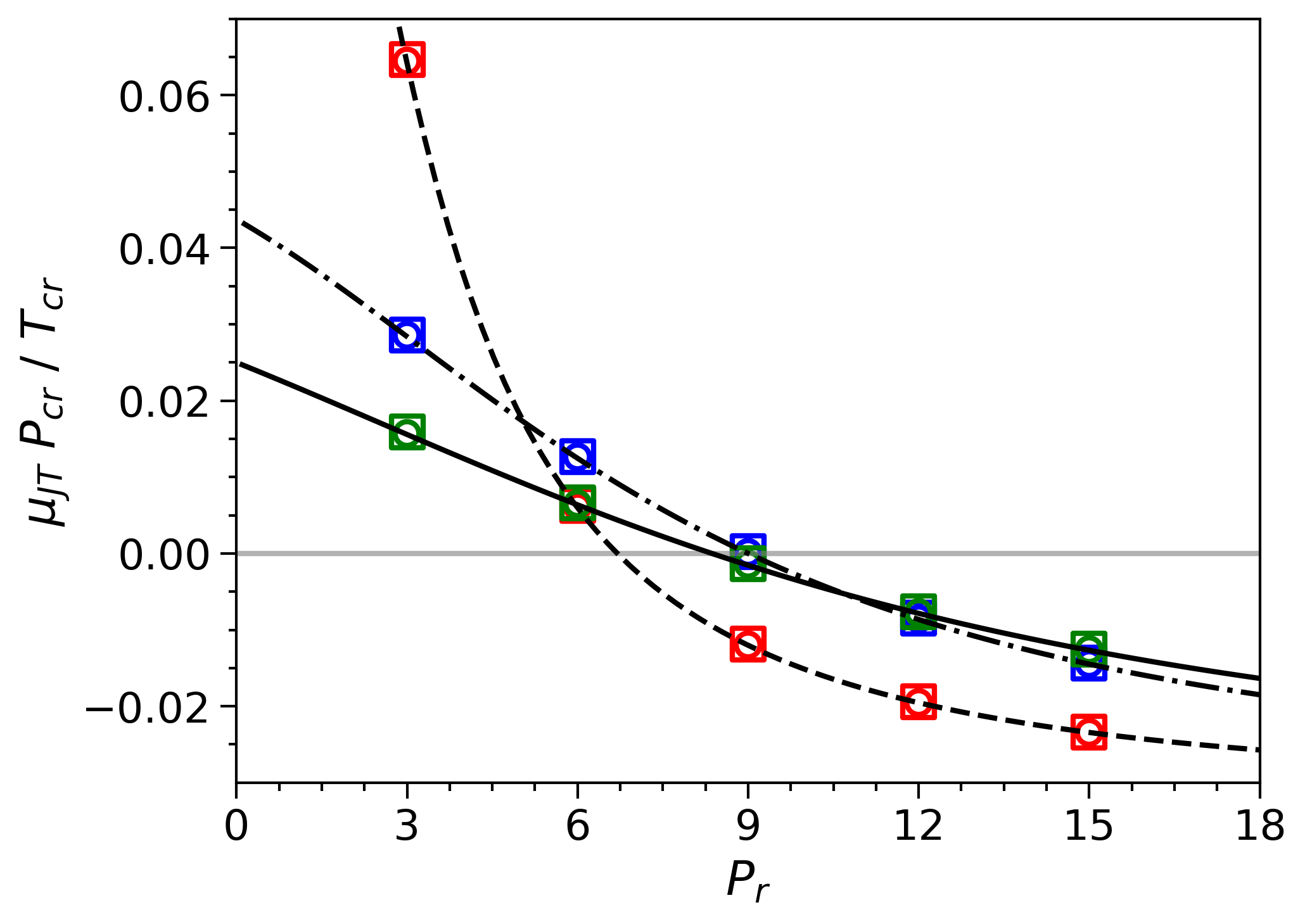}%
                \caption{Simulation results compared to theoretical predictions from the van der Waals equation of state. 
                (Left) The thermodynamic states in the $P-T$ plane, with the inversion line ($\mu_{JT}=0$) indicated by the solid gray curve. (Right) The corresponding Joule--Thomson coefficient. 
                The dashed, dash-dot, and solid curves denote iso-contours of $\hat{h} = 5$, $11.25$, and $15$, respectively.
                Circular markers represent values obtained from thermodynamic derivative evaluations (Sec.~\ref{sec:JTthrottling1}), while square markers correspond to direct throttling simulations (Sec.~\ref{sec:JTthrottling2}).}
                \label{fig:Forcing_NI_InversionLine}%
            \end{figure*}
            
            \begin{figure*}[htbp!]
                \centering
                \includegraphics[width=0.49\linewidth]{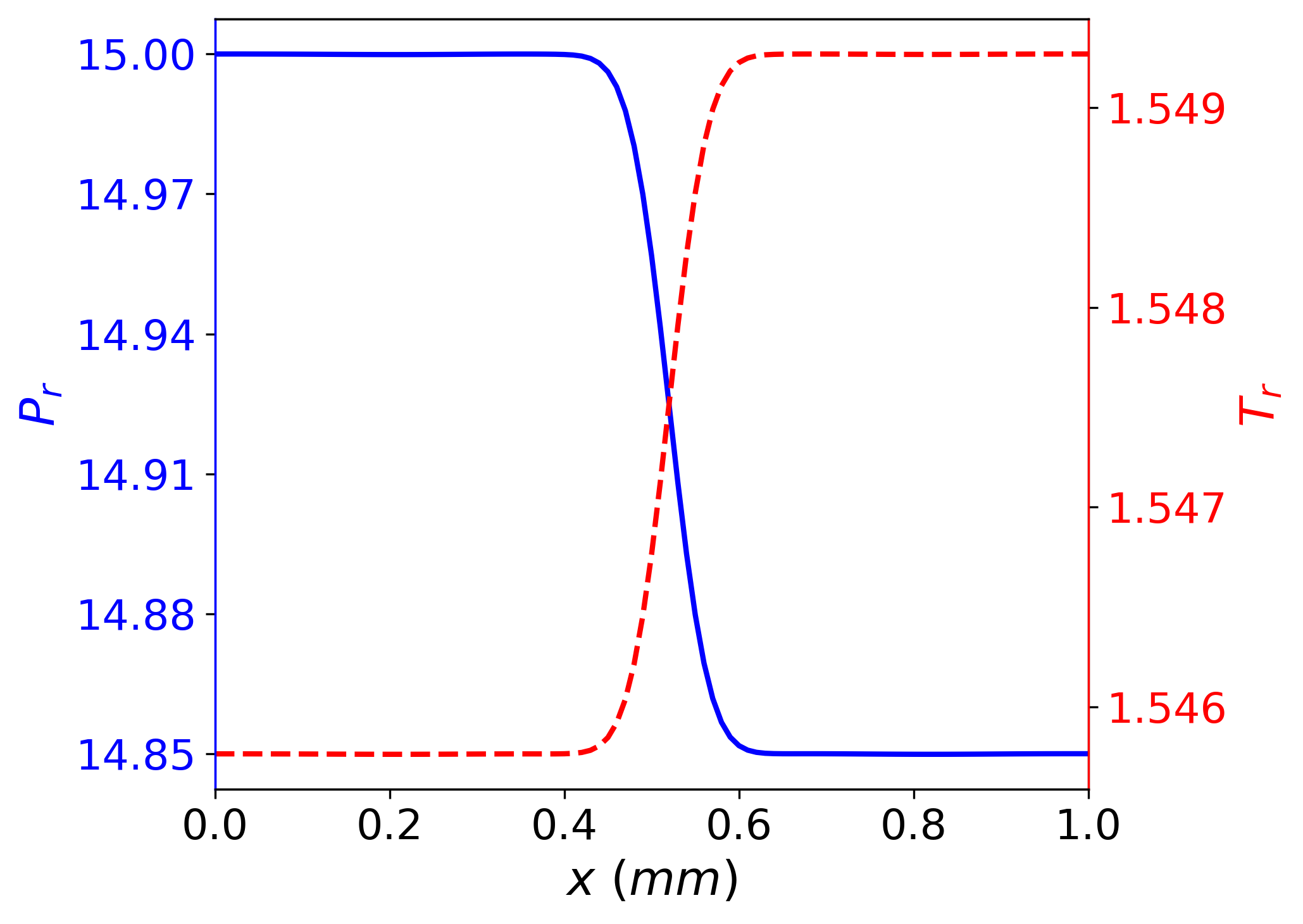}\phantom{a}%
                \includegraphics[width=0.49\linewidth]{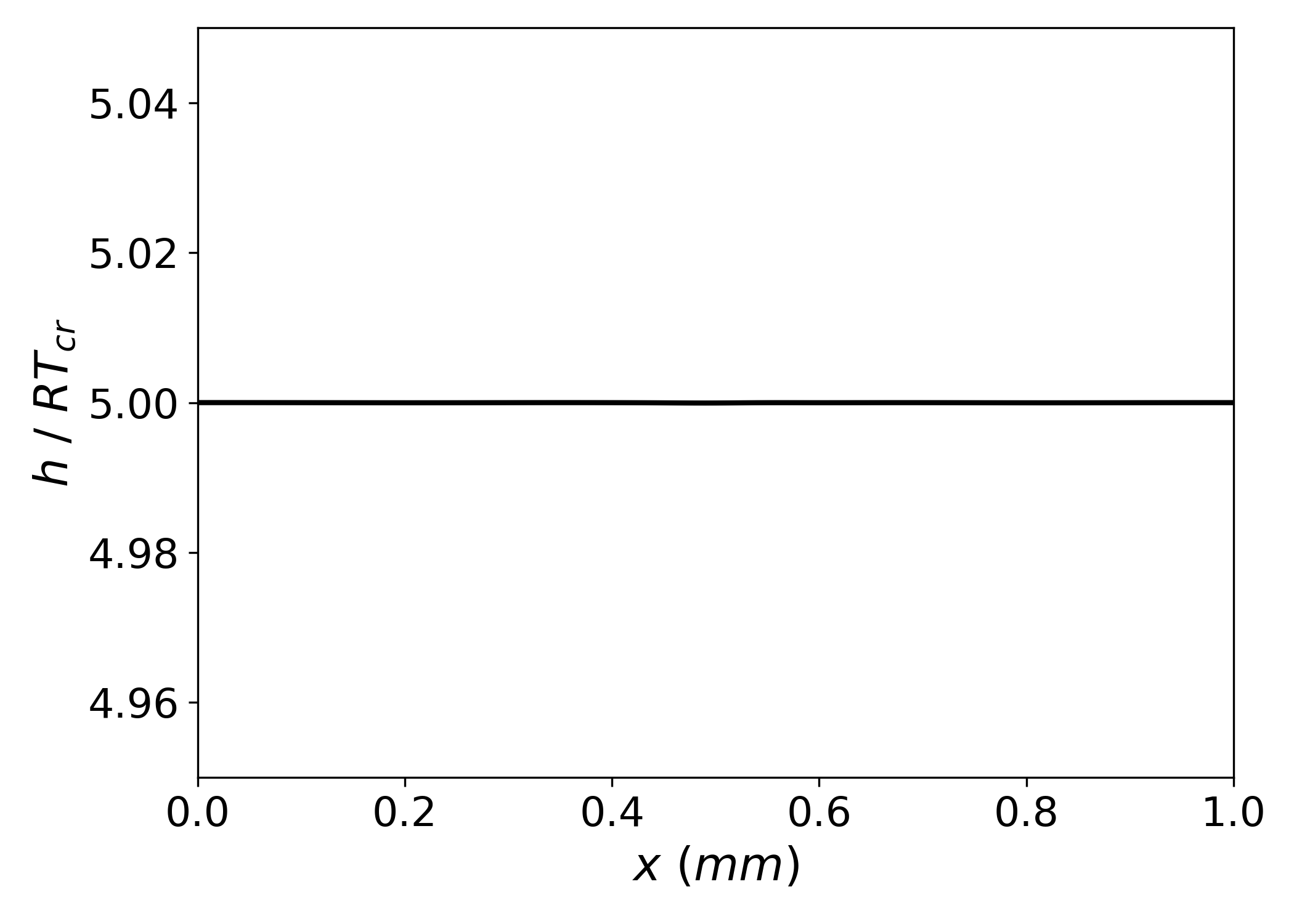}%
                \caption{Results for a throttling simulation at steady state with $P_{r,\rm in} = 15$, $\hat{h}_{\rm in} = 5$, and $P_{r,\rm out} = 0.99\,P_{r,\rm in}$. (left) Reduced pressure (solid line) and temperature (dashed line) profiles. (right) Reduced enthalpy profile.}
                \label{fig:Forcing_NI_Throttling}%
            \end{figure*}
        
            Having verified the thermodynamic derivatives entering the Joule--Thomson coefficient independently, we next simulate the throttling process directly and measure the resulting temperature change. This provides a more stringent validation of the coupling between the momentum and energy equations, as well as the corresponding forcing terms. 

            The simulations are performed in a one-dimensional domain of length $L_x = 1$ mm, discretized using a  lattice spacing of $\delta x = L_x/100$. Fixed thermodynamic conditions $(P_{\rm in}, T_{\rm in})$ are prescribed at the inlet, while the outlet pressure is set to $P_{\rm out} = 0.99 P_{\rm in}$, thereby introducing a pressure drop in the domain. To model the throttling device, a localized momentum sink is introduced through a Darcy-type resistance force,  
            \begin{equation}
                F_{x} = -K \ \phi(x) \ \rho u_x,
            \end{equation}
            where $K$ is a throttle-strength parameter and $\phi(x)$ is a prescribed shape function. The latter is chosen as a Gaussian centered in the domain,
            \begin{equation}
                \phi(x) = \exp\left[-\left(\frac{x-L_x/2}{\sigma}\right)^2\right],
            \end{equation}
            where $\sigma$ controls the spatial extent of the throttling region. In the present simulations, $K=0.01$ and $\sigma=5$ lattice units are used. Similar to the Fanno-flow simulations, mechanical energy dissipated by the throttling force needs to be converted into internal energy, which is modeled by a compensating volumetric heat source $Q = -\bm{F} \cdot \bm{u}$. 

            The simulations are performed for the same thermodynamic states considered in the previous section (Sec.~\ref{sec:JTthrottling1}). 
            Each case is initialized with uniform conditions equal to the inlet state and evolved until a steady state is reached, as indicated by a spatially uniform and time-independent mass flux through the domain. Figure~\ref{fig:Forcing_NI_Throttling} shows the pressure, temperature and enthalpy distributions at the end of a representative simulation. A pressure drop accompanied by a corresponding change in temperature is observed across the throttle region located at the center of the domain. The enthalpy remains uniform throughout the domain, confirming the consistency of the forcing and heat-source formulations and the conservation of enthalpy across the throttling process. The Joule--Thomson coefficient is then obtained directly from the simulated temperature and pressure changes as $\Delta T / \Delta P$, where $\Delta T$ is the temperature difference between the inlet and outlet and $\Delta P$ is the imposed pressure drop. Figure~\ref{fig:Forcing_NI_InversionLine} compares the resulting values of $\mu_{JT}$ with both the theoretical predictions of the van der Waals equation of state and the values obtained from the thermodynamic-derivative approach described in the previous section (Sec.~\ref{sec:JTthrottling1}). The results from the direct throttling simulations are indistinguishable from those obtained using the independently measured derivatives and exhibit excellent agreement with the theoretical predictions.
            
        \subsubsection{Interface consistency and liquid-vapor co-existence}
                                      
            The addition of the Korteweg force in non-ideal multiphase regimes is assessed next. 
            For that, we examine the liquid-vapor two-phase coexistence as a validation of thermodynamic and mechanical consistency of the interface, where we add  $\bm{F} = \bm{F}_{\rm K}$, cf. \eqref{eqn:korteweg_force}.
            Simulations are performed for reduced temperatures $T_r \in [0.6, 0.7, 0.8, 0.9]$ in a one-dimensional periodic domain of size $L_x = 0.5$ mm with $\delta x = 0.1$ \textmu{}m. The domain is initially filled with a liquid column in $x \in [L_x/4, 3 L_x/4]$ and vapor on either side. The reference liquid and vapor densities are obtained using Maxwell's equal area construction. The capillarity coefficient is set to $\kappa = 10^{-10}$ m$^7$ kg$^{-1}$ s$^{-2}$. Simulations are evolved until the density field converges. 
    
\begin{figure}[h]
                \centering
                \includegraphics[width=0.98\linewidth]{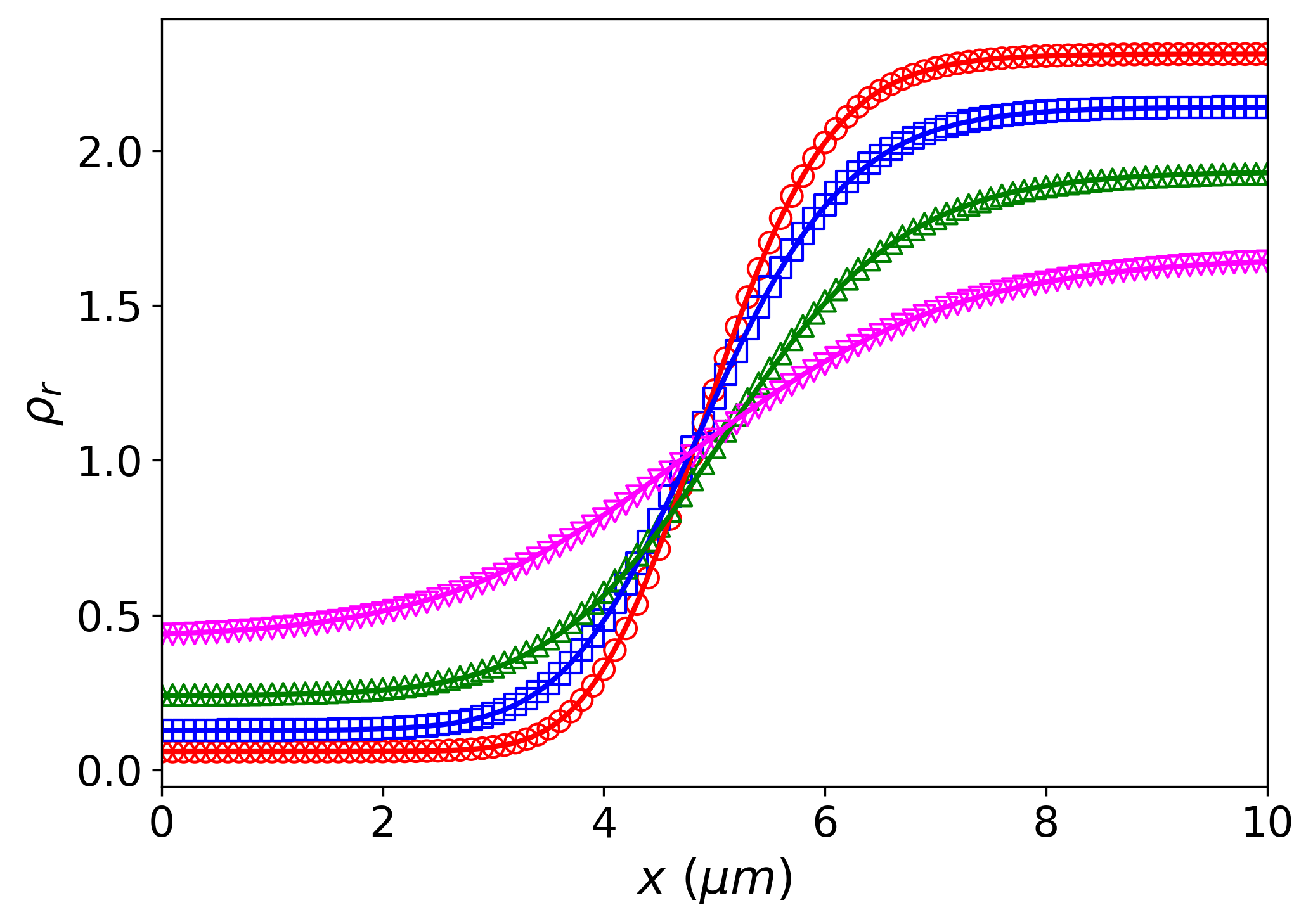}
                \caption{Solutions of reduced coexistence densities (markers) at different reduced temperatures compared to the reference solution (solid lines) obtained from a shooting method solution to the boundary value problem \ref{eqn:coexistence_bvp}. Markers $\mcirc$, $\square$, $\triangle$, and $\mtriangledown$ represent the results for $T_r \in \{0.6, 0.7, 0.8, 0.9\}$ respectively. }
                \label{fig:Kortewegforce_CoexistenceDensities}%
            \end{figure}

            The reference solution is obtained by solving for the mechanical equilibrium condition
            \begin{equation}
                \partial_x P = \kappa \rho \partial_x^3\rho
                \label{eqn:coexistence_bvp}
            \end{equation}
            on the half-domain $x \in [0, L_x/2]$, subject to $\rho = \rho_v$ and $\partial_x\rho = 0$ at $x = 0$, and $\partial_x\rho=0$ at $x = L_x/2$, using a high-resolution shooting method. The solution is subsequently mirrored about $x = L_x/2$ to recover the density profile of the full-domain. The results obtained from the simulations are compared to this reference solution in Fig. \ref{fig:Kortewegforce_CoexistenceDensities}. Excellent agreement can be found for the entire range of reduced temperatures.

\section{Summary, conclusions and outlook\label{sec:conslusions}}

    A consistent framework for incorporating body-force and heat-source terms into a lattice Boltzmann model for compressible flows of generic fluids has been developed and validated. Building upon a thermodynamically consistent double-distribution-function kinetic framework, the proposed formulation introduces forcing and heating through shifted quasi-equilibrium states while retaining the consistent recovery of the Navier--Stokes--Fourier equations. The resulting discrete-velocity realization preserves the required conservation properties and enables independent control of the thermodynamic and transport coefficients, including shear viscosity, bulk viscosity, and thermal conductivity, for both ideal and non-ideal fluid regimes.
    
    The hydrodynamic consistency of the formulation was established through Chapman--Enskog analysis and subsequently assessed using a broad hierarchy of numerical benchmarks covering ideal-gas compressible flows as well as non-ideal and multiphase compressible flows. The model accurately reproduces non-classical shock-tube dynamics, thermal Couette flow, and force-driven Poiseuille and Womersley flows, thereby validating the treatment of compressibility, viscous and thermal transport, and body-force effects. The consistent incorporation of heat sources and external forcing was further demonstrated for Rayleigh and Fanno flows, the Joule--Thomson effect, and throttling processes. For non-ideal and multiphase conditions, the formulation correctly captures interface properties and liquid--vapor coexistence. Across all investigated cases, excellent agreement with analytical solutions and reference data was obtained, while spatio-temporal grid-refinement studies confirmed the expected second-order accuracy of the scheme.
    
    The present framework provides a robust and efficient foundation for lattice Boltzmann simulations of highly compressible flows of generic fluids with physical and numerical source terms for complex engineering and physics applications. In particular, the consistent incorporation of body forces and heat sources is essential for force- and heat-driven flows, thermodynamic processes, and immersed-boundary simulations in complex geometries, which shall be the focus of future work.

\section*{Acknowledgments}
This work was supported by the European Research Council (ERC) Advanced Grant No. 834763-PonD and by the Swiss National Science Foundation (SNSF) Grant Nos. 200021-228065 and 200021-236715.
Computational resources at the Swiss National Supercomputing Centre (CSCS) were provided under Grant Nos. s1286, sm101 and s1327.
Open access funding was provided by the Swiss Federal Institute of Technology Zürich (ETH Zürich).

\section*{Author Declarations}

\noindent\textbf{\textit{Conflict of interest}}\\
The authors have no conflicts to disclose.

\noindent\textbf{\textit{Ethical approval}}\\
The work presented by the authors herein did not require ethics approval or consent to participate.

\noindent\textbf{\textit{AI-assisted technologies}}\\
The authors declare that no AI or AI-assisted technologies were used in the research process, including conceptualization and design of the study, methodology development, data analysis, or interpretation. 
AI-assisted tools were used during the preparation of this manuscript for basic language improvement and writing support.

\noindent\textbf{Author contributions}\\
S.C.:
Conceptualization of the study, 
formal analysis,
derivation and implementation of the solver, 
evaluation of the solver, 
data analysis,
writing- initial manuscript and revised versions. 
\\
R.M.S.:
Conceptualization of the study, 
formal analysis,
writing- initial manuscript and revised versions. 
\\
S.A.H.:
Conceptualization of the study, 
formal analysis,
writing- initial manuscript and revised versions. 
\\
I.V.K.:
Conceptualization of the study, 
funding acquisition, resources,
writing- initial manuscript and revised versions.
\\
All authors approved the final manuscript.

\noindent\textbf{Data availability statement}\\
The data that support the findings of this study are available within the article or from the corresponding author(s) upon reasonable request.

\noindent\textbf{Open access}\\
This article is licensed under a Creative Commons Attribution 4.0 International License, which permits use, sharing, adaptation, distribution and reproduction in any medium or format, as long as you give appropriate credit to the original author(s) and the source, provide a link to the Creative Commons licence, and indicate if changes were made. The images or other third party material in this article are included in the article’s Creative Commons licence, unless indicated otherwise in a credit line to the material. If material is not included in the article’s Creative Commons licence and your intended use is not permitted by statutory regulation or exceeds the permitted use, you will need to obtain permission directly from the copyright holder. To view a copy of this licence, visit http://creativecommons.org/licenses/by/4.0/.

\section*{Appendix}
\appendix

\section{Multi-scale analysis\label{app:CE}}

The first step in the multi-scale analysis is a Taylor expansion of the lattice Boltzmann equations,
\begin{multline}
    \{f_i,g_i\}(\bm{x}+\bm{c}_i\delta t, t+\delta t)
    \\ = \{f_i,g_i\}(\bm{x}, t) + 2\beta\left(\{f_i^{\rm eq},g_i^{\rm eq}\}(\bm{x},t) - \{f_i,g_i\}(\bm{x},t)\right) \\+ \left(1-\beta\right) \left(\{f_i^{\star},g_i^{\star}\}(\bm{x},t) - \{f_i^{\rm eq},g_i^{\rm eq}\}(\bm{x},t)\right),
\end{multline}
around $(\bm{x},t)$, leading to the following space and time-evolution equations,
\begin{multline}
    \delta t\mathcal{D}_t \{f_i,g_i\} + \frac{\delta t^2}{2}{\mathcal{D}_t}^2 \{f_i,g_i\} + \mathcal{O}\left(\delta t^3\right) \\= 2\beta\left(\{f_i^{\rm eq},g_i^{\rm eq}\} - \{f_i,g_i\}\right) + \left(1-\beta\right)\left(\{f_i^\star,g_i^\star\} - \{f_i^{\rm eq},g_i^{\rm eq}\}\right).
\end{multline}
Introducing the flow characteristic size and time, $\mathcal{L}$ and $\mathcal{T}$ the equations are made non-dimensional as,
\begin{multline}
    \frac{\delta x}{\mathcal{L}}\mathcal{D}'_t \{f_i,g_i\} + \frac{\delta x^2}{2\mathcal{L}^2}{\mathcal{D}'_t}^2 \{f_i,g_i\} = 2\beta\left(\{f_i^{\rm eq},g_i^{\rm eq}\} - \{f_i,g_i\}\right) \\ + \left(1-\beta\right)\left(\{f_i^{\star},g_i^{\star}\} - \{f_i^{\rm eq},g_i^{\rm eq}\}\right),
\end{multline}
where,
\begin{equation}
    \mathcal{D}'_t = \frac{\mathcal{L}/\mathcal{T}}{\delta x/\delta t}\left(\partial'_t + \bm{c}'_i\cdot\bm{\nabla}'\right).
\end{equation}
Assuming acoustic scaling and hydrodynamic scaling, {$\varepsilon\sim\delta x/\mathcal{L} \sim\delta t/\mathcal{T}$} and dropping the primes,
\begin{multline}
    \varepsilon\mathcal{D}_t \{f_i,g_i\} + \frac{\varepsilon^2}{2}{\mathcal{D}_t}^2\{f_i,g_i\}  = 2\beta\left(\{f_i^{\rm eq},g_i^{\rm eq}\} - \{f_i,g_i\}\right) 
    \\+ \left(1-\beta\right)\left(\{f_i^{\star},g_i^{\star}\} - \{f_i^{\rm eq},g_i^{\rm eq}\}\right).
\end{multline}

Next, we introduce the following multi-scale expansions: 
\begin{subequations}
\begin{align}
    f_i &= f_i^{(0)} + \varepsilon f_i^{(1)} + \varepsilon^2 f_i^{(2)} + O(\varepsilon^3), \\
    g_i &= g_i^{(0)} + \varepsilon g_i^{(1)} + \varepsilon^2 g_i^{(2)} + O(\varepsilon^3), \\
    {f_i^\star} &= {f_i^\star}^{(0)} + \varepsilon {f_i^\star}^{(1)} + \varepsilon^2 {f_i^\star}^{(2)} + O(\varepsilon^3), \\
    {g_i^\star} &= {g_i^\star}^{(0)} + \varepsilon {g_i^\star}^{(1)} + \varepsilon^2 {g_i^\star}^{(2)} + O(\varepsilon^3), \\
    \partial_t &= \varepsilon \partial_t^{(1)} + \varepsilon^2 \partial_t^{(2)} + O(\varepsilon^3).\label{expansion-time}
\end{align}
\end{subequations}
Noting that for the definition of ${f_i^\star}$, deviation from the equilibrium will arise only at the $\varepsilon^1$ level (i.e. ${f_i^\star}^{(0)} = f^{\rm eq}_i$), we separate by orders of the smallness parameter,

	\begin{subequations}
	\begin{align}
    \varepsilon^0 : &\{f_i^{(0)},g_i^{(0)}\} = \{f_i^{\rm eq},g_i^{\rm eq}\},\\
	\varepsilon^1 : &\mathcal{D}_{t}^{(1)} \{f_i^{(0)},g_i^{(0)}\} \\&= -2\beta \{f_i^{(1)},g_i^{(1)}\} + \left(1-\beta\right)\{{f^\star}_i^{(1)},{g^\star}_i^{(1)}\},\label{eq:CE_order_1}\\
	\varepsilon^2 : &\partial_t^{(2)}\{f_i^{(0)},g_i^{(0)}\} \\ &+ \mathcal{D}_{t}^{(1)}(1-\beta) \left(\{f_i^{(1)},g_i^{(1)}\} + {\frac{1}{2}} \{{f^\star}_i^{(1)},{g^\star}_i^{(1)}\}\right) =  \nonumber \\ &-2\beta \{f_i^{(2)},g_i^{(2)}\} + \left(1-\beta\right)\{{f^\star}_i^{(2)},{g^\star}_i^{(2)}\}.
	\end{align}
    \label{Eq:CE_Eq_orders_LB}
    \end{subequations}
The following solvability conditions apply,
\begin{subequations}
    \begin{gather}
		\sum_{i=1}^Q {f}_i^{(k)} = 0,\ \forall k>0, \label{Eq:CE_solvability_LB_1}\\
		\sum_{i=1}^Q \bm{c}_i {f}_i^{(1)}  + \frac{1}{2}\sum_{i=1}^Q \bm{c}_i {f^\star}_i^{(1)}  = 0, \label{Eq:CE_solvability_LB_2}\\
        \sum_{i=1}^Q \bm{c}_i {f}_i^{(k)}  = 0,\ \forall k>1, \label{Eq:CE_solvability_LB_3}\\
        \sum_{i=1}^Q {g}_i^{(1)} + \frac{1}{2} \sum_{i=1}^Q {g^\star}_i^{(1)} = 0,
	    \label{Eq:CE_solvability_LB_4}\\
        \sum_{i=1}^Q {g}_i^{(k)} = 0,\ \forall k>1.
	    \label{Eq:CE_solvability_LB_5}
    \end{gather}
\end{subequations}

Taking the zeroth-order moment of Eq.~\eqref{eq:CE_order_1} {for $f_i$},
\begin{equation}
    \partial_t^{(1)}\rho + \bm{\nabla}\cdot\rho\bm{u} = 0,
\end{equation}
where we have used,
\begin{equation}
    \sum_{i=1}^{Q} {f_i^\star}^{(1)} = 0,
\end{equation}
and solvability condition \eqref{Eq:CE_solvability_LB_1}. 

For the first-order moment of \eqref{eq:CE_order_1} {for $f_i$},
\begin{multline}
    \partial_t^{(1)}\rho\bm{u} + \bm{\nabla}\cdot\rho\bm{u}\otimes\bm{u} + \bm{\nabla} P = \\-2\beta\underbrace{\left(\sum_{i=1}^Q \bm{c}_if_i^{(1)}+\frac{1}{2}\bm{c}_i{f_i^\star}^{(1)}\right)}_{=0}  + \underbrace{\sum_{i=1}^Q \bm{c}_i{f_i^\star}^{(1)}}_{=\bm{F}},
\end{multline}
where we used \eqref{Eq:CE_solvability_LB_2}.

Taking the zeroth-order moment of \eqref{eq:CE_order_1} for $g_i$,
\begin{multline}
    \partial_t^{(1)}\rho E + \bm{\nabla}\cdot \bm{u}\left(\rho E + P\right) \\= -2\beta\underbrace{\left(\sum_{i=1}^Q g_i^{(1)}+\frac{1}{2}{g_i^\star}^{(1)}\right)}_{=0} + \underbrace{\sum_{i=1}^Q {g_i^\star}^{(1)}}_{=\bm{u}\cdot\bm{F} 
        {+Q}
    }.
\end{multline}
Summarizing the balance equations at order $\varepsilon$,
	\begin{gather}
	    \partial_t^{(1)}\rho + \bm{\nabla}\cdot \rho \bm{u} = 0,\label{eq:LB_Euler_density_balance}
        \\
	    \partial_t^{(1)}\rho \bm{u} + \bm{\nabla}\cdot \left(\rho \bm{u}\otimes\bm{u} + P\bm{I} \right) - \bm{F} = 0,\label{eq:LB_Euler_momentum_balance}
        \\
        \partial_t^{(1)}\rho E + \bm{\nabla}\cdot \rho \bm{u}\left(E + P/\rho\right) - \bm{u}\cdot\bm{F} - Q= 0. \label{eq:LB_Euler_energy_balance}
	\end{gather}
    
The last equation Eq.~\eqref{eq:LB_Euler_energy_balance} can be transformed into a balance equation for internal energy, using ,
\begin{equation}
    \partial_t^{(1)} \mathcal{K} + \bm{\nabla}\cdot(\bm{u} \mathcal{K}) + \bm{u}\cdot\bm{\nabla}P - \bm{u}\cdot\bm{F} = 0,
\end{equation}
as,
{\begin{equation}
    \partial_t^{(1)} \rho e + \bm{\nabla}\cdot\rho \bm{u} e + P\bm{\nabla}\cdot\bm{u} 
        {-Q}  
    = 0.
\end{equation}}
Furthermore, using
\begin{equation}
    de = c_v dT - \left(T \left(\frac{\partial P}{\partial T}\right)_\rho- P\right)\frac{d\rho}{\rho^2},
\end{equation}
and Eq.~\eqref{eq:LB_Euler_density_balance} a balance equation for temperature can be derived as,
\begin{equation}
    \partial_t^{(1)}T + \bm{u}\cdot\bm{\nabla}T + \frac{T}{\rho c_v}\left(\frac{\partial P}{\partial T}\right)_\rho\bm{\nabla}\cdot\bm{u} 
        {- \frac{Q}{\rho c_v}} 
    = 0.
\end{equation}
Finally, using
\begin{equation}
    dP = \left(\frac{\partial P}{\partial \rho}\right)_T d\rho + \left(\frac{\partial P}{\partial T}\right)_\rho dT,
\end{equation}
we can also write a balance equation for pressure as,
\begin{equation}
    \partial_t^{(1)} P + \bm{u}\cdot\bm{\nabla}P + \rho c_s^2\bm{\nabla}\cdot\bm{u} 
        {-\left(\frac{\partial P}{\partial T}\right)_\rho \frac{Q}{\rho c_v}} 
    = 0,
\end{equation}
where,
\begin{equation}
    c_s^2 = \left(\frac{\partial P}{\partial \rho}\right)_T + \frac{T}{c_v\rho^2} \left(\frac{\partial P}{\partial T}\right)_\rho^2.
\end{equation}

At order $\varepsilon^2$, the zeroth order moment of $f_i$ leads to,
\begin{equation}
\partial_t^{(2)} \rho = 0.
\end{equation}

The first order moments lead to
\begin{multline}
    \partial_t^{(2)}\rho\bm{u} + \bm{\nabla}\cdot\left(1-\beta\right)\left[\left(\sum_{i=1}^Q \bm{c}_i\otimes\bm{c}_i {f_i}^{(1)}\right) \right.\\\left. + \frac{1}{2}\left(\sum_{i=1}^Q \bm{c}_i\otimes\bm{c}_i {f^\star}_i^{(1)}\right)\right] = 0.
\end{multline}
Here we can use Eq.~\eqref{eq:CE_order_1} to obtain,
\begin{multline}
    \partial_t^{(2)}\rho\bm{u} + \bm{\nabla}\cdot\left(\frac{1}{2}-\frac{1}{2\beta}\right)\left(\partial_t^{(1)}\Pi_2(f^{(0)}) 
    \right.\\\left. + \bm{\nabla} \cdot \Pi_3(f^{(0)})  - \sum_{i=1}^Q \bm{c}_i\otimes\bm{c}_i {f^\star}_i^{(1)}\right) = 0.\label{eq:epsilon2_mom_1_1}
\end{multline}
Here, $\Pi_2(f^{(0)})$ and $\Pi_3(f^{(0)})$ are the second and third order moment tensor of the equilibrium distribution function,
	\begin{gather}
	    \Pi_2(f^{(0)}) = \rho\bm{u}\otimes\bm{u} + P\bm{I},\\
	    \Pi_3(f^{(0)}) = \rho\bm{u}\otimes(\bm{u}\otimes\bm{u}+P\bm{I})\circ(\bm{1}-\bm{J}) + 3\rho \varsigma^2 \bm{u}\otimes\bm{I}\circ\bm{J}. 
	\end{gather}
Using Eqs.~\eqref{eq:LB_Euler_density_balance} and \eqref{eq:LB_Euler_momentum_balance} we can write,
\begin{multline}
    \partial^{(1)}_t \left(\rho\bm{u}\otimes\bm{u} + P\bm{I}\right) \\= \bm{u}\otimes\bm{F} + \bm{F}\otimes\bm{u}  -\bm{\nabla}\cdot\rho\bm{u}\otimes\bm{u}\otimes\bm{u} - \bm{\nabla}P\bm{u} - (\bm{\nabla}P\bm{u})^\dagger \\  
        {+} 
    P\left(\bm{\nabla}\bm{u} + \bm{\nabla}\bm{u}^\dagger\right)  + \partial^{(1)}_t P \bm{I}.
\end{multline}
For the last term, i.e. $\partial^{(1)}_t P$, we write a balance equation for $P$,
\begin{equation}
    \partial_t^{(1)} P  = \left(P - \rho c_s^2\right)\bm{\nabla}\cdot\bm{u} - \bm{\nabla}\cdot P\bm{u}      {+ \left(\frac{\partial P}{\partial T}\right)_\rho \frac{Q}{\rho c_v} }.
\end{equation}
We also have
\begin{multline}
    \bm{\nabla}\cdot\Pi_3(f^{(0)}) = \left[\bm{\nabla}\cdot\rho\bm{u}\otimes\bm{u}\otimes\bm{u} + \bm{\nabla}P\bm{u} + \bm{\nabla}P\bm{u}^\dagger\right] \\+ \rho\Psi +\bm{I}\bm{\nabla}\cdot P\bm{u},
\end{multline}
where
\begin{equation}
    \Psi_{\alpha\alpha} = -\frac{1}{\rho}\partial_\alpha u_\alpha\left(u_\alpha^2 + 3(P-\rho\varsigma^2)\right).
\end{equation}
Adding up all the terms,
\begin{multline}
    \partial_t^{(1)}\Pi_2(f^{(0)}) + \bm{\nabla}\cdot\Pi_3(f^{(0)}) = \bm{u}\otimes\bm{F} + {\bm{F}\otimes\bm{u}} + P\left(\bm{\nabla}\bm{u} + \bm{\nabla}\bm{u}^\dagger\right) \\+ \left(P - \rho c_s^2\right)\bm{\nabla}\cdot\bm{u}\bm{I}  
        {+ \left(\frac{\partial P}{\partial T}\right)_\rho \frac{Q}{\rho c_v} \bm{I}} 
    +  \rho\Psi.
\end{multline}
Setting,
\begin{multline}
    \sum_{i=1}^Q \bm{c}_i\otimes\bm{c}_i {f^\star}_i^{(1)} = \Bigg(\bm{u}\otimes\bm{F} + \bm{F}\otimes\bm{u} 
        {+ \left(\frac{\partial P}{\partial T}\right)_\rho \frac{Q}{\rho c_v} \bm{I}} 
    \Bigg.\\\Bigg. + \rho\Psi + P\left(\frac{D+2}{D}-\frac{\rho c_s^2}{P} -\frac{\eta}{\mu}\right)\bm{\nabla}\cdot\bm{u}\bm{I}\Bigg) 
\end{multline}
and plugging it back into Eq.~\eqref{eq:epsilon2_mom_1_1},
\begin{equation}
    \partial_t^{(2)}\rho\bm{u} + \bm{\nabla}\cdot \bm{T}_{\rm NS}  = 0,
\end{equation}
where the Navier-Stokes stress tensor is 
\begin{equation}
\bm{T}_{\rm NS} =  -\mu\left(\bm{\nabla}\bm{u} + \bm{\nabla}\bm{u}^\dagger - \frac{2}{D}\bm{\nabla}\cdot\bm{u} \bm{I}\right) - \eta \bm{\nabla}\cdot\bm{u}\bm{I},
\end{equation}
and we used
\begin{equation}
    \mu = \left(\frac{1}{2\beta}-\frac{1}{2}\right) P \delta t.
\end{equation}

For the second population at order $\varepsilon^2$, the zeroth order moment leads to
\begin{multline}
    \partial_t^{(2)}\rho E + \bm{\nabla}\cdot\left(\frac{1}{2}-\frac{1}{2\beta}\right)\Bigg( \partial_t^{(1)}\Pi_1(g_i^{(0)}) \Bigg.\\\Bigg. + \bm{\nabla}\cdot\Pi_2(g_i^{(0)})  - \sum_{i=1}^Q \bm{c}_i {g_i^\star}^{(1)}\Bigg) = 0,
\end{multline}
where,
\begin{gather}
    \Pi_1(g_i^{(0)}) = \bm{u}(\rho E + P),\\
    \Pi_2(g_i^{(0)}) = \bm{u}\otimes\bm{u}\left(\rho E+2P\right) + \left(E + P/\rho\right)P\bm{I}.
\end{gather}
Here we can use,
\begin{multline}\label{eq:pu_balance}
    \partial_t^{(1)}P\bm{u} + \bm{\nabla}\cdot P\bm{u}\otimes\bm{u} + \frac{P}{\rho}\bm{\nabla}P -\frac{P}{\rho}\bm{F} + \bm{u}\left(\rho c_s^2 - P\right)\bm{\nabla}\cdot\bm{u} 
    \\ 
        {- \bm{u} \left(\frac{\partial P}{\partial T}\right)_\rho \frac{Q}{\rho c_v}} 
    = 0,
\end{multline}
and
\begin{multline}\label{eq:Eu_balance}
    \partial_t^{(1)}\rho E\bm{u} + \bm{\nabla}\cdot (\rho E + P)\bm{u}\otimes\bm{u} \\ + E\bm{\nabla}P - E\bm{F} - P\bm{u}\cdot\bm{\nabla}\bm{u} - \bm{u}(\bm{u}\cdot\bm{F} 
        {+ Q}
    )= 0.
\end{multline}
Adding up both contributions,
\begin{multline}
    \partial_t^{(1)}\Pi_1(g_i^{(0)}) + \bm{\nabla}\cdot\Pi_2(g_i^{(0)}) \\ = \underbrace{-2\bm{\nabla}\cdot P \bm{u}\otimes\bm{u} + 2\bm{\nabla}\cdot P \bm{u}\otimes\bm{u}}_{=0} \underbrace{+ \bm{\nabla}\cdot \rho E \bm{u}\otimes\bm{u} - \bm{\nabla}\cdot \rho E \bm{u}\otimes\bm{u}}_{=0}
     \\ \underbrace{+ \bm{\nabla} P(P/\rho + E) - (E+P/\rho)\bm{\nabla}P}_{=P\bm{\nabla}(E+P/\rho)} + P\bm{u}\cdot\bm{\nabla}\bm{u} - \bm{u}\left(\rho c_s^2 - P\right)\bm{\nabla}\cdot\bm{u}
     \\ 
        {+ \bm{u} \left(\frac{\partial P}{\partial T}\right)_\rho \frac{Q}{\rho c_v}}
    +(P/\rho+E)\bm{F} + \bm{u}(\bm{u}\cdot\bm{F} 
        {+Q}
    ).
\end{multline}
Further expanding,
\begin{multline}
    \partial_t^{(1)}\Pi_1(g_i^{(0)}) + \bm{\nabla}\cdot\Pi_2(g_i^{(0)}) \\= 
     P\bm{\nabla}h +\underbrace{ P\bm{\nabla}(\bm{u}^2/2) + P\bm{u}\cdot\bm{\nabla}\bm{u}}_{=P\bm{u}\cdot(\bm{\nabla}\bm{u} + \bm{\nabla}\bm{u}^\dagger)} + \bm{u}\left(P - \rho c_s^2\right)\bm{\nabla}\cdot\bm{u}
     \\ 
        {+ \bm{u} \left(\frac{\partial P}{\partial T}\right)_\rho \frac{Q}{\rho c_v}} 
    +(P/\rho+E)\bm{F} + \bm{u}(\bm{u}\cdot\bm{F} 
        {+Q}
    ),
\end{multline}
where $h = e + P/\rho$. Plugging this back into the balance equation,
\begin{multline}
    \partial_t^{(2)}\rho E + \bm{\nabla}\cdot\underbrace{\bm{u}\cdot\left[-\mu\left(\bm{\nabla}\bm{u} + \bm{\nabla}\bm{u} - \frac{2}{D}\bm{\nabla}\cdot\bm{u}\bm{I}\right) - \eta\bm{\nabla}\cdot\bm{u}\bm{I}\right]}_{=\bm{u}\cdot\bm{T}_{\rm NS}} \\ + \bm{\nabla}\cdot\left(\frac{1}{2}-\frac{1}{2\beta}\right)\left(P\bm{\nabla}h  + \bm{u}\cdot P\left(\frac{D+2}{D}-\frac{\rho c_s^2}{P}-\frac{\eta }{\mu}\right)\bm{\nabla}\cdot\bm{u}\bm{I}
    \right. \\ \left. + (P/\rho+E)\bm{F} + \bm{u} \left(\bm{u}\cdot\bm{F} 
        {+ \left(\frac{\partial P}{\partial T}\right)_\rho \frac{Q}{\rho c_v}+ Q}
    \right) \right.\\\left.- \sum_{i=1}^Q \bm{c}_i {g_i^{\star}}^{(1)}\right) = 0.
\end{multline}
Using the definition of the shifted equilibrium,
\begin{equation}
    \Pi_0(g_i^{\star(1)}) = \bm{u}\cdot\bm{F} 
        {+ Q},
\end{equation}
and 
\begin{multline}
    \Pi_1(g_i^{\star(1)}) = \bm{u}\left(\bm{u}\cdot\bm{F} 
        {+ \left(\frac{\partial P}{\partial T}\right)_\rho \frac{Q}{\rho c_v}+ Q}
    \right) + \bm{F} \left(\frac{P}{\rho} + E\right) 
    \\
    + P\left(\bm{\nabla}h - \frac{k}{\mu}\bm{\nabla}T + \bm{u} \cdot\left(\frac{D+2}{D}-\frac{\rho c_s^2}{P}-\frac{\eta}{\mu}\right)\bm{\nabla}\cdot\bm{u}\bm{I}\right),
\end{multline}
we recover
\begin{equation}
    \partial_t^{(2)}\rho E + \bm{\nabla}\cdot \bm{u}\cdot\bm{T}_{\rm NS} + \bm{\nabla}\cdot \underbrace{(-k\bm{\nabla}T)}_{\bm{q}_{\rm F} } = 0,
\end{equation}
where the Fourier heat flux is
\begin{equation}
    \bm{q}_{\rm F} = - k \bm{\nabla}T.
\end{equation}
Finally, summing up balance equations at order $\varepsilon$ and $\varepsilon^2$ and using Eq. \eqref{expansion-time} yields
\begin{gather}
    \partial_t\rho + \bm{\nabla}\cdot \rho \bm{u} = 0,
    \\
    \partial_t\rho \bm{u} + \bm{\nabla}\cdot \left(\rho \bm{u}\otimes\bm{u} + P\bm{I} + \bm{T}_{\rm NS} \right) - \bm{F}  = 0,
    \\
    \partial_t\rho E + \bm{\nabla}\cdot\left( \rho \bm{u}\left(E + P/\rho\right) + \bm{u}\cdot\bm{T}_{\rm NS} +  \bm{q}_{\rm F} \right) - \bm{u}\cdot\bm{F} - Q= 0.
\end{gather}
This concludes the multiscale analysis of the proposed lattice Boltzmann model.

\section{Model consistency without external body force and heat source\label{app:Modeconsistency}}

    A brief validation of the dispersion and dissipation modes of the compressible lattice Boltzmann model in the non-ideal single-phase regime, without addition of body force and heat source terms, is conducted hereafter for the sake of completeness; the reader is referred to \cite{hosseini_2025_compressiblenonideal} for more detailed evaluations.
    For all benchmarks in the following, we consider a van der Waals fluid fitted to the critical properties of nitrogen (N\textsubscript{2}) (cf. Table \ref{tab:nitrogen_properties}).

    \begin{figure*}[htbp!]
        \centering
        \includegraphics[width=0.49\linewidth]{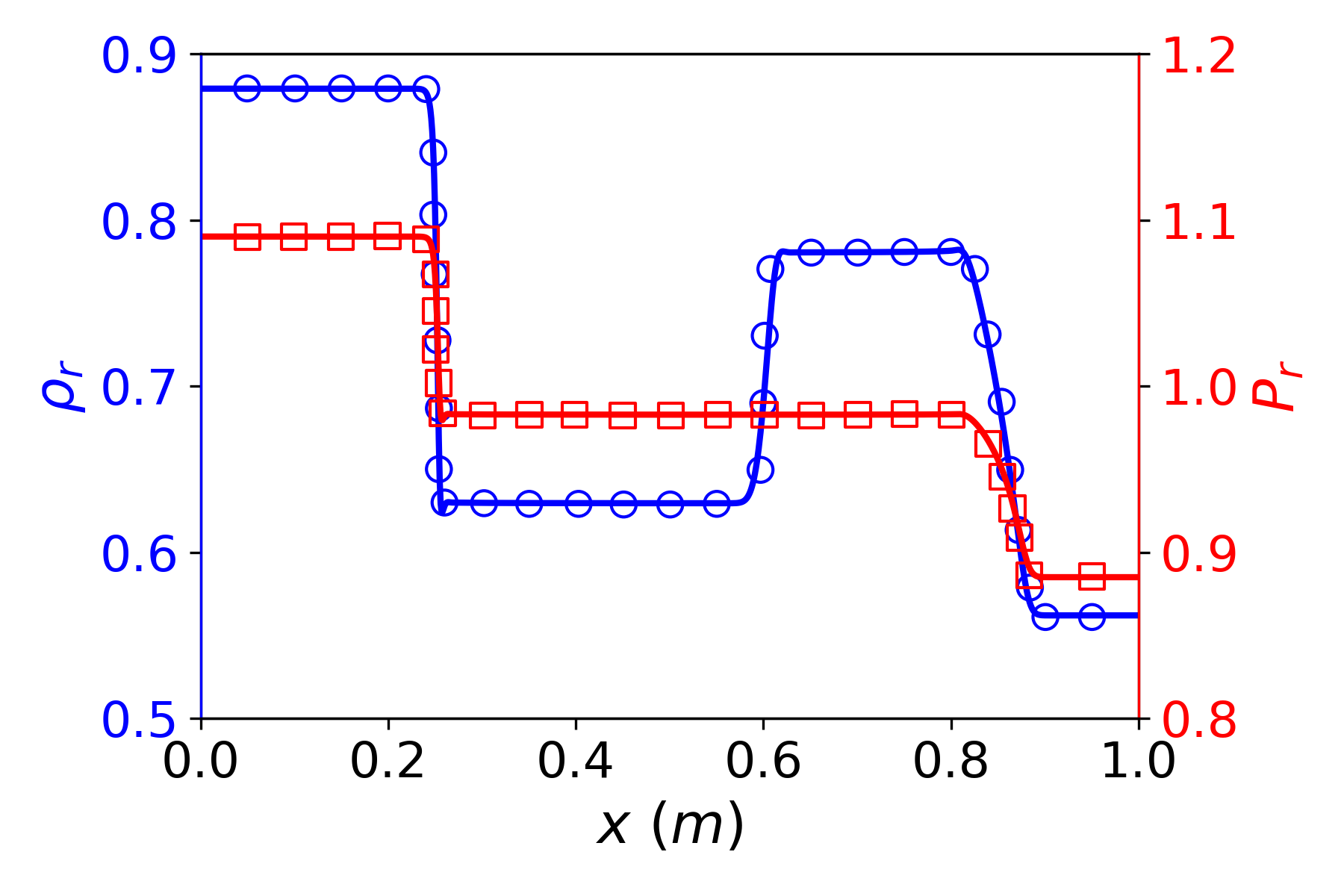}\phantom{a}%
        \includegraphics[width=0.49\linewidth]{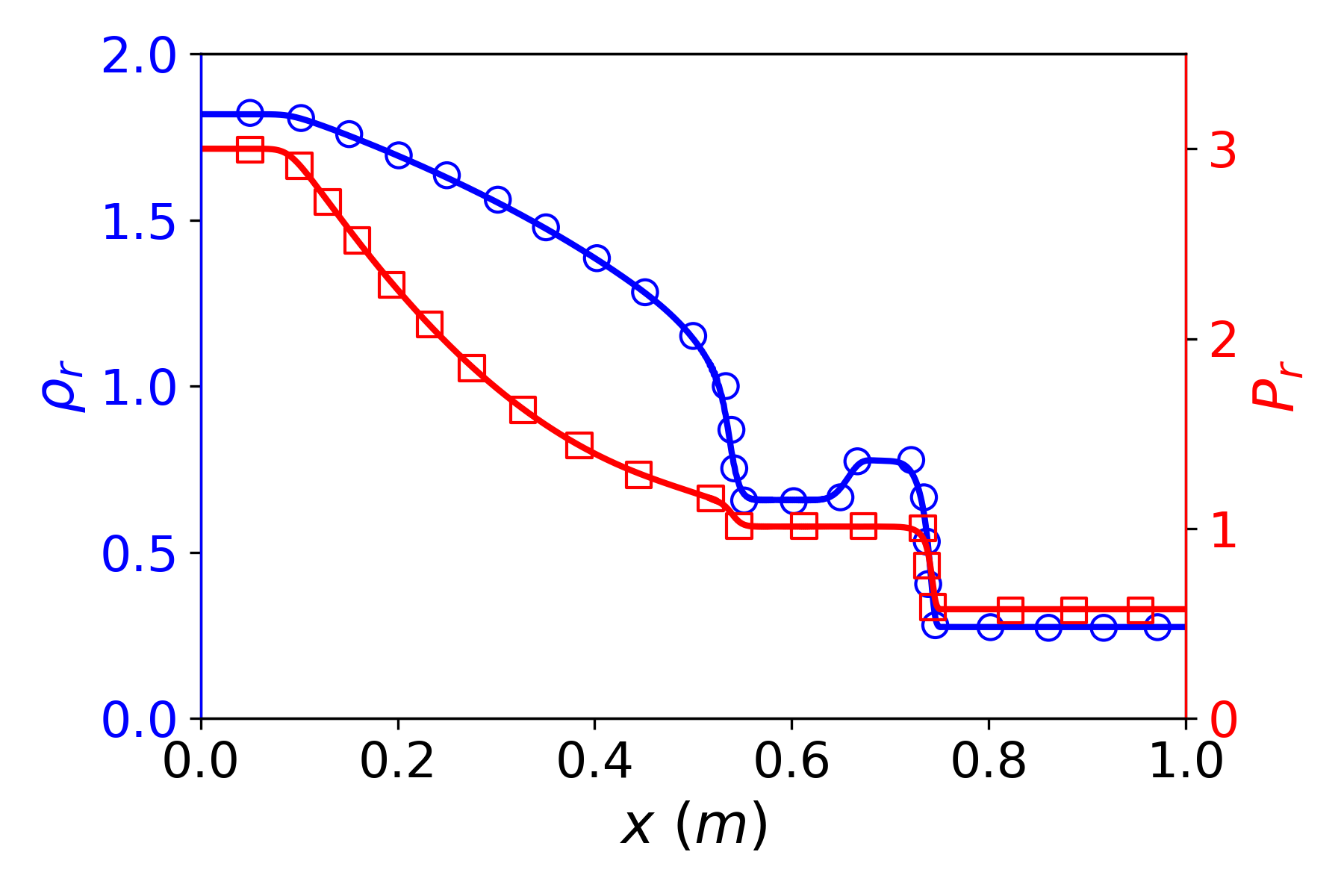}%
        \caption{Solutions of reduced density and pressure (solid lines) compared to the reference data (markers) from \cite{GUARDONE_2002_Roe} for (left) the shock tube $\mathrm{I}$ at time $t = 0.45 \ L_x \rho_{cr} / P_{cr}$ and (right) for the shock tube $\mathrm{II}$ at time $t = 0.15 \ L_x \rho_{cr} / P_{cr}$. Markers $\mcirc$ and $\square$ represent the reduced density and pressure fields respectively.}
        \label{fig:NoForce_NIShockTubes}%
    \end{figure*}
        
    \subsection{Dispersion: Non-ideal shock tubes with non-classical wave fields}
       
        First, we probe the proper recovery of the dispersion rates together with accurate recovery of dispersion dominated systems.
        For this, we investigate non-classical wave fields using two non-ideal shock tube cases in which the flow is either fully or partially within the regime of negative nonlinearity of the fundamental derivative of gas dynamics \cite{fundamentalderivative_Thomson}:
        \begin{equation}
            \Gamma = 1 + \frac{\rho}{c_s}\biggl(\frac{\partial \rho}{\partial x}\biggr)_s.
        \end{equation}
        Note that, for $\Gamma > 0$, a flow exhibits positive nonlinearity, in which disturbances steepen in the forward direction and eventually form compression shocks. In contrast, for $\Gamma < 0$, the flow exhibits negative nonlinearity, in which disturbances steepen in the opposite direction, leading to the formation of expansion shocks. In these regimes, gases exhibit a range of non-classical phenomena.
             
        In both shock tube cases, the simulations are performed on a one-dimensional domain of size $L_x = 1$ m with a spatial resolution of $\delta x = 1$ mm. The simulations are initialized with zero velocity everywhere and initial discontinuity in pressure and density at $x = L_x/2$, with the parameters listed in Table \ref{tab:shock_tube_parameters}. 
        \begin{table}[h]
            \caption{List of initial conditions for the shock-tube cases.}
            \centering
            \setlength{\tabcolsep}{10pt}
            \begin{tabular}{cccc}
            \hline
            Case & $R/c_v$ & $(P_r, \rho_r)_{\text{left}}$ & $(P_r, \rho_r)_{\text{right}}$    \\ \hline
            I    & 0.0125  & (1.09, 0.879)                  & (0.885, 0.562)                     \\
            II   & 0.0125  & (3, 1.818)                     & \multicolumn{1}{c}{(0.575, 0.275)} \\ \hline
            \end{tabular}
            \label{tab:shock_tube_parameters}
        \end{table}
        
        Fig. \ref{fig:NoForce_NIShockTubes} compares the simulation results with reference data from \cite{GUARDONE_2002_Roe}. Case I represents a typical configuration in non-classical gas dynamics, in which a compression fan and an expansion shock form instead of the classical combination of a compression shock and an expansion wave. In Case II, a non-classical rarefaction front propagates leftwards; it initially resembles a rarefaction fan but progressively sharpens near $x = 0.5$ m. Simultaneously, a compression front propagates rightwards and steepens into a shock. Both cases show excellent agreement with the reference data. For a more detailed physical interpretation of these results, see \cite{hosseini_2025_compressiblenonideal} for instance.
      
    \subsection{Dissipation: Thermal Couette flow in single-phase regime}

        \begin{figure*}[htbp!]
            \centering
            \includegraphics[width=0.49\linewidth]{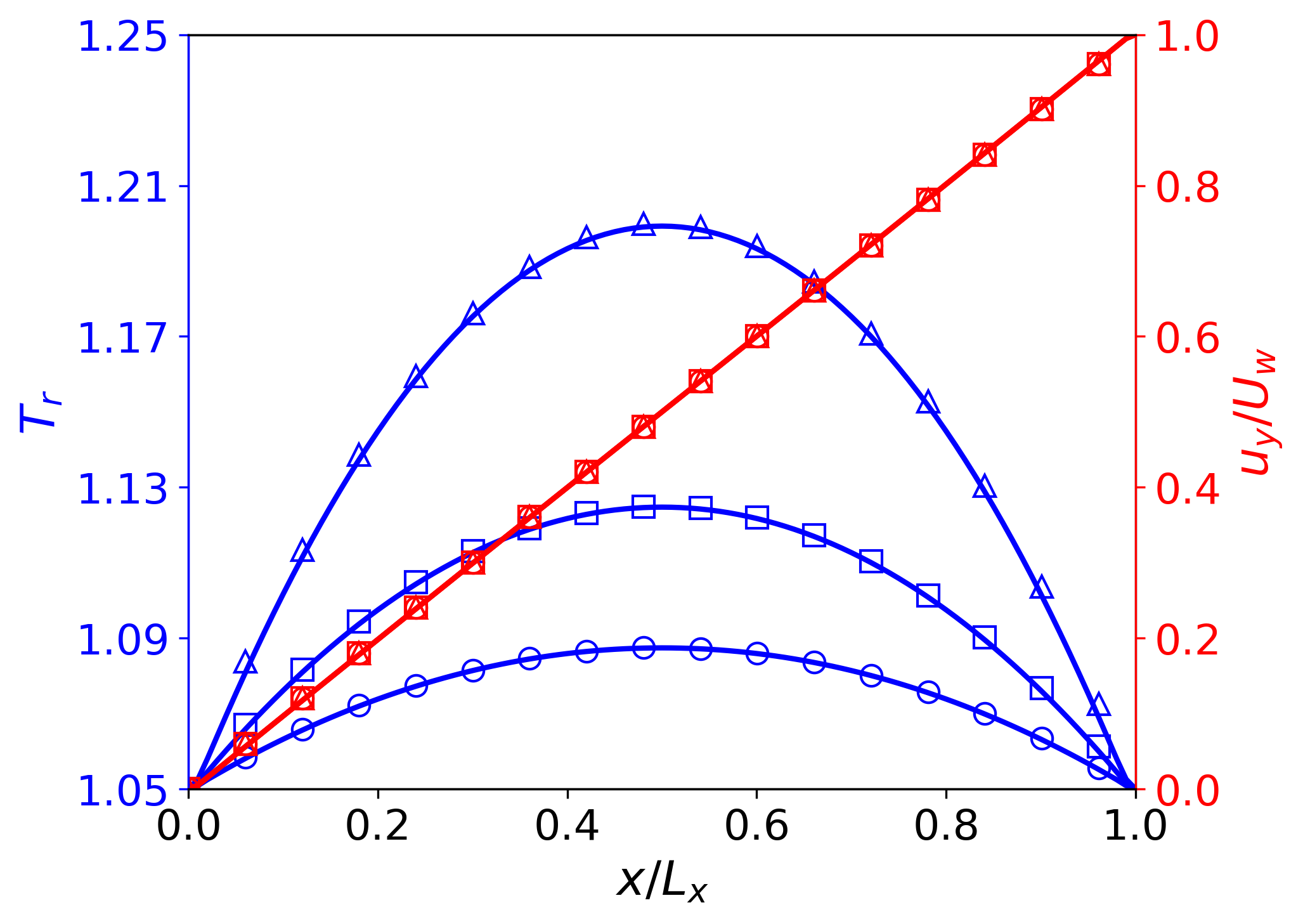}\phantom{a}%
            \includegraphics[width=0.49\linewidth]{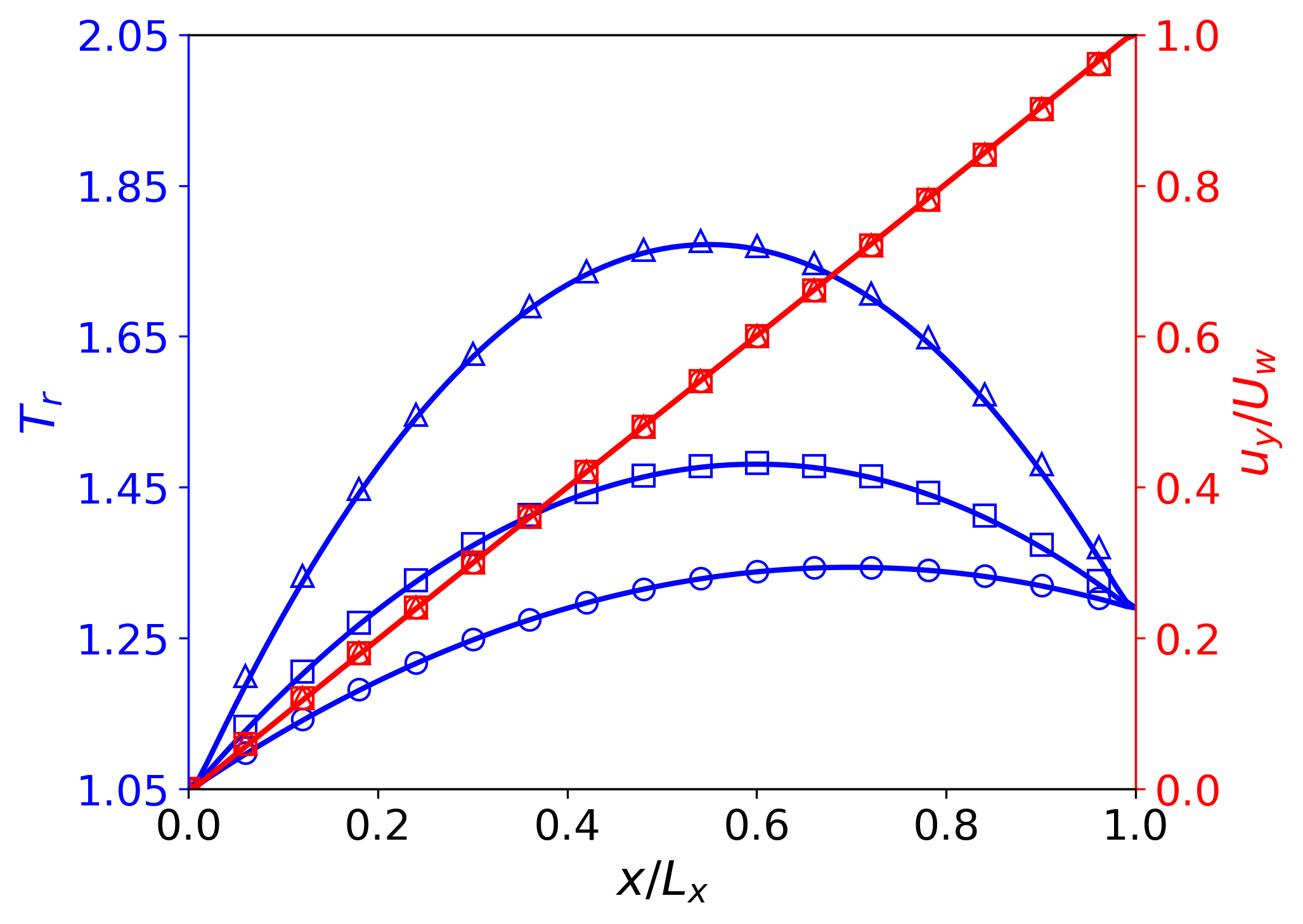}\\
            \caption{Solutions of reduced velocity and reduced temperature for the Couette problem at (left) $\mathrm{Ma} = 0.75, \mathrm{Ec} \to \infty$ and (right) $\mathrm{Ma} = 1.5, \mathrm{Ec} = 10$. Markers $\mcirc$, $\square$ and $\triangle$ are analytical solutions for $\mathrm{Pr} \in \{0.5, 1.0, 2.0\}$ respectively.}
            \label{fig:NoForce_Couette}%
        \end{figure*}
        
        Next, we probe the proper recovery of dissipation rates and their interplay together with accurate recovery of dissipation dominated systems.
        For this, the thermal Couette flow is the applied benchmark problem, as it captures the combined effects of viscous momentum dissipation, viscous heating, and thermal diffusion, governed by the corresponding Prandtl number. 
        A no-slip wall with a higher temperature $T_{\rm W} = T_{\rm H}$ on the right side of the domain $x \in [0,L_x]$ is in motion with a constant speed in $y$ direction $u_{y,\mathrm{W}} = U_\mathrm{W}$, while a left no-slip wall is at rest, $u_{y,\mathrm{W}} = 0$, with a temperature $T_{\mathrm{W}} = T_{\rm C}$.
        
        While periodic BCs were enforced in $y$-direction, the BC at the no-slip walls were implemented following the setup in \cite{strässle2025consistent-compressible}: the non-equilibrium extrapolation approach \cite{Bhaduria23, Guo_2002_BCextrapolation} was used by splitting the boundary populations into an equilibrium and non-equilibrium part (see also \cite{KarlinTwoPop, Saadat2019, ProbingDoubleDist2024} for other valuable approaches in combination with the thermal Couette case). This treatment is used for all wall boundary conditions considered in this work.
        The approach consists of the following steps, with the indices $\mathrm{B}$, $\mathrm{N}$ and $\mathrm{W}$ denoting the boundary node, neighboring node and the wall respectively.
        First, the density at the boundary node is determined by simply extrapolating from the neighboring node as $\rho_{\mathrm{B}} = \rho_{\mathrm{N}}$.
        Thereafter, the equilibrium is computed using the imposed values of velocity and temperature at the wall, and the non-equilibrium part is extrapolated from the neighboring nodes, yielding the expressions for both $f_i$ and $g_i$ populations as
        \begin{align}
            f_{i,\mathrm{B}} &= f_i^{\rm eq}(\rho_{\rm B}, \bm{u}_{\rm W}, T_{\rm W}) + f_i^{\rm neq}(\rho_{\rm N}, \bm{u}_{\rm N}, T_{\rm N}),
            \\
            g_{i,\mathrm{B}} &= g_i^{\rm eq}(\rho_{\rm B}, \bm{u}_{\rm W}, T_{\rm W}) + g_i^{\rm neq}(\rho_{\rm N}, \bm{u}_{\rm N}, T_{\rm N}).
        \end{align}
        
        Simulations were run for a domain of length $L_x = 1$ mm with $\delta x = L_x / 100$. The system pressure was set to $P_0 = 3 \ P_{cr}$, the lower wall temperature to $T_{\rm C} = 1.05 \ T_{cr}$, and the dynamic viscosity to $\mu = 10^{-3}$ Pa$\cdot$s.
        The cases differ by the Mach and Eckert number, as $\mathrm{Ma}  = U_{\rm W}/ c_s(T_{\rm C}, P_0) = \{ 0.75, 1.5\}$ and $\mathrm{Ec} = U_{\rm W^2}/(C_P(T_{\rm C}, P_0) \ [T_{\rm H} - T_{\rm C}]) = \{\infty, 10 \}$.
        The simulations were initialized with a uniform pressure and linear profiles in temperature and velocity as
        \begin{equation}
          (u_y, T) = (U_{\rm W}\frac{x}{L_x},  T_{\rm C} + (T_{\rm H} - T_{\rm C})\frac{x}{L_x}), \quad 0 \leq x \leq L_x,
        \end{equation}
        and were run until the solution converged.

        Fig. \ref{fig:NoForce_Couette} depicts the velocity and reduced temperature profiles in each case of $\{\mathrm{Ma}, \mathrm{Ec}\}$ for three different Prandtl numbers, $\mathrm{Pr} \in \{0.5, 1.0, 2.0\}$.
        The analytical solution for the temperature can be written as \cite{Roshko1957ElementsOG}
        \begin{equation}
            \frac{T-T_{\rm C}}{T_{\rm H}-T_{\rm C}} = \frac{x}{L_x} + \frac{\mathrm{Pr} \ \mathrm{Ec}}{2} \frac{x}{L_x} \left(1-\frac{x}{L_x}\right).
        \end{equation} 
        It can clearly be seen that the results are in excellent agreement with the analytical reference solutions.

\section*{References}

%

\end{document}